\def\Nc{N_{\rm c}}
\def\q{{\bm q}}
\def\x{{\bm x}}
\def\y{{\bm y}}
\def\j{{\bm j}}
\def\A{{\bm A}}
\def\Q{{\bm Q}}
\def\tr{\operatorname{tr}}
\def\sgn{\operatorname{sign}}
\def\gammaE{\gamma_{\rm E}^{}}

\def\Im{\operatorname{Im}}
\def\grad{{\bm\nabla}}
\def\Tan{\operatorname{Tan}}
\def\Tanh{\operatorname{Tanh}}
\def\Ai{\operatorname{Ai}}
\def\hyperF{\operatorname{\it F}}
\def\Res{\operatorname{Res}}

\def\gYM{g_{\rm YM}}
\def\gSG{g_{\rm SG}}
\def\envelope{\Lambda}
\def\pol{\bar\varepsilon}
\def\kbig{\bar k}
\def\wbig{E}
\def\Aamp{{\cal N}_A}
\def\Acl{A_{\rm cl}}
\def\Aclplus{A_{\rm cl(+)}}
\def\Charge{{\cal Q}}
\def\Chargebar{\bar\Charge}
\def\charge{\Charge^{(3)}}
\def\chargec{\Charge^{(c)}}
\def\chargebar{{\Chargebar}^{(3)}}

\def\dep{\Theta}
\def\depint{\Sigma_\dep}
\def\formL{\Psi}
\def\num{c}
\def\numpole{c_1}
\def\numres{c_2}
\def\calGE{{\cal G}_{\rm E}}
\def\calGR{{\cal G}_{\rm R}}
\def\calGA{{\cal G}_{\rm A}}
\def\calGEup{{\cal G}^{\rm E}}
\def\calGRup{{\cal G}^{\rm R}}
\def\calGAup{{\cal G}^{\rm A}}
\def\hatcalGRup{{\hat{\cal G}}^{\rm R}}
\def\hathatcalGRup{{\bar{\cal G}}^{\rm R}}
\def\uB{u_{\rm B}}

\def\resp{\left\langle j^{(3)\mu} \right\rangle}
\def\respox{\left\langle j^{(3)0}(x) \right\rangle}
\def\respx{\left\langle j^{(3)\mu}(x) \right\rangle}
\def\responseo{\left\langle j^{(3)0}(x) \right\rangle_{\Acl}}
\def\responsec{\left\langle j^{(c)\mu}(x) \right\rangle_{\Acl}}
\def\response{\left\langle j^{(3)\mu}(x) \right\rangle_{\Acl}}
\def\jmu{\left\langle j^{\mu}(x) \right\rangle}
\def\jo{\left\langle j^{0}(x) \right\rangle}
\def\ub{\bar u}
\def\Field{{\cal A}}
\def\econj{*} 
\def\NR{{\rm NR}}
\def\finv{{\cal F}}
\def\tauo{\tau}
\def\umatch{u_{\rm match}}

\documentclass[prd,preprint,eqsecnum,nofootinbib,amsmath,amssymb,
               tightenlines,dvips]{revtex4}
\usepackage{graphicx}
\usepackage{bm}
\usepackage{amsfonts}

\def\funG{{\mathfrak G}}

\begin {document}



\title
    {
      Jet quenching in hot strongly coupled gauge theories revisited:
      3-point correlators with gauge-gravity duality
    }

\author{
  Peter Arnold and Diana Vaman
}
\affiliation
    {%
    Department of Physics,
    University of Virginia, Box 400714,
    Charlottesville, Virginia 22904, USA
    }%

\date {\today}

\begin {abstract}%
{%
   Previous studies of high-energy jet stopping in strongly-coupled
   plasmas have lacked a clear gauge-theory specification of the
   initial state.  We show how to set up a well-defined gauge theory
   problem to study jet stopping
   in pure ${\cal N}{=}4$ super Yang Mills theory (somewhat analogous
   to Hofman and Maldacena's studies at zero temperature) and solve
   it by using gauge-gravity duality for real-time, finite-temperature
   3-point correlators.
   Previous studies have found that the stopping distance scales
   with energy as $E^{1/3}$ (with disagreement on the gauge
   coupling dependence).
   We do find that none of the jet survives beyond this scale, but
   we find that {\it almost}\/ all of our jet stops at a parametrically
   smaller scale proportional to $(E L)^{1/4}$, where $L$ is the
   size of the space-time region
   where the jet is initially created.
}%
\end {abstract}

\maketitle
\thispagestyle {empty}


\section {Introduction and Results}
\label{sec:intro}

  How far does a localized, high-energy excitation travel in a
quark-gluon plasma before slowing down, stopping, and thermalizing?
This question is of phenomenological interest for heavy ion collisions,
and it has long been a problem of theoretical interest
to calculate the result in various idealized situations.
In a weakly-coupled gauge theory with massless partons,
the stopping distance scales
with energy as $E^{1/2}$, up to powers of logarithms, where $E$ is the
initial energy of a high-energy parton.%
\footnote{
  A specific calculation for QCD
  of the stopping distance at weak coupling in
  the high-energy limit may be found in Ref.\ \cite{stop}.
  However, the scaling of this result was
  implicit in the early pioneering work of
  Refs.\ \cite{BDMPS,Zakharov} on
  bremsstrahlung and energy loss rates in QCD plasmas.
}
At the other extreme,
investigations
\cite{GubserGluon,HIM,CheslerQuark}
of strongly-coupled, large-$\Nc$, supersymmetric
versions of QCD, using gauge-gravity duality
\cite{duality1,Witten,duality3,dualityT},
have indicated that the maximum stopping distance scales
like $E^{1/3}$.
In this paper, we revisit this gauge-gravity duality result.
In particular, previous calculations have always specified the
high-energy initial state using the gravity description (and
in some cases relegated other matters of interpretation to
the gravity description): there has not been a complete specification
of the problem, from beginning to end, solely in
terms of 4-dimensional gauge theory.
We will investigate what happens if a localized, high-energy excitation
is created in the gauge theory, and the response later measured
in the gauge theory.
For our method of creating the initial excitation,
we find that there is an additional scale characterizing the response:
almost all of the
excitation's conserved charge is deposited at a distance that
scales with energy as $E^{1/4}$ rather than $E^{1/3}$ and
is sensitive to the initial spatial size of the excitation.
Nonetheless, we will still see $E^{1/3}$ emerge as the furthest
distance that any non-negligible fraction of the charge propagates
before stopping and thermalizing.

To be more concrete, we need to
explain more precisely what we calculate.


\subsection {The problem}

The specific theory we study is pure ${\cal N}{=}4$ super Yang Mills
theory in the large $\Nc$ and large $\lambda\equiv \gYM^2\Nc$ limit.
Readers needing a general introduction to the use of gauge-gravity
duality to study finite-temperature physics in this and similar
strongly-coupled theories should refer to Ref.\ \cite{ViscReview}.

\begin {figure}
\begin {center}
  \includegraphics[scale=0.4]{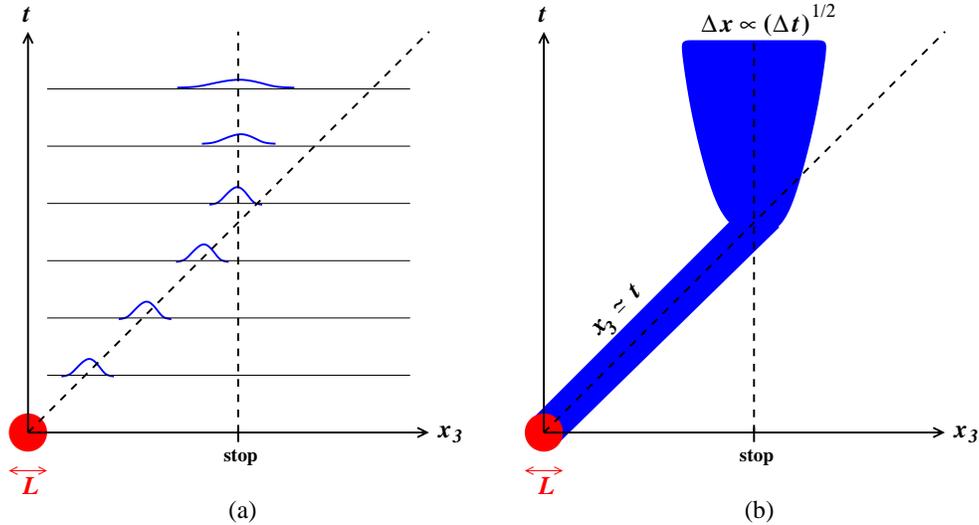}
  \caption{
     \label{fig:evolution}
     The space-time development of a conserved charge density carried by an
     initial high-energy excitation that is moving along the light cone
     and interacting with the thermal medium.  The development
     transitions between a ballistic trajectory at early times to
     diffusion at late times.  (a) shows sketches of density vs. $x$
     at a sequence of larger and larger times; (b) depicts the space-time
     region where the density is non-negligible.  The red circle at
     the origin of space-time denotes the region of size $L$ where the
     source (\ref{eq:source}) that creates the initial excitation
     is non-negligible.
  }
\end {center}
\end {figure}

We will follow the general philosophy of Chesler et al.\
\cite{CheslerQuark,CJK} that
the way to study stopping distances is to locally create
a high-energy excitation and then measure the subsequent evolution
of conserved charge densities such as energy or momentum density.
A cartoon of this evolution is shown in fig.\ \ref{fig:evolution}.
In contrast to previous studies at finite temperature $T$,
we will give an explicit gauge
theory prescription for creating the initial excitation.
We excite the gauge theory plasma by turning on external
sources, localized in space-time, that produce a
high-energy state with nearly-definite energy and momentum.
One could in principle use most any type of source that has a simple
translation to the gravity dual theory under the AdS/CFT correspondence,
but in this paper we will focus on an example where the sources
couple to the global R-charge currents of the gauge theory.
It will also simplify the analysis to use a source that is
translation invariant in the two spatial dimensions transverse
to the motion of the excitation, but localized in time and
the third space direction.  Specifically, we modify the
4-dimensional field theory
Lagrangian by
\begin {equation}
   {\cal L} \to {\cal L} + j_\mu^a \Acl^{a\mu} ,
\label {eq:Lsource}
\end {equation}
where $j_\mu^a$ are the SU(4) R-charge currents of the
theory and $\Acl$ is a classical external source.
We choose the external source to have the form of (i) a high-energy
plane wave $e^{i\kbig\cdot x}$ times (ii) a smooth, slowly varying,
real-valued envelope
function $\envelope_L(x)$ localizing the source to a space-time
region of size $L$.  Specifically,
\begin {equation}
   \Acl^{\mu}(x)
   = \pol^\mu \Aamp \Bigl[
       \frac{\tau^+}{2} \, e^{i \kbig\cdot x} +
       {\rm h.c.}
      \Bigr] \, \envelope_L(x) ,
\label{eq:source}
\end {equation}
where
\begin {equation}
   \kbig^\mu = (\wbig,0,0,\wbig)
\label {eq:kbig}
\end {equation}
is a very large light-like 4-momentum with frequency
$\wbig \gg T$;
$\Aamp$ is an arbitrarily small source amplitude;
$\pol$ is a transverse linear polarization, such as
\begin {equation}
   \pol^\mu = (0,1,0,0) ;
\end {equation}
and $\tau^i$ are Pauli matrices
for any SU(2) subgroup of the
SU(4) R-symmetry, with
\begin {equation}
   \tau^\pm = \tau^1 \pm i \tau^2 .
\label {eq:taupm}
\end {equation}
(The motivation for the $\tau^+$ factor will be discussed below).
A simple example of an appropriate envelope function would be
\begin {equation}
   \envelope_L(x)
   = e^{-\frac12 (x_0/L)^2} e^{-\frac12 (x_3/L)^2} .
\label {eq:Genvelope}
\end {equation}
$L$ should be chosen large compared to $1/\wbig$, so that the
momentum components in the source are all close to
(\ref{eq:kbig}), but small compared to the large stopping distance
that we wish to study.  To avoid the mental clutter of an
over-abundance of scales, it is convenient (but not necessary)
to consider $L$ to be of order $1/T$ in what follows.

\begin {figure}
\begin {center}
  \includegraphics[scale=0.6]{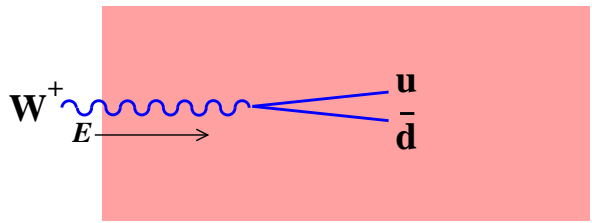}
  \caption{
     \label{fig:W}
     A very high energy $W^+$ boson decaying inside a standard-model
     quark-gluon plasma and producing high-energy partons moving to
     the right with net 3rd component of isospin, $\tau^3/2$.
     In the context of ${\cal N}{=}4$ super Yang Mills,
     the $u$ and $\bar d$ above
     represent adjoint-color fermions or scalars carrying R charge
     and, for strong coupling, should not be pictured
     perturbatively as in this picture.
  }
\end {center}
\end {figure}

The source (\ref{eq:source}) creates an excitation that carries
energy, momentum, and R charge.  We could subsequently track
the densities of any of these conserved charges to study the
evolution of the excitation.  In this paper, we have chosen
to study the evolution of the R charge density, specifically
the large-time behavior ($t \gg$ both $T^{-1}$ and $L$) of
\begin {equation}
  \responseo
\label {eq:measurement}
\end {equation}
if the system starts in thermal equilibrium
at $t=-\infty$.
Here, the superscript ``$(3)$'' indicates the R charge current associated
with $\tau^3/2$ in the SU(2) subgroup referenced by
(\ref{eq:taupm}), and the subscript ``$\Acl$'' indicates that the
expectation is taken with the source term (\ref{eq:Lsource}) present
in the Lagrangian.
Because of the $\tau^+$ in (\ref{eq:source}),
the situation is analogous to the interaction of
a quark-gluon plasma with
an external, high-energy $W^+$ boson in a wavepacket of size
$L$ (and with a decay time of order $L$), as depicted in
fig.\ \ref{fig:W}.%
\footnote{
   One could imagine strengthening this analogy by gauging an
   SU(2) subgroup of the ${\cal N}{=}4$ super Yang Mills R
   symmetry with a very weak coupling constant $g_{\rm w} \ll 1$.
   The full SU(4) R symmetry is anomalous and so cannot be
   consistently gauged unless one adds yet other fields to the model
   to cancel the anomaly.  But an SU(2) subgroup is not anomalous and
   could be gauged, provided one defines the currents of that subgroup
   appropriately.  The currents that the gauge bosons would have
   to couple to would be slightly different than the usual
   currents defined in the AdS/CFT correspondence with
   holographic regularization because the latter
   treat all the R currents on an equal footing.  This difference
   in currents reflects the difference between the covariant and
   consistent anomalies \cite{anomalies}.  None of these
   distinctions actually matter in the current problem with
   source (\ref{eq:source}), but we will simply avoid gauging
   any of the R currents so that we do not need to ponder these issues.
}
The $W^+$ boson will leave behind an excitation that carries
electric charge and
the third component of isospin $\tau^3/2$.
Subsequently measuring the latter is analogous to
(\ref{eq:measurement}).

\subsection {The result}

We will take the source amplitude $\Aamp$ to be arbitrarily small
so that we can treat the source term (\ref{eq:Lsource}) as a small
perturbation in our later analysis.  A small-amplitude source will
most of the time have no effect at all on the system, producing no
excitation and no R charge.  We can normalize away this case simply
by dividing the average charge density distribution
$\responseo$
by the average total charge produced by the source,
\begin {equation}
   \charge \equiv
   \int d^3x \> \responseo
   \biggl|_{x^0 \gg L}
   .
\end {equation}
In the case of a transverse-translational invariant source, such as
will be studied in this paper, it is more appropriate to consider
the charge per unit transverse area
$\chargebar \equiv \charge/V_\perp$.

The simplest way to express our final result
is to give a charge-deposition function $\dep(x)$ which
represents how much thermalized charge the high energy
excitation leaves behind at each space-time point $x$.
More specifically, $\dep(x)$ is the source term for the
diffusion equation for the subsequent evolution of that
charge, so that the late-time charge density is given by%
\footnote{
   The idea of defining and investigating $\Theta(x)$ has been
   taken from Chesler, Jensen, and Karch \cite{CJK}.
   However, in that paper, they applied it only to scales large
   compared to the stopping distance of the high-energy excitations.
   Here, we resolve all scales where hydrodynamics is
   applicable.
}
\begin {equation}
   (\partial_t - D \grad^2)
   \responseo
   \simeq
   \chargebar \, \dep(x) ,
\label {eq:depdef}
\end {equation}
where the $\simeq$ here indicates that we are only resolving
structure on distance and times scales large compared to the
thermal wavelength $\sim 1/T$.  For a strongly-coupled plasma,
that is the hydrodynamic limit---the limit where the diffusion equation is
applicable.  The value of the R-charge diffusion constant $D$ is
\cite{Rdiffusion}
\begin {equation}
   D = \frac{1}{2\pi T} \,.
\end {equation}

Our result is that, if one contents oneself with only resolving
details on distance scales large compared to both the source size
$L$ and the thermal wavelength $1/T$, then
\begin {equation}
   \dep(x) \simeq
   2 \,\delta_L(x^-) \, \theta(x^+)
   \begin{cases}
      \frac{(4 \num^4\wbig L)^2}{(2\pi T)^8(x^+)^9} \,
      \formL\Bigl(-\frac{\num^4\wbig L}{(2\pi T x^+)^4}\Bigr) ,
      & x^+ \ll \wbig^{1/3} / (2\pi T)^{4/3} ;
   \\
      \frac{(2\pi T)^4 2 (\numres L)^2}{\wbig} \, \formL(0) \,
      \exp\left( -\frac{\numpole (2\pi T)^{4/3}x^+}{\wbig^{1/3}} \right) ,
      & x^+ \gg \wbig^{1/3} /  (2\pi T)^{4/3} .
   \end {cases}
\label {eq:final}
\end {equation}
where $\theta(x^+)$ is the step function;
$x^\pm \equiv x^3 \pm x^0$;
and
$\formL(y)$ is a source-dependent function that suppresses
$|y| \gg 1$, causing suppression of $x^+ \ll (\wbig L)^{1/4}/(2\pi T)$
above.
In the case of the Gaussian source (\ref{eq:Genvelope}),
\begin {equation}
   \formL(y) = e^{-2 y^2} .
\end {equation}
The subscript $L$ on $\delta_L(x^-)$ indicates that
$\delta_L(x^-)$ is only an
approximate delta function, with a width of order $L$.
Approximating it as a true delta function is good enough if we
are only interested in the hydrodynamic response on scales large
compared to $L$.
In (\ref{eq:final}), the $c$'s are numerical constants, given by
\begin {equation}
   \num \equiv \frac{\Gamma^2(\tfrac14)}{(2\pi)^{1/2}} \,,
\quad
   \numpole \simeq \, 0.927 \,,
\quad
   \numres \simeq 3.2 \,.
\label {eq:num}
\end {equation}
We shall see later that $\numpole$ and $\numres$ are determined by the
first quasi-normal mode in the gravity description.

\begin {figure}
\begin {center}
  \includegraphics[scale=0.4]{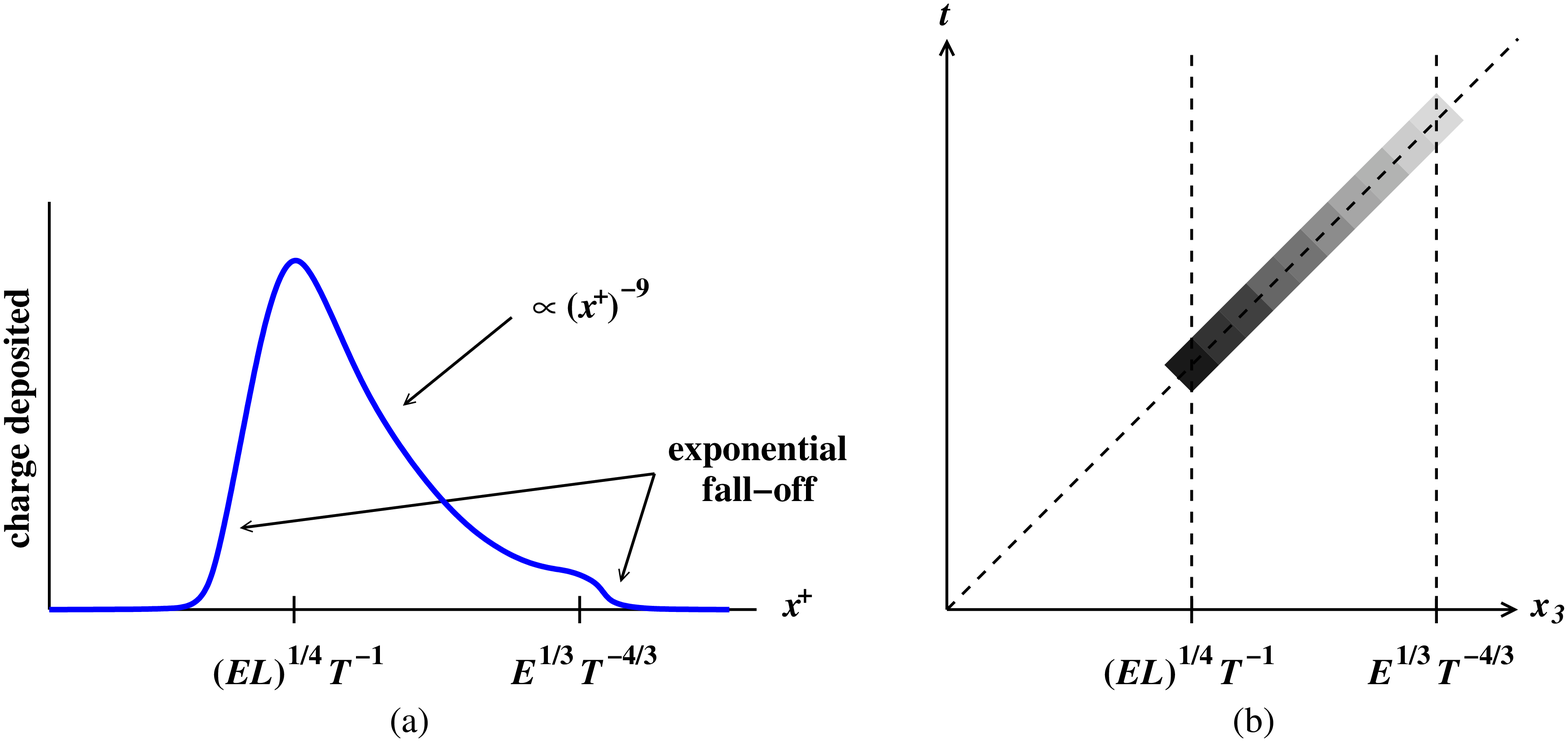}
  \caption{
     \label{fig:deposit}
     The deposition of charge.  (a) shows the coefficient of the
     $\delta_L(x^-)$ in (\ref{eq:final}) for $\dep(x)$ as a function
     of $x^+$.  (b) depicts the space-time points where $\dep(x)$
     is not exponentially suppressed, with lighter and lighter
     shading representing the algebraic $(x^+)^{-9}$ fall-off
     of the strength of $\dep(x)$.
     In both figures, the axis are not represented linearly.
  }
\end {center}
\end {figure}

A qualitative summary of the the $x^+$ dependence of $\dep(x)$ is
shown in fig.\ \ref{fig:deposit}.  The deposition of charge furthest
from the origin that is not exponentially suppressed is at
\begin {equation}
   (x_3)_{\rm max} \sim \frac{E^{1/3}}{T^{4/3}} \,,
\label {eq:xmax}
\end {equation}
which scales with energy as $E^{1/3}$ like the various results of
Refs.\ \cite{GubserGluon,HIM,CheslerQuark}.  However, at least on
average, only a tiny $O\bigl([L/(x_3)_{\rm max}]^2\bigr)$ fraction of the
total charge is deposited at this distance if we keep the source
size $L$ small compared to  $(x_3)_{\rm max}$ itself.
Most of the charge is deposited at the much
smaller distance scale
\begin {equation}
   (x_3)_{\rm dominant} \sim \frac{(EL)^{1/4}}{T} \,.
\label {eq:xdominant}
\end {equation}

There is an important difference between (\ref{eq:xmax}) and
the subset \cite{GubserGluon,CheslerQuark} of earlier results
which have studied jet stopping by studying the dynamics of classical
strings on the gravity side.  Ref.\ \cite{CheslerQuark}
(see also \cite{CJK}) 
added massless ${\cal N}{=}2$ fundamental-charge matter
to supersymmetric Yang Mills and argued that one could study the
stopping of excitations carrying the analog of baryon number by
studying the stopping of moving classical strings in the gravity dual.
Ref.\ \cite{GubserGluon} studied the pure ${\cal N}{=}4$ Yang-Mills
theory and modeled gluon jets by the evolution of folded pieces of
string in the gravity dual.  In both references, the stopping distance
was found to be of order
\begin {equation}
   (x_3)_{\rm string}
   \sim \frac{E_{\rm string}^{1/3}}{\lambda^{1/6} T^{4/3}} \,,
\end {equation}
which is parametrically smaller
than our corresponding scale
(\ref{eq:xmax}) by a factor of $\lambda^{-1/6}$ in the strong-coupling
limit $\lambda\to\infty$.
Formally, the origin of the factor of
$\lambda^{-1/6}$ in the calculations
based on classical strings is the fact that the string tension, and
therefore the energy of the state represented by the string, is
proportional to $\sqrt{\lambda}$, so that
$E_{\rm string}^{1/3} \propto \lambda^{1/6}$.
In our calculation, in contrast, we will not consider classical
strings at all.  We will just use the most basic, original elements of the
AdS/CFT dictionary for relating gauge theory operators to classical
boundary sources in the gravitational dual.

Our maximal scale (\ref{eq:xmax}) does agree (including the absence
of $\lambda$) with the scale previously found by Hatta, Iancu, and
Mueller \cite{HIM}, who specifically studied R charge excitations like
we do.  Gauge-gravity duality relates the SU(4) R currents to classical
5-dimensional SU(4) gauge fields on the gravity side.  They studied how a
wave solution of the gravity-theory fields would
fall into the black brane horizon,
and then they used rough, qualitative arguments to relate this
behavior back to what happens in the 4-dimensional gauge theory.
In this paper, we precisely relate the field theory problem we have
outlined to a calculation in the gravity dual.
Along the way of
solving the gravity dual problem,
we will eventually encounter the same sort of
problem studied by Hatta, Iancu, and Mueller. However, we will find
another scale to the problem, which was missed in their qualitative
interpretation, and which corresponds to the scale (\ref{eq:xdominant})
at which almost all of the charge is deposited.  In fact, we will
see that the appearance of this scale from the gravity calculation
is intimately related to the conservation of R charge in
the 4-dimensional gauge theory problem after the source turns off
($t \gg L$).

Before moving on, we record the generalization of our result
(\ref{eq:final}) to the case of a generic source envelope, which is that
\begin {equation}
   \formL(\bar q_+ L) =
   \frac{\int dq_- \> |\tilde\envelope_L^{(2)}(\bar q_+,q_-)|^2}
        {4L^2 \int dq_+ \> dq_- \> \theta(-q_+) \,
           |q_+|  \bigl|\tilde\envelope_L^{(2)}(q_+,q_-)\bigr|^2} ,
\label {eq:formL}
\end {equation}
where $\theta$ is the step function and
$\tilde\Lambda^{(2)}$ is the two-dimensional Fourier transform
\begin {equation}
   \tilde\envelope_L^{(2)}(q_+,q_-) =
   \int \frac{dx^+\,dx^-}{2} \> \envelope_L(x) \,
   e^{-i(q_+ x^+ + q_- x^-)} .
\label {eq:envelope2}
\end {equation}
Provided the envelope function $\envelope_L(x)$ is smooth on the
scale $L$ and falls sufficiently rapidly for
$|x^\mu| \gg L$, the qualitative conclusions are the same as
for the Gaussian envelope.


\subsection {Jets in strong coupling}

We loosely use the term ``jet'' to refer to a
spatially localized, high energy excitation that initially moves
at nearly the speed of light through the plasma.
There is a potential for confusion
because sometimes people
loosely summarize the (zero temperature) results
of Hofman and Maldacena \cite{HofmanMaldacena}
as indicating that there are no jets in
strongly-coupled ${\cal N}{=}4$ super Yang Mills.
Hofman and Maldacena considered the case of an isotropic source localized
in all four space-time dimensions, with 4-momentum narrowly peaked around
$\kbig=(\wbig,0,0,0)$.
In weak coupling, this source would predominantly produce
two back-to-back partons, flying in opposite directions, and
so produce a non-trivial angular distribution in the late-time energy-energy
correlation function far away from the source.  In the strong
coupling limit, they found the opposite: there was no such angular
correlation.  Their source did not create a jet-like structure.

However, at zero temperature, one may change the appearance of
something simply by boosting to a different reference frame.
An expanding spherical shell of energy in the original frame looks
like a single, slowly spreading, localized jet if one boosts by
a very large amount in the $x_3$ direction.  Effectively, this is
what our source (\ref{eq:source}) does when we choose
$\kbig=(\wbig,0,0,\wbig)$ in (\ref{eq:kbig}).  The envelope
function $\envelope_L(x)$ causes a narrow spread in momentum $k$
around $(\wbig,0,0,\wbig)$, and we will see later that it is
the time-like subset  ($k^\mu k_\mu < 0$) of these momenta
that produce our result
(see also \cite{HIM}).  So the physics of the creation of
our initial ``jet'' state is essentially an extremely boosted version of
the zero-temperature problem studied by Hofman and Maldacena.

Hofman and Maldacena related the measurement of
one-point correlations $\langle {\cal E}(x) \rangle$ of energy density to
the calculation of three-point correlations
$\langle 0| O_{\kbig}^\dagger \, {\cal E}(x)
 O_{\kbig} |0 \rangle$,
where $O_{\kbig}$ was the operator that created their initial state.
In order to measure charge densities such as (\ref{eq:measurement})
in this paper, we will similarly investigate three-point correlators
(in our case at finite temperature)
between (i) the measured charge density and
(ii) operators associated with the creation of
the source.  However, for reasons of
calculational simplicity that we
will explain later, we have set up the problem in a slightly
different way than Hofman and Maldacena did and so evaluate a
slightly different type of 3-point correlator ordering.  Specifically, we
cast the problem in terms of {\it retarded}\/ 3-point correlators.


\subsection {What follows}

In the next section, we set up the basic integrals that we will have
to evaluate to obtain the charge density $\respox$
in terms of bulk-to-boundary propagators in AdS${}_5$-Schwarzschild
space.  Then we warm up to the task of evaluating these integrals
in section \ref{sec:zeroT} by applying our method to the simpler
case of zero temperature.  In particular, we will make comparison
with Hatta, Iancu, and Mueller \cite{HIM} in section \ref{sec:HIM}.
We move on to the finite-temperature
case in section \ref{sec:finiteT}, where we derive our final
result (\ref{eq:final}).  Finally, we conclude with some
suggestions for future work in section \ref{sec:conclusion}.


\section {General Set-up}

\subsection {Notational Preliminaries}

We will use the form of the AdS${}_5$-Schwarzschild metric given
by
\begin {align}
  ds^2
       &=  \frac{R^2}{4} \left[ \frac{1}{\ub}(-f \, dt^2 + d\x^2)
         + \frac{1}{\ub^2 f} \, d\ub^2 \right]
\nonumber\\
       &=  \frac{R^2}{4} \left[ \frac{(2\pi T)^2}{u}(-f \, dt^2 + d\x^2)
         + \frac{1}{u^2 f} \, du^2 \right] ,
\label {eq:metric}
\end {align}
where
\begin {equation}
   f \equiv 1 - (2\pi T)^4 \ub^2 \equiv 1 - u^2 ,
\end {equation}
$R$ is the radius of the AdS space-time (which results will not
depend upon), $u{=}0$ corresponds to the 4-dimensional boundary,
and $u{=}1$ corresponds to the horizon.
For space-time indices, we will use capital roman letters
($I$, $J$, ...) for indices in 5-dimensional
space-time and Greek letters ($\mu$, $\nu$, ...) for four-dimensional
space-time.  When we write a lower Greek index on a 4-momentum $Q$
or polarization $\pol$, we
will always mean that the index is lowered with the 4-dimensional
metric $\eta_{\mu\nu}$ and not the 5-dimensional metric $g_{IJ}$;
so
\begin {equation}
   Q_\mu \equiv \eta_{\mu\nu} Q^\nu ,
   \qquad
   \pol_\mu \equiv \eta_{\mu\nu} \pol^\nu ,
\end {equation}
where $\eta=\operatorname{diag}(-1,1,1,1)$.

Our conventions for light cone coordinates will be
\begin {equation}
   x^\pm = x^3 \pm x^0 ,
   \qquad
   x_+ = \tfrac12 \, x^- = \frac{x^3 - x^0}{2} \,,
   \qquad
   x_- = \tfrac12 \, x^+ = \frac{x^3 + x^0}{2} \,,
\end {equation}
and similarly
\begin {equation}
   q^\pm = q^3 \pm q^0 ,
   \qquad
   q_+ = \tfrac12 \, q^- = \frac{q^3 - q^0}{2} \,,
   \qquad
   q_- = \tfrac12 \, q^+ = \frac{q^3 + q^0}{2} \,,
\end {equation}
and so $q_\mu x^\mu = q_+ x^+ + q_- x^- + \q_\perp\cdot\x_\perp$.
When integrating over 4-momenta $q$, we will use the short-hand notation
\begin {equation}
   \int_q \cdots
   \equiv \int \frac{d^4q}{(2\pi)^4} \cdots
   = \int \frac{2\>d q_+ \> dq_- \> d^2q_\perp}{(2\pi)^4} \cdots
   .
\end {equation}

Where there is no opportunity for confusion, we will abbreviate
$\response$ as $\jmu$ and sometimes even as $j^\mu(x)$.


\subsection {Field Theory: 3-point functions}
\label {sec:FT3point}

To relate the response of 1-point functions such as
energy density or R charge density $\respox$
to equilibrium $n$-point correlation functions,
for small source amplitudes $\Aamp$,
one follows the same steps as in derivations of the
fluctuation-dissipation theorem or Kubo formulas.
Since it's relatively simple, we'll take
a moment to review it here.  Write the Hamiltonian as
$H(t) = H_0 + \delta H(t)$, where $\delta H(t)$
are the small-amplitude source
terms and $H_0$ is everything else in the full Hamiltonian of the
theory.  If the system initially starts in equilibrium, before the
sources turn on, then the later evolution of an observable ${\cal O}$ is
given by
\begin {equation}
   \langle {\cal O}(t) \rangle_H
   = Z_0^{-1}
     \tr\left( e^{-\beta H_0} 
              [U(t,-\infty)]^\dagger {\cal O} U(t,-\infty) \right) ,
\end {equation}
where
\begin {equation}
   U(t,t_0) = {\cal T} \exp\left(-i\int_{t_0}^t dt'\>H(t') \right)
\end {equation}
is the evolution operator under $H$,
and ${\cal T}$ is time ordering.  Working in the interaction
picture and expanding in powers of the small $\delta H$, one finds
\begin {equation}
   \langle {\cal O}(t) \rangle_H
   -
   \langle {\cal O} \rangle_{H_0}
   =
   \int dt_1 \> G_{\rm R}(t_1;t)
   +
   \frac{1}{2!}
   \int dt_1 \> dt_2 \> G_{\rm R}(t_1,t_2;t)
   +
   \cdots
\label {eq:fd}
\end {equation}
where the various $G_{\rm R}$ are the equilibrium
$n$-point retarded correlation functions, given in this case by%
\footnote{
   Readers familiar with the (r,a) formalism may know these
   retarded Green functions as
   $G_{\rm ar}(t_1,t)$, $G_{\rm aar}(t_1,t_2,t)$, etc.
   See, for example, the review of notation in
   Ref.\ \cite{WangHeinz}.
}
\begin {align}
   iG_{\rm R}(t_1;t) &=
     \theta(t-t_1) \,
     \bigl\langle [{\cal O}(t),\delta H(t_1)] \bigr\rangle_{H_0} ,
\\
   i^2 G_{\rm R}(t_1,t_2;t) &=
     \theta(t-t_2) \, \theta(t_2-t_1) \,
     \bigl\langle [[{\cal O}(t),\delta H(t_2)],\delta H(t_1)]
            \bigr\rangle_{H_0}
\nonumber\\ &
     +
     \theta(t-t_1) \, \theta(t_1-t_2) \,
     \bigl\langle [[{\cal O}(t),\delta H(t_1)],\delta H(t_2)]
            \bigr\rangle_{H_0} ,
\label {eq:G3generic}
\end {align}
etc.
In our problem, the first non-vanishing term is the one involving
the 3-point correlator
because the operator $[j^{(3)0}(x),\j^a\cdot\A_{\rm cl}^a(x_1)]$ has
non-zero R-charge, causing the 2-point correlator
$G_{\rm R}(x_1;x)$ to vanish.  However, the conclusion is more general
than our specific example involving R charge currents.
To create a high-energy excitation, the source $\delta H$ should
have large momentum $k$.  To measure later hydrodynamic behavior after
the jet stops in the medium, one wants to examine relatively
{\it low}-wavenumber components $q$ of the late-time diffusing
density ${\cal O}(x)$.  Because of this momentum mismatch between
source and observable, the equilibrium
two-point function $G_{\rm R}(k;q)$ will vanish by
momentum conservation.  It is only when we get to the
three-point function that we first find a non-vanishing result.%
\footnote{
   Hydrodynamic quantities like viscosity and charge diffusion constants
   can be studied using two-point correlators
   \cite{Rdiffusion,ViscReview,AdSviscosity}
   because one may measure
   them by studying the low-wavenumber response of the system to
   a low-wavenumber source.
   The reason one has to go to 3-point functions here (and in Hofman
   and Maldacena \cite{HofmanMaldacena}) is that we are specifically
   interested in a high-momentum source in order to study ``jets.''
}

In our problem, (\ref{eq:fd}) manifests in detail as
\begin {align}
  \response
  &=
  \tfrac12
  \int d^4x_1 \> d^4x_2 \>
  G_{\rm R}^{(ab3)\alpha\beta\mu}(x_1,x_2;x) \,
  A^a_{\alpha,\rm cl}(x_1) \,
  A^b_{\beta,\rm cl}(x_2) \,
\nonumber\\
  &=
  \tfrac12
  \int_{Q_1Q_2Q}
  G_{\rm R}^{(ab3)\alpha\beta\mu}(Q_1,Q_2;Q) \,
  A^{a\econj}_{\alpha,\rm cl}(Q_1) \,
  A^{b\econj}_{\beta,\rm cl}(Q_2) \,
  e^{iQ\cdot x}
  (2\pi)^4 \delta^{(4)}(Q_1+Q_2+Q)
\label {eq:Rf0}
\end {align}
in the limit of arbitrarily small source amplitude $\Aamp$.

Now comes a crucial argument that we will use repeatedly.
In studying the response (\ref{eq:Rf0}) to determine how
far the jet travels, we will not care about the detailed
structure on small distance and time scales such
as $1/\wbig$ or even $1/T$.  It would be perfectly adequate to
look at a {\it smeared}\/ response such as
\begin {equation}
   \respx_{\rm smeared}
   \equiv
   \int d^4(\Delta x)
   \left\langle j^{(3)\mu}(x+\Delta x) \right\rangle_{\Acl}
   \frac{
     e^{-(\Delta x^0)^2/\ell_{\rm smear}^2}
     e^{-|\Delta\x|^2/\ell_{\rm smear}^2}
   }{\pi \ell_{\rm smear}^2}
   ,
\label {eq:smear}
\end {equation}
where the smearing distance $\ell_{\rm smear}$ is chosen
large compared to microscopic scales such as $1/\wbig$ and $1/T$ but
small compared to scales we're interested in resolving, such as
the stopping distances (\ref{eq:xmax}) and (\ref{eq:xdominant}).
In momentum space, the smearing (\ref{eq:smear}) retains only small
wavenumbers $Q$, by which we mean $Q$ whose components are all
$\lesssim 1/\ell_{\rm smear}$.
If we make an approximation to (\ref{eq:Rf0}) that
changes the integrand for large $Q$ but not for small $Q$
(where $Q$ is conjugate to the
point $x$ where we measure the charge density), then the
smeared response (\ref{eq:smear}) containing all the information
we are interested in will not change.
In the rest of this paper, we will not again explicitly reference
the smeared response (\ref{eq:smear}), but we will feel free to
make approximations that are only valid
when the components of $Q$ have magnitudes small
compared to $\wbig$ and $T$.  Note that this only applies to the
$Q$ conjugate to the measurement point $x$; no such approximation
would be acceptable for the source momenta $Q_1$ and $Q_2$ in
(\ref{eq:Rf0}), which are both large.

In particular, using the explicit source (\ref{eq:source}) in
(\ref{eq:Rf0}) gives
\begin {align}
  & \response
  \simeq
\nonumber\\ & \quad
  \Aamp^2
  \int_{Q_1Q_2} \pol_\alpha \pol_\beta \,
  G_{\rm R}^{({-}{+}3)\alpha\beta\mu}(Q_1,Q_2;Q) \,
  \tilde\envelope_L^\econj(Q_1-\kbig) \,
  \tilde\envelope_L^\econj(Q_2+\kbig) \,
  e^{-iQ_1\cdot x}
  e^{-iQ_2\cdot x}
  \biggl|_{Q=-Q_1-Q_2} ,
\label {eq:Rfstart}
\end {align}
where we have ignored terms involving
$\envelope^\econj_L(Q_1-\kbig) \, \envelope^\econj_L(Q_2-\kbig)$ and
$\envelope_L^\econj(Q_1+\kbig) \, \envelope_L^\econj(Q_2+\kbig)$ because
these contribute only to to very large momenta
$Q = -Q_1-Q_2 \simeq \pm 2\kbig$.

Before moving on to the gravity side of the calculation of
retarded correlators,
it will be useful to review the fact that retarded real-time
correlators are related to time-ordered imaginary-time correlators
by analytic continuation in frequency.
For two-point correlators, we are used to seeing this in the
form%
\footnote{
   The imaginary-time $n$-point Green function is defined here as
   $(-)^{n-1}$ times the imaginary-time time-ordered correlator of
   fields.  So, for instance, $G_{\rm E}(q) = -1/(q^2+m^2)$ for
   a free massless scalar.
}
\begin {equation}
   G_{\rm R,A}(\omega)
   = G(\omega\pm i\epsilon) ,
\label {eq:G2continue}
\end {equation}
where $G$ is the analytic continuation of the imaginary-time
Green function $G_{\rm E}$ to
real-time frequencies;%
\footnote{
  $\omega{+}i\epsilon$ ($\omega{-}i\epsilon$) indicates that one continues
  from positive (negative) imaginary frequencies.
}
the upper and lower signs are for the retarded (R) and
advanced (A) Green function respectively; and we have
suppressed showing the spatial momentum $\q$.
To understand the generalization to $n$-point functions, it
is useful to write the 2-point function in terms of
two momenta, trivially related by momentum conservation:
\begin {equation}
   G(Q_1;Q)
   \equiv \int d^4x_1 \> d^4x \> G(x_1;x) \, e^{-iQ_1\cdot x_1}
                                          \, e^{-iQ\cdot x}
   = G(Q) \, (2\pi)^4\delta(Q_1+Q) .
\end {equation}
Since $Q_1 = -Q$, the prescription (\ref{eq:G2continue}) is
in this language
\begin {equation}
   G_{\rm R,A}(\omega_1;\omega)
   = G(\omega_1\mp i\epsilon; \omega\pm i\epsilon) .
\end {equation}
That is, the frequency associated with the response has a
$\pm i\epsilon$ prescription and that associated with the
source has the opposite.  This generalizes to the higher-point
functions, so that \cite{Evans}%
\footnote{
  One may check that the $i\epsilon$ prescriptions (\ref{eq:G3continue})
  enforce vanishing of the retarded correlator $G_{\rm R}(t_1,t_1;t)$
  unless $t$ is the largest of the three times, just as in the
  equivalent but more explicit formula (\ref{eq:G3generic}).
  (i) Fourier transform (\ref{eq:G3continue})
  back to $t_1$, $t_2$, and $t$, (ii)
  use frequency conservation $\delta(\omega_1+\omega_2+\omega)$
  for the $\omega$ integration to rewrite
  $e^{-i\omega_1 t_1} e^{-i\omega_2 t_2} e^{-i \omega t}$
  as $e^{-i\omega_1 (t_1-t)} e^{-i\omega_2 (t_2-t)}$,
  and (iii) close the $\omega_1$ and/or $\omega_2$ integration contours
  in the lower half plane if $t_1 > t$ and/or $t_2 > t$
  [realizing that the prescription
  (\ref{eq:G3continue}) makes
  $G_{\rm R}(\omega_1,\omega_2;\omega)$ analytic in the
  lower half planes of $\omega_1$ and/or $\omega_2$ (corresponding to
  the upper-half plane of $\omega = -\omega_1-\omega_2$)].
}
\begin {equation}
   G_{\rm R}(\omega_1,\omega_2;\omega)
   = G(\omega_1 - i\epsilon, \omega_2 - i\epsilon; \omega + 2 i\epsilon) .
\label {eq:G3continue}
\end {equation}


\subsection {Gravity Dual: 3-point functions}

\subsubsection {Basics}

\begin {figure}
\begin {center}
  \includegraphics[scale=0.3]{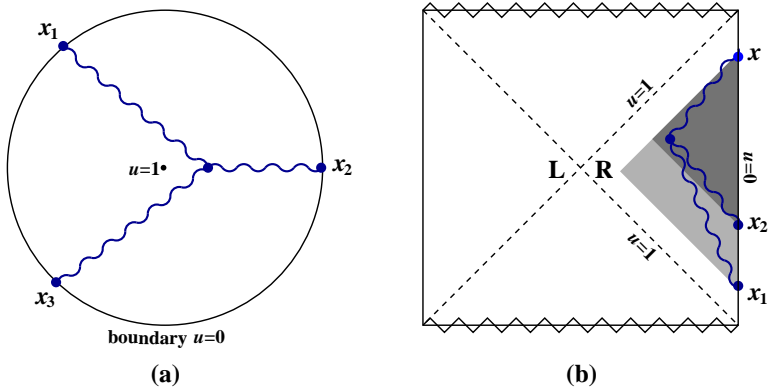}
  \caption{
     \label{fig:Witten}
     Witten diagram for (a)
     3-point boundary correlator in imaginary-time
     AdS${}_5$-Schwarzschild and (b) retarded 3-point boundary correlator
     $G_{\rm R}(x_1,x_2;x)$ in
     real-time AdS${}_5$-Schwarzschild.  The darker shaded region shows
     the region of bulk vertex position that gives non-vanishing
     contribution to the retarded
     correlator, which is the intersection of the causal future of
     $x_1$, the causal future of $x_2$, and the causal past of $x$.
     We have taken artistic license when drawing the boundary of
     the Penrose diagram with all four sides straight
     \cite{SquarePenrose}.
  }
\end {center}
\end {figure}

The AdS/CFT correspondence translates the problem of computing
the 3-point Green function of R charge currents
in strongly-coupled gauge theory to
the problem of computing 3-point boundary correlators of
classical gauge fields living in
AdS${}_5$-Schwarzschild space.
If we were interested in {\it imaginary-time}\/ correlators,
the boundary correlator would be given by the Witten diagram
of fig.\ \ref{fig:Witten}a, with the circle representing the boundary
of imaginary-time AdS${}_5$-Schwarzschild space.

To start with a simpler example, if we were studying a 3-point
boundary correlator of 5-dimensional scalar fields in
the gravity theory with 3-point vertex $\lambda$, then the
Witten diagram would give
\begin {equation}
   G_{\rm E}(Q_1,Q_2,Q_3) =
   \lambda \int_{\uB}^1 du \sqrt{g_{\rm E}} \>
      \calGE(Q_1,u) \, \calGE(Q_2,u) \, \calGE(Q_3,u) ,
\label {eq:G3EgravityS}
\end {equation}
where the $Q_i$ are 4-momenta in the boundary theory,
$u$ is the coordinate for the 5th dimension,
and $\calGE(Q,u)$ is the bulk-to-boundary propagator
which solves the 5-dimensional imaginary-time
classical equation of motion for the scalar field,
appropriately normalized on the boundary.
$\uB \to 0$ is the usual boundary regulator.
In our calculation, there will turn out to be no divergences as
$u \to 0$, and so we will set $\uB = 0$ for simplicity.

Imaginary-time AdS${}_5$-Schwarzschild spacetime is smooth,
and we can integrate over the entire space-time without
worrying about horizons or singularities.
In real time, one in principle has to worry about such issues,
but for the retarded 3-point propagator one can find the
correct prescription simply
by the analytic continuation (\ref{eq:G3continue})
of the imaginary-time result:
\begin {align}
   G_{\rm R}(Q_1,Q_2;Q)
   &= \lambda \int_0^1 du \sqrt{-g} \>
      {\cal G}(\omega_1-i\epsilon,\Q_1,u) \,
      {\cal G}(\omega_2-i\epsilon,\Q_2,u) \,
      {\cal G}(\omega+2i\epsilon,\Q,u)
\nonumber\\
   &= \lambda \int_0^1 du \sqrt{-g} \>
      \calGA(Q_1,u) \, \calGA(Q_2,u) \, \calGR(Q,u) ,
\label {eq:G3RgravityS}
\end {align}
where $\calGR$ and $\calGA$ solve the linearized 5-dimensional real-time
equations of motion with retarded or advanced boundary conditions
respectively.
The result (\ref{eq:G3RgravityS}) was found in Ref.\ \cite{paper1},
where more discussion of both retarded and other
3-point correlators may be found.
(See also the related discussion in Ref.\ \cite{Rees}.)

As a matter of convention,
note that our 4-momenta $Q_1$, $Q_2$ and $Q$ are momenta in the
gauge theory, and therefore they are the momenta conjugate to
the {\it boundary}\/ points in the bulk-to-boundary propagators.
(The momenta conjugate to the 4-position of the bulk point
are correspondingly $-Q_1$, $-Q_2$, and $-Q$.)
As a result, our convention is that
a retarded bulk-to-boundary propagator refers to the case
where information flows from the bulk to the boundary, and so
corresponds to a solution where waves flow out of the horizon.
Similarly, our
advanced bulk-to-boundary propagator is
the solution where waves flow into the horizon.%
\footnote{
  Let us relate this to the notation of Son and Starinets
  \cite{ViscReview}.  Let $q$ be the 4-momentum conjugate to
  the boundary position and $p=-q$ the 4-momentum conjugate to
  the bulk position.  Then, in the scalar case, our
  $\calGR(q,u)$ corresponds to their $f^*_p(u)$.  Note
  that $f^*_p = f^*_{-q} = f_q$.  In position space,
  our retarded bulk-to-boundary propagator from a
  bulk point $(y,u)$ to a boundary point $x$ is
  $\calGR(y,u;x) = \int_q f_q(u) \, e^{iq\cdot(x-y)}
  = \int_p f_p^*(u) \, e^{ip\cdot(y-x)}$.
  The last is the same as the advanced boundary-to-bulk propagator
  $\funG_{\rm A}(x;y,u)$ defined by (\ref{eq:funG}).
  If one wished to describe a bulk wave falling from the
  boundary into the horizon, it would be given by
  $\phi(y,u) = \int d^4x \> \funG_{\rm R}(x;y,u) \, \phi(x,0)
             = \int d^4x \> \calGA(y,u;x) \, \phi(x,0)$
  or equivalently
  $\phi(p,u) = \funG_{\rm R}(p,u) \, \phi(p,0) 
             = \calGA(-p,u) \, \phi(p,0)
             = \calGA(q,u) \, \phi^*(q,0) $.
}
The two are related by
\begin {equation}
   \calGA(q,u) = [\calGR(q,u)]^* = \calGR(-q,u) .
\label{eq:AvR}
\end {equation}
Readers who find it awkward or confusing
to think in terms of bulk-to-boundary flow
rather than boundary-to-bulk flow may, if desired, rewrite
our equations in terms of boundary-to-bulk propagators
$\funG(p,u)$ defined by
\begin {equation}
   \funG_{\rm R}(p,u) = \calGA(-p,u) ,
   \qquad
   \funG_{\rm A}(p,u) = \calGR(-p,u) ,
\label {eq:funG}
\end {equation}
where $p$ is the 4-momentum conjugate to the bulk position.

The real-time Witten diagram associated with (\ref{eq:G3RgravityS})
corresponds to Figure \ref{fig:Witten}b, and the bulk vertex lives
exclusively in the right-hand quadrant of the Penrose diagram.
This can be understood from the causality properties of the
retarded and advanced bulk-to-boundary propagators in the
analytic continuation (\ref{eq:G3RgravityS}) of the imaginary-time
result (\ref{eq:G3EgravityS}).  $\calGA$ only has support when the
bulk point is in the causal future of the boundary point, and
$\calGR$ only when the bulk point is in the causal past of the
boundary point.  Taking the boundary points to all be on the
right-hand boundary, the combination of these causality constraints
requires the bulk vertex to be in the right-hand quadrant.
This argument is similar to the discussion
by Caron-Huot and Saremi \cite{HydroTails}
of the ``causal diamond'' in their
analysis of long-time hydrodynamic tails from
one-loop gravity corrections to the 2-point retarded
correlator.

It is the simple form (\ref{eq:G3RgravityS}) of the retarded correlator
that led us to choose to set up the problem of studying jet evolution
in terms of retarded correlators, rather than the type of  correlators
$\langle O_q^\dagger \, {\cal O}(x) \, O_q \rangle$ considered
by Hofman and Maldacena, which would be more complicated to analyze
at finite temperature.%
\footnote{
   Such correlators can be related to the retarded correlator.
   Since the source is localized, and we are only interested in
   measurements at times $x^0$ after the source turns off, we can
   rewrite the correlator
   $\langle O_q^\dagger \, {\cal O}(x) \, O_q \rangle$
   in Schwinger-Keldysh (closed time path) notation as
   $G_{211}(O_q^\dagger,{\cal O},O_q)$.
   Each subscript in $G_{211}$ designates
   whether the corresponding operator appears
   on (1) the first
   leg of the Schwinger-Keldysh integration contour
   ($t{=}{-}\infty$ to $+\infty$) or (2) the second leg
   ($+\infty$ to $-\infty$), and operators are ordered accordingly.
   Specifically,
   $G_{211}^{ABC} = \langle A \, ({\cal T} B C)\rangle$, where ${\cal T}$
   represents ordinary time ordering.  (Since
   $x^0$ is our largest time,
   $G_{211} = G_{221} = G_{2{\rm r}1}$ here, where ${\rm r}$ is the average
   of using 1 and 2.)  3-point Schwinger-Keldysh propagators
   $G_{\alpha_1 \alpha_2 \alpha_3}$
   can all be (non-locally) related to retarded Green functions $G_{\rm aar}$,
   $G_{\rm ara}$, and $G_{\rm raa}$ (the distinction being which of
   the three operators is the latest time when defining ``retarded'')
   and their complex conjugates.
   Explicit formulas (and more explanation of the notation)
   are given in Ref.\ \cite{WangHeinz}.
   Alternatively, see Ref.\ \cite{paper1} for a discussion of different
   3-point correlators directly in terms of integrating over both
   right and left quadrants in the gravity theory.
   (Beware that Ref.\ \cite{paper1}, following Ref.\ \cite{HerzogSon},
   uses a different convention for
   the Schwinger-Keldysh contour than Ref.\ \cite{WangHeinz}.
   This introduces factors of $e^{-\sigma \omega_i}$ into the definition
   of $G_{\alpha_1 \alpha_2 \alpha_3}$, where $\sigma = \beta/2$.)
}


\subsubsection {The vector-vector-vector vertex}

The generalization of (\ref{eq:G3RgravityS}) from scalar interaction to
vector interaction is that we need to use the
vector-vector-vector vertex in the super-gravity (SUGRA) theory:
\begin {equation}
   (G_{\rm R})^{(abc)}_{\alpha\beta\mu}
   =
   \int_0^1 du \sqrt{-g}
        ~\mbox{vertex}^{(abc)IJK}\bigl(
           \calGAup_{I\alpha}(Q_1,u),
           \calGAup_{J\beta}(Q_2,u),
           \calGRup_{K\mu}(Q,u)
         \bigr) ,
\label{eq:Gform}
\end {equation}
where the advanced and retarded bulk-to-boundary propagators
${\cal G}_{I\alpha}$ have
one 4-dimensional vector index ($\alpha$) associated with the boundary point
and one 5-dimensional vector index ($I$) associated with the bulk point
and are normalized on the boundary so that
\begin {equation}
   {\cal G}_{\beta\alpha}(Q,0) = \eta_{\beta\alpha}
   \qquad \mbox{and} \qquad
   {\cal G}_{5\alpha}(Q,0) = 0 .
\end {equation}
The vertex function is extracted from the cubic terms in the
SUGRA interaction \cite{Witten,Freedman}
\begin {equation}
   - \frac{1}{4\gSG^2 R} \int d^5x \sqrt{-g} \, F^{IJ a} F_{IJ}^a
   - \frac{k}{96\pi^2} \int d^5x \,
     \left[
       d^{abc} \varepsilon^{IJKLM} A_I^a (\partial_J A_K^b) (\partial_L A_M^c)
       + \cdots
     \right] ,
\label {eq:action}
\end {equation}
where we work in real time,
have only shown explicitly the cubic term in the
Chern-Simons interaction, and
\begin {equation}
   \gSG = \frac{4\pi}{\Nc}
   \qquad \mbox{and} \qquad
   k = \Nc^2 - 1.
\label {eq:gSG}
\end {equation}
The $F^2$ term produces a 3-point vertex with R charge structure
$f^{abc}$, whereas the Chern-Simons term (which reproduces the
R charge current anomaly) yields $d^{abc}$.
Here $f^{abc}$ and $d^{abc}$ are defined in terms of
SU(4) Hermitian generators $T^a$ by
$\tr(T^a T^b T^c) = \frac14(d^{abc} + i f^{abc})$, normalized
by
$\tr(T^a T^b) = \frac12\,\delta^{ab}$.

Recall that the currents in our source (\ref{eq:source}) and
measurement (\ref{eq:measurement}) were chosen to lie in an
SU(2) subgroup of SU(4).  Since $d^{abc}$ vanishes for SU(2),
the Chern-Simons term will not contribute in our application.%
\footnote{
   An independent reason that the Chern-Simons term
   does not contribute in our problem
   is our choice of linear polarization $\pol$ for our source: the two $A$'s
   which attach to the source points on the boundary will give
   factors of $\pol$ and $\pol^* \propto \pol$ contracted with the
   $\varepsilon^{IJKLM}$, giving zero.
}
So the only relevant contribution to the vertex comes from the
cubic interaction
\begin {equation}
   - \frac{f^{abc}}{2\gSG^2 R} \int d^5x \sqrt{-g} \,
   g^{IM} g^{JN}
   (\partial_I A_J^a - \partial_J A_I^a) A_M^b A_N^c .
\label {eq:vertex}
\end {equation}
The factors of the AdS radius
$R$ all cancel [the explicit $R^{-1}$ above with the
factors in $\sqrt{-g}\,g^{IM} g^{JN}$ from (\ref{eq:metric})].
If desired, one could simply set $R=1$ in the rest of the paper.

To get the vertex function in (\ref{eq:Gform}), substitute
the three bulk-to-boundary propagators ${\cal G}$ for the
three $A$'s in (\ref{eq:vertex}) in all possible permutations.
Since the problem studied in this paper is invariant under
translation in the transverse directions, we will restrict
attention to 5-dimensional gauge choices that respect this
invariance.  In particular, the bulk-to-boundary propagators
will preserve
transversality of polarization,
\begin {equation}
   \pol^\alpha {\cal G}_{I\alpha} \propto \pol_I ,
\end {equation}
and transverse derivatives will vanish,
\begin {equation}
   \pol_\alpha g^{\alpha I} \partial_I \cdots = 0 .
\end {equation}
Here it is convenient to define a 5-dimensional $\pol_I$ in terms
of 4-dimensional $\pol_\alpha$ by
\begin {equation}
   \pol_I \equiv (\pol_0,\pol_1,\pol_2,\pol_3,0) = (0,1,0,0,0).
\end {equation}
Putting everything together, the piece of (\ref{eq:Gform})
that contributes in our problem is then
\begin {align}
   &
   \pol^\alpha \pol^\beta (G_{\rm R})^{(abc)}_{\alpha\beta\mu}
   =
\nonumber\\ & \qquad
   - \frac{f^{abc}}{\gSG^2 R} \int d^4x'\>du\> \sqrt{-g} \,
   g^{IM} g^{JN}
   \biggl\{
   [- \pol_I \partial'_J \, \calGAup_{\perp}(x',u;x_1)] \,
   \pol_M \calGAup_{\perp}(x',u;x_2) \,
   \calGRup_{N\mu}(x',u;x)
\nonumber\\ & \qquad
   - (x_1 \leftrightarrow x_2) \biggr\} ,
\label {eq:GfRinit}
\end {align}
where
${\cal G}_\perp \equiv
\pol^\rho {\cal G}_{\rho\sigma}  \pol^\sigma =
\pol_\mu \eta^{\mu\rho} {\cal G}_{\rho\sigma}  \eta^{\sigma\nu} \pol_\nu$
is the transverse piece of the
bulk-to-boundary propagator.
(Other than as an attempt to save space in
equations, there is no significance
to whether we write the R/A for retarded/advanced as
subscripts or superscripts.)
Switching to four-dimensional momentum space, and using
the notation
$f\tensor\partial g \equiv f\partial g - (\partial f) g$,
\begin {align}
   \pol^\alpha \pol^\beta (G_{\rm R})^{(abc)}_{\alpha\beta\mu}
   &=
   \frac{f^{abc}}{\gSG^2 R} \int du \> \sqrt{-g} \,
   (\pol_I g^{IJ} \pol_J)
\nonumber\\ & \qquad \times
   \biggl[
     \calGAup_{\perp}(Q_1,u) \,
     \calGAup_{\perp}(Q_2,u) \,
     i(Q_2-Q_1)_\rho g^{\rho\sigma} \calGRup_{\sigma\mu}(Q,u)
\nonumber\\ & \qquad\quad
   +
     \calGAup_{\perp}(Q_1,u) \,
     \tensor\partial_5 \,
     \calGAup_{\perp}(Q_2,u) \,
     g^{55} \calGRup_{5\mu}(Q,u)
   \biggr] .
\label {eq:epsGR1}
\end {align}


\subsubsection{Summary in $A_5{=}0$ gauge}

The last formula (\ref{eq:epsGR1}) is simplest in $A_5{=}0$ gauge, where
the last term vanishes.  We will specialize to $A_5{=}0$ gauge in the
remainder of the paper.  Define $G^{\rm R}_{\perp\perp\mu}$ by
\begin {equation}
   \pol^\alpha \pol^\beta (G_{\rm R})^{(abc)}_{\alpha\beta\mu}
   = f^{abc} G^{\rm R}_{\perp\perp\mu} .
\end {equation}
Putting everything together and using
$f^{{-}{+}3} = 2i$, we then have
\begin {subequations}
\label {eq:basic}
\begin {multline}
  \jmu
\\
  \simeq
  2 i \Aamp^2 \eta^{\mu\nu}
  \int_{Q_1Q_2}
  G^{\rm R}_{\perp\perp\nu}(Q_1,Q_2;Q) \,
  \tilde\envelope^\econj_L(Q_1-\kbig) \,
  \tilde\envelope^\econj_L(Q_2+\kbig) \,
  e^{-iQ_1\cdot x}
  e^{-iQ_2\cdot x}
  \biggl|_{Q=-Q_1-Q_2} ,
\label {eq:Rf}
\end {multline}
with
\begin {equation}
   G^{\rm R}_{\perp\perp\mu}
   =
   i(Q_2-Q_1)_\rho \,
   \frac{1}{\gSG^2 R} \int du \> \sqrt{-g} \,
   (\pol_I g^{IJ} \pol_J) \, g^{\rho\sigma} 
     \calGAup_{\perp}(Q_1,u) \,
     \calGAup_{\perp}(Q_2,u) \,
     \calGRup_{\sigma\mu}(Q,u)
\label {eq:epsGR}
\end {equation}
\end {subequations}
in $A_5{=}0$ gauge.

As it currently stands, (\ref{eq:basic}) is
challenging to evaluate.  Our crucial approximation in what
follows will be to replace $\calGRup(Q,u)$ in (\ref{eq:epsGR})
by its small-$Q$ approximation valid for $Q \ll T$.
As discussed earlier
in section \ref{sec:FT3point}, such approximations
are adequate for the long-distance physics that we wish to
study.  As a simple example, for a massless bulk scalar field,
the bulk-to-boundary propagator in the small-$Q$ limit
is \cite{AdSminkowski,ViscReview}%
\footnote{
   To have an expression valid all the way to the horizon, it
   is important not to expand in powers of the exponent as
   $(1 - u^2)^{-i\omega/4\pi T} \simeq
   1 - \frac{i\omega}{4\pi T} \, \ln(1-u^2)$.
   For any fixed $\omega \ll T$, the corrections to this
   truncation are not small when $u$ is close enough to the horizon
   that $(\omega/T)\ln(1-u) \gg 1$.
}
\begin {equation}
   \calGRup_{\rm scalar}(Q,u) =
   (1 - u^2)^{-i\omega/4\pi T}
   + O\Bigl( \frac{\omega^2}{T^2}, \frac{|\Q|^2}{T^2} \Bigr) ,
\label {eq:GSapprox}
\end {equation}
where $\omega \equiv Q^0$.
The approximation (\ref{eq:GSapprox}) has the nice property that
it factorizes for $Q=-Q_1-Q_2$:
\begin {equation}
   \calGRup_{\rm scalar}(Q,u) \simeq
   (1 - u^2)^{i\omega_1/4\pi T}
   (1 - u^2)^{i\omega_2/4\pi T}
   .
\end {equation}
If this were the propagator to use for $\calGRup(Q,u)$ in
(\ref{eq:epsGR}), then the $Q_1$ and $Q_2$ integrals would
factorize in (\ref{eq:Rf}), greatly simplifying the
calculation by allowing us to evaluate them
independently.
We will see in section \ref{sec:finiteT}
that the issue of factorization is a little
more complicated for the vector bulk-to-boundary propagator
$\calGRup_{\sigma\mu}(Q,u)$ at finite temperature,
but we will be able to use a
variant of this trick to factorize the calculation.


\section{The zero-temperature calculation}
\label{sec:zeroT}

In this section, we warm up to the calculation by applying to the case
of zero temperature the methods that we will later use for finite
temperature.
At zero temperature, the metric (\ref{eq:metric}) becomes
\begin {equation}
  ds^2
        =  \frac{R^2}{4} \left[ \frac{1}{\ub}(-dt^2 + d\x^2)
         + \frac{1}{\ub^2} \, d\ub^2 \right]
\end {equation}
for AdS${}_5$,
where $\ub$ runs from zero (at the boundary) to infinity.
This is related to the Poincare metric
\begin {equation}
   ds^2 = R^2 \, \frac{ \eta^{\mu\nu} dx_\mu dx_\nu + dz^2}{z^2}
\label {eq:metricz}
\end {equation}
by $\ub = \frac14 \, z^2$.
Eq.\ (\ref{eq:epsGR}) for the 3-point function becomes
\begin {equation}
   G^{\rm R}_{\perp\perp\mu}
   =
   i(Q_2-Q_1)_\rho \eta^{\rho\sigma} \,
   \frac{1}{2\gSG^2} \int_{0}^\infty \frac{d\ub}{\ub} \,
     \calGAup_{\perp}(Q_1,\ub) \,
     \calGAup_{\perp}(Q_2,\ub) \,
     \calGRup_{\sigma\mu}(Q,\ub) .
\label {eq:epsGRT0}
\end {equation}

At zero temperature, it is notationally a little more convenient to
use the coordinate $z$ than $\ub$.  However, we will use
the coordinate $u$ for the finite-temperature calculations, and so
we stick to $\ub$ here in order to make the two cases as easy to
compare as possible.

In imaginary time, the regular solution to the linearized 5-dimensional
vector equation of motion $d \, {}^\star \! F=0$ is
\begin {subequations}
\label {eq:Azero}
\begin {equation}
   A_\mu(q,\ub)
   = {\cal G}_{\mu\nu}(q,\ub) \, A_\nu(q,0)
\end {equation}
in momentum space in $A_5{=}0$ gauge,
where%
\footnote{
   For the momentum-space solution in covariant conformal gauge
   ($\nabla^I A_I = 0$), see, for example, Ref.\ \cite{Mueck}
   specialized to the case $d{=}4$ and $m{=}0$.
   This is the gauge transformation $A_I \to A_I - \partial_I\Lambda$
   of (\ref{eq:Azero}) with
   $\Lambda(q,u) =
    i \bigl[ 2 \ub q^2 \, K_2(\sqrt{4\ub q^2}) - 1\bigr]
    (q_\nu/q^2) \, A_\nu(q,0)$.
}
\begin {equation}
   \calGEup_{\mu\nu}(q,\ub)
   =
   \sqrt{4\ub q^2}\,  K_1\bigl(\sqrt{4\ub q^2}\bigr)
       \left( \delta_{\mu\nu} - \frac{q_\mu q_\nu}{q^2} \right)
   + \frac{q_\mu q_\nu}{q^2}
\end {equation}
\end {subequations}
and $K_n$ is the modified Bessel function of the second kind.
The real-time version is then simply
\begin {equation}
   {\cal G}_{\mu\nu}(q,\ub)
   =
   \sqrt{4\ub q^2}\,  K_1\bigl(\sqrt{4\ub q^2}\bigr)
       \left( \eta_{\mu\nu} - \frac{q_\mu q_\nu}{q^2} \right)
   + \frac{q_\mu q_\nu}{q^2}
\label {eq:calGzero}
\end {equation}
where $q^2 = \eta^{\alpha\beta} q_\alpha q_\beta$
with $\omega \to \omega \pm i\epsilon$ for retarded and advanced.
For time-like momenta ($q^2 < 0$), it can be useful to
recast $K_1$ in terms of Hankel functions,
\begin {equation}
   \sqrt{4 \ub q^2}\, K_1\bigl(\sqrt{4 \ub q^2}\bigr)
   =
   \begin {cases}
      \phantom{-}
      i\pi \, \sqrt{-\ub q^2}\,
         H_1^{(1)}\bigl(\sqrt{- 4 \ub q^2}\bigr)
         & \mbox{for retarded $q^0>0$};
      \\
      -
      i\pi \, \sqrt{-\ub q^2}\,
         H_1^{(2)}\bigl(\sqrt{- 4 \ub q^2}\bigr)
         & \mbox{for advanced $q^0>0$}.
   \end {cases}
\end {equation}


\subsection {The crucial approximation}
\label {sec:crucial}

Now we will make the same approximation that we suggested for
the finite temperature limit: take the small-$Q$ approximation
of $\calGRup_{\sigma\mu}(Q,\ub)$.  At zero temperature, however,
we do not have temperature to define a natural scale $Q \ll T$,
and so the discussion of the small-$Q$ limit is a bit more complicated.
The zero-temperature ${\cal G}_{\sigma\mu}(Q,\ub)$ given by
(\ref{eq:calGzero}) simplifies
when $\ub Q^2 \ll 1$, in which limit it is simply
$\eta_{\sigma\mu}$.
\begin {quote}
   {\it The crucial approximation:}
   \begin {equation}
     \calGRup_{\sigma\mu}(Q,\ub) \simeq \eta_{\sigma\mu},
   \label {eq:crucial}
   \end {equation}
   where $Q=-Q_1-Q_2$ is the momentum conjugate to the measurement point
   $x$.
\end {quote}
We will discuss the validity of this approximation in a moment, but
first let's see what it gives.
Making the zero-temperature version (\ref{eq:crucial}) of the crucial
approximation in (\ref{eq:epsGRT0}) and using the relation
(\ref{eq:AvR}) between advanced and retarded propagators, we may
approximate (\ref{eq:Rf}) and (\ref{eq:epsGRT0}) as
\begin {subequations}
\label {eq:approx1}
\begin {equation}
  \jmu
  \simeq
  - \eta^{\mu\rho} \, \frac{\Aamp^2}{\gSG^2}
  \int_{0}^\infty \frac{d\ub}{\ub} \,
  \Field(x,\ub)^* i \tensor\partial_\rho \Field(x,\ub) ,
\label {eq:RfF0}
\end {equation}
where
\begin {equation}
  \Field(x,\ub)
  \equiv
  \int_q 
  \calGRup_{\perp}(q,\ub) \,
  \tilde\envelope_L(q-\kbig) \,
  e^{iq\cdot x} .
\label {eq:F0}
\end {equation}
\end {subequations}
(Note: $q\cdot x$ is a flat-space dot product
$\eta^{\mu\nu} q_{\mu} x_\nu$ here.)
The combination $- \Field^* i \tensor\partial_\rho \Field$ has
the same form as the expression for the
``current'' associated with a charged bosonic field in field theory.

The definition (\ref{eq:F0}) of $\Field$ is simply (up to normalization
factors)
a convolution of the retarded bulk-to-boundary propagator and
the positive-energy piece of the
classical source (\ref{eq:source}) in the 4-dimensional gauge
theory:
\begin {equation}
   \eta_{\mu\nu} \pol^\nu \Field(x,\bar u) \propto
   \int d^4y \> \calGRup_{\mu\alpha}(y,\bar u;x) \, \Aclplus^{\alpha}(y)
\end {equation}
where
\begin {equation}
  \Aclplus^\alpha(y)
  \equiv \pol^\alpha \Aamp e^{i\kbig\cdot y} \, \envelope_L(y) ,
\end {equation}
That is, $\Field(x,\ub)$ is a retarded solution to the
linearized 5-dimensional vector equation of motion that is proportional to
the source $\Aclplus$ on the boundary.
As time $x^0$ progresses, the bulk excitation represented by
$\Field$  will fall away from the boundary, into the fifth dimension.%
\footnote{
  Our nomenclature gets a bit convoluted here.  Earlier we said that
  our convention was that $\calGRup$ corresponds to waves propagating
  from the bulk to the boundary.  So why do we say here that
  $\Field(x,\ub)$ falls {\it into}\/ the horizon as $x^0$ increases?
  Note that $x$ is the {\it boundary}\/ point for
  $\calGRup(y,\ub;x)$, whereas $(y,\ub)$ is the bulk point.
  $x^0-y^0\to\infty$ is equivalent to $y^0-x^0 \to -\infty$:
  the bulk point becomes further and further back in time compared
  to the boundary point and so closer to the horizon.
  It might have been clearer in this respect to use notation like
  $\Field(\ub;x)$ or $\Field(\ub\to x)$ instead of $\Field(x,\ub)$
  [and perhaps similarly for $\calGRup(q,\ub)$],
  but we decided that would be too cumbersome.
}
This evolution of $\Field$ is roughly the type of problem studied in the
context of jet evolution by Hatta, Iancu, and Mueller \cite{HIM}.%
\footnote{
   A very minor difference is that for most of their paper
   they chose to focus on the
   evolution of the $A_0$ and $A_3$ components of the 5-dimensional
   field, whereas we have found it convenient to choose a transverse
   source, and so our $\Field$ tracks the evolution of the transverse
   components.
}
They made qualitative interpretations concerning the initial conditions
on $\Field$ and the result of its evolution.
Here,
we have seen the question of $\Field$'s time evolution
arise step by step from a problem posed completely in
the 4-dimensional field theory.  Consequently, we also have a
quantitative way to interpret the solution for $\Field$: use
it in (\ref{eq:RfF0}) [or the appropriate generalization to finite
temperature coming in section \ref{sec:finiteT}]
to find the current density response.

When does one expect the approximation (\ref{eq:crucial})
to be valid in the zero temperature
case?  It is correct in the limit
$R^2 g^{\mu\nu} Q_\mu Q_\nu = 4 \ub Q^2 \ll 1$,
where $Q^2$ means $\eta^{\mu\nu} Q_\mu Q_\nu = 4 Q_+ Q_-$.
Writing $Q_\pm \sim 1/\Delta x^\pm$ where $\Delta x^\pm$ are the
desired resolutions of $x^+$ and $x^-$ in the response, this suggest
that the approximation is valid when
\begin {equation}
   \Delta x^+ \, \Delta x^- \gg \ub .
\label {eq:resolve}
\end {equation}
The question then boils down to knowing the natural scale for
$\ub$ in this calculation.

Later on we'll see that one
of the natural scales for $\ub$ that arises in the analysis,
identified by Hatta et al.\ \cite{HIM},
is
\begin{equation}
   \ub \sim \frac{x^+}{\wbig} \,.
\label {eq:uscale1}
\end {equation}
However, we will find that a more important and
parametrically larger scale is
\begin {equation}
  \ub \sim \frac{(x^+)^2}{\wbig L} \,.
\label {eq:uscale2}
\end {equation}
The approximation (\ref{eq:crucial}) is then valid provided we only
apply the result to resolve questions on scales
\begin {equation}
   \Delta x^+ \, \Delta x^- \gg \frac{(x^+)^2}{\wbig L} \,.
\label {eq:resolve2}
\end {equation}
Since $\wbig L \gg 1$, there is no problem simultaneously
resolving $\Delta x^+$ and $\Delta x^-$ to scales small compared
to $x^+$.

If we want to resolve $\Delta x^-$ all the way down
to $\Delta x^- {\sim} L$ (as we shall implicitly do later)
and simultaneously resolve $\Delta x^+$ to better than
$x^+$ itself, then (\ref{eq:resolve2}) requires
$x^+ \ll \wbig L^2$.  These constraints are special to the
zero-temperature problem---at finite temperature, choosing
a resolution distance large compared to $1/T$ will be all that we
will need.

 Thanks to the approximation (\ref{eq:crucial}),
we now have just a {\it single}\/ 4-momentum integral
(\ref{eq:F0}) to evaluate or approximate, followed by a one-dimensional
$\ub$ integration (\ref{eq:RfF0}).
In Appendix \ref{app:conservation}, we verify that the
approximation (\ref{eq:RfF0}) obeys current conservation outside
of the space-time region of the external source.


\subsection {The high-energy approximation}
\label{sec:highEapprox}

Our next task is to evaluate $\Field$.
Using $q^2=4q_+ q_-$ and rewriting $q_-$ as $\wbig + \Delta q_-$,
the formula (\ref{eq:F0}) for $\Field$ becomes
\begin {equation}
  \Field(x,\ub)
  =
  e^{i\wbig x^-}
  \int \frac{2 \> dq_+ \> d(\Delta q_-)}{(2\pi)^2} \,
  \calGRup_{\perp}\bigl(4\ub(\wbig+\Delta q_-)q_+\bigr) \,
  \tilde\envelope_L^{(2)}(q_+,\Delta q_-) \,
  e^{iq_+x^+} e^{i\, \Delta q_-x^-}
  ,
\label {eq:F0explicit}
\end {equation}
where the transverse bulk-to-boundary propagator extracted from
(\ref{eq:calGzero}) depends only on the combination $\ub q^2$:
\begin {equation}
   {\cal G}_\perp(\ub q^2)
   \equiv {\cal G}_\perp(q,\ub)
   = \sqrt{4\ub q^2}\,  K_1\bigl(\sqrt{4\ub q^2}\bigr) .
\label {eq:Gperpzero}
\end {equation}
$\tilde\envelope_L^{(2)}(q_+,q_-)$ is the two-dimensional Fourier
transform (\ref{eq:envelope2}) of the source envelope, which for a
Gaussian envelope (\ref{eq:Genvelope}) would give
\begin {equation}
   \tilde\envelope_L^{(2)}(q_+,\Delta q_-)
   = 2\pi L^2 \, e^{-(q_+ L)^2} e^{-(\Delta q_- \, L)^2} .
\label {eq:Genvelopetilde}
\end {equation}
The smooth envelope function restricts support for the integral
(\ref{eq:F0explicit}) to $\Delta q_- \lesssim 1/L$ and therefore
$\Delta q_- \ll \wbig$.  We may therefore approximate
\begin {subequations}
\label {eq:HEapprox}
\begin {equation}
   \calGRup_{\perp}\bigl(4\ub(\wbig+\Delta q_-)q_+\bigr)
   \simeq
   \calGRup_{\perp}\bigl(4\ub\wbig q_+\bigr)
\label {eq:second}
\end {equation}
in (\ref{eq:F0explicit}).

With this approximation, the $\Delta q_-$ integration gives
\begin {equation}
  \Field(x,\ub)
  \simeq
  e^{i\wbig x^-}  \,
  \int \frac{dq_+}{2\pi} \,
  \calGRup_{\perp}(4\ub\wbig q_+) \,
  \envelope_L^{(2)}(q_+;x^-) \,
   e^{iq_+x^+} ,
\label {eq:F0approx}
\end {equation}
\end {subequations}
where
\begin {equation}
   \envelope_L^{(2)}(q_+;x^-)
   \equiv
   \int dx^+ \> \envelope_L(x) \, e^{-iq_+ x^+}
\label{eq:envelope2xm}
\end {equation}
is insignificant unless $q_+ \lesssim 1/L$ and $x^- \lesssim L$.
For the Gaussian envelope (\ref{eq:Genvelope}),
\begin {equation}
   \envelope_L^{(2)}(q_+;x^-) =
   2\pi^{1/2}L \, e^{-(q_+ L)^2} e^{-(x^-)^2/4L^2} .
\label {eq:Genvelope2}
\end {equation}

\begin {figure}
\begin {center}
  \includegraphics[scale=0.4]{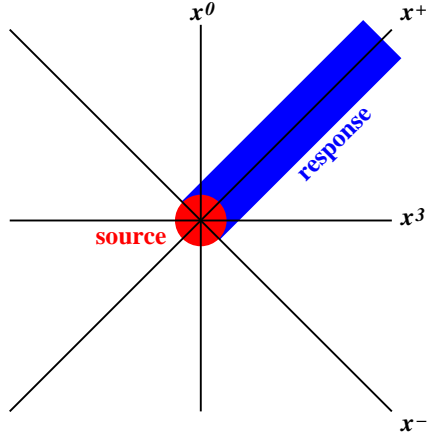}
  \caption{
     \label{fig:xminus}
     The space-time development of a high-energy excitation at
     zero temperature.
  }
\end {center}
\end {figure}

As depicted in fig.\ \ref{fig:xminus}, the response (\ref{eq:F0approx})
is localized in $x^-$.  Up to the matters of resolution discussed at the end
of section \ref{sec:crucial}, the high energy excitation produced
by the source simply propagates along
the lightcone to the right of the 1+1 dimensional source region.

Now return to eq.\ (\ref{eq:RfF0}) for the current response:
\begin {equation}
  \jmu
  \simeq
  - \eta^{\mu\rho} \, \frac{\Aamp^2}{\gSG^2}
  \int_{0}^\infty \frac{d\ub}{\ub} \,
  \Field(x,\ub)^* i \tensor\partial_\rho \Field(x,\ub) ,
\end {equation}
The derivative $\partial_\rho\Field$ of (\ref{eq:F0approx}) will be
dominated by the term where the derivative hits the highly-oscillating
factor of $e^{i\wbig x^-}$:
\begin {equation}
  \jmu
  \simeq
  2 \kbig^\mu \, \frac{\Aamp^2}{\gSG^2}
  \int_{0}^\infty \frac{d\ub}{\ub} \,
  |\Field(x,\ub)|^2 .
\label{eq:Rapprox2}
\end {equation}

In this approximation,
the statement $\partial_\mu j^\mu = 0$ of current conservation
(outside of the source region) is
$\partial_+ \langle j^{+} \rangle = 0$, which means that
(for zero temperature)
\begin {equation}
   \int_{0}^\infty \frac{d\ub}{\ub} \> |\Field(x,\ub)|^2
\label {eq:djzero}
\end {equation}
should be independent of $x^+$ outside of the source region.


\subsection {An approximation that doesn't quite work}
\label {sec:approx3}

We now discuss a further approximation, which will be flawed.
But it will be close to what we eventually need, and how it fails will be
instructive.  We will fix it up afterward.

\begin {figure}
\begin {center}
  \includegraphics[scale=0.4]{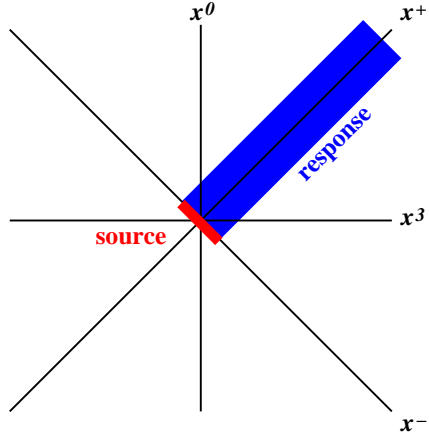}
  \caption{
     \label{fig:xminus2}
     Picture of the source and response
     after the (flawed) approximation (\ref{eq:Gapprox3})
     or (\ref{eq:approx3}).
  }
\end {center}
\end {figure}

In the case of a Gaussian envelope (\ref{eq:Genvelope2}),
the evolution (\ref{eq:F0approx}) of $\Field$ is
\begin {equation}
  \Field(x,\ub)
  \simeq
  2\pi^{1/2}L \, 
  e^{i\wbig x^-}  \,
  e^{-(x^-)^2/4L^2}
  \int \frac{dq_+}{2\pi} \,
  \calGRup_{\perp}(4\ub\wbig q_+) \,
  e^{-(q_+ L)^2}
  e^{iq_+x^+} .
\label {eq:F0approxG}
\end {equation}
The integral determines the $x^+$ dependence of the
response.
There are three factors in the integrand, associated with three
different scales for $q_+$: the corresponding wavelengths
$1/q_+$ are
$4\ub\wbig$, $L$, and $x^+$ respectively.
To study the response far away from the source, where
$L \ll x^+$,
one might hope one could treat $L$ as arbitrarily small in
the integrand of (\ref{eq:F0approxG}), replacing it by a
convergence factor:
\begin {subequations}
\label {eq:Gapprox3}
\begin {equation}
     e^{-(q_+ L)^2}
     \rightarrow
     e^{-\epsilon q_+^2} ,
\end {equation}
where $\epsilon$ is infinitesimal.
Mathematically, this approximation corresponds to replacing the
Gaussian source envelope
\begin {equation}
   \envelope_L(x) =
   e^{-\frac14 (x^+/L)^2} e^{-\frac14 (x^-/L)^2}
   \to
   2\pi^{1/2}\,L \, e^{-\frac14 (x^+/L)^2} \delta(x^-)
\end {equation}
\end {subequations}
and so corresponds to replacing fig.\ \ref{fig:xminus} by
fig.\ \ref{fig:xminus2}.
The generalization to generic source envelopes would be
\begin {equation}
  \Field(x,\ub)
  \to
  e^{i\wbig x^-}  \,
  \envelope_L^{(2)}(0;x^-)
  \int \frac{dq_+}{2\pi} \,
  \calGRup_{\perp}(4\ub\wbig q_+) \,
   e^{-\epsilon q_+^2} e^{iq_+x^+} .
\label {eq:approx3}
\end {equation}

Changing integration variables from $q_+$ to
\begin {equation}
   \kappa \equiv 4 \ub \wbig q_+ ,
\end {equation}
we can rewrite (\ref{eq:approx3}) as
\begin {equation}
  \Field(x,\ub)
  \to
  e^{i\wbig x^-}  \,
  \frac{\envelope_L^{(2)}(0;x^-)}{4\ub\wbig} \,
  I\Bigl( \frac{x^+}{4\ub\wbig} \Bigr) ,
\label {eq:Fplusapprox}
\end {equation}
where
\begin {equation}
  I(s) \equiv
  \int \frac{d\kappa}{2\pi} \,
  \calGRup_{\perp}(\kappa) \,
   e^{-\epsilon \kappa^2} e^{i\kappa s} .
\label {eq:I}
\end {equation}
Note that the only parameter in the definition of $I(s)$ is its argument
$s$, and so a natural scale of the problem is $s \sim 1$.  From
(\ref{eq:Fplusapprox}), this
corresponds to
\begin {equation}
   \ub \sim \frac{x^+}{\wbig} ,
\end {equation}
which was the first of the two scales for $\ub$ previewed back in
(\ref{eq:uscale1}).


\subsubsection* {An apparent paradox}

We can now see the origin of a
problem.  The response (\ref{eq:Rapprox2}) after all our approximations
becomes
\begin {equation}
  \jmu
  \to
  \frac{2\kbig^\mu}{(x^+)^2} \, \frac{\Aamp^2}{\gSG^2} \,
  \bigl| \envelope_L^{(2)}(0;x^-) \bigr|^2
  \int_{0}^\infty ds \> s |I(s)|^2 ,
\label {eq:paradox}
\end {equation}
where
\begin {equation}
   s = \frac{x^+}{4\ub\wbig}
\end {equation}
is $\infty$ on the boundary $\ub{=}0$.
This result appears to {\it depend}\/ on $x^+$ as
$1/(x^+)^2$, which is
inconsistent with current conservation, as discussed for
(\ref{eq:djzero}).
The loophole to this paradox is that the integral in (\ref{eq:paradox})
turns out to be $s{\to}0$ divergent (corresponding to contributions
far away from the boundary).
Specifically, explicit evaluation of the integral
$I(s)$ defined by (\ref{eq:I}) yields
(see Appendix \ref{app:I})
\begin {equation}
  I(s) = -\frac{i}{s^2} \, e^{i/s} \, \theta(s) ,
\label {eq:Iresult}
\end {equation}
where $\theta(s)$ is the step function, giving
\begin {equation}
   \int_{0}^\infty ds \> s |I(s)|^2 \propto
   \int_0^\infty \frac{ds}{s^3} \,.
\end {equation}
As we will now see, this divergence arises from ignoring the
width $L$ of the source in $x^+$ in order to make the last approximation
(\ref{eq:approx3}).


\subsection {Fixing the last approximation}
\label {sec:fix}

We will now step away from the last approximation (\ref{eq:approx3})
but will still treat $L$ as relatively ``small'' in a sense to be
made precise in a moment.
We know from the previous analysis that what will be important are
small values of $s = x^+/4\ub\wbig$.  Since the
full integral in eq.\ (\ref{eq:F0approx}) for $\Field$
is more complicated than
$I(s)$, we will make our lives easier by approximating it in the
small $s$ (large $\ub$) limit instead of attempting to evaluate it exactly for
general $s$.  As we'll see, the appropriate approximation in this
limit is the method of steepest descent.


\subsubsection{Steepest descent evaluation of $\Field$}
\label {sec:steep0}

For the sake of concreteness, we focus for the moment
on the case of the Gaussian envelope function,
which gives (\ref{eq:F0approxG}) for $\Field$, proportional to
\begin {equation}
  {\cal I}(x^+,\ub) \equiv
  \int \frac{dq_+}{2\pi} \,
  \calGRup_{\perp}(4\ub\wbig q_+) \,
  e^{-(q_+ L)^2}
  e^{iq_+x^+} .
\end {equation}
The first thing to note is that the
contribution to this integral from space-like momenta
($q_+{>}0$) is small
in the limit $\ub \to \infty$.  That's because, in that case,
$\calGRup_\perp(4\ub\wbig q_+)$ falls rapidly
with $q_+$.  The dominant large $\ub$ behavior therefore comes
from the time-like ($q_+{<}0$) region of integration, where
$\calGRup$ is oscillatory.  Writing
$q_+ = -k-i\epsilon$ (retarded prescription), the
integral then gives
\begin {equation}
  {\cal I}(x^+,\ub) \simeq
  \int_0^\infty \frac{dk}{2\pi} \,
  {\cal G}_\perp(e^{-i\pi}4\ub\wbig k) \,
  e^{-(k L)^2}
  e^{-ik x^+}
  .
\end {equation}
Now use the large argument approximation to eq.\ (\ref{eq:Gperpzero})
for ${\cal G}_\perp$ (which we will verify later is appropriate):
\begin {equation}
   {\cal G}_{\perp}(\xi) \simeq
   \sqrt{\pi} \, \xi^{1/4} e^{-2\sqrt{\xi}} ,
\end {equation}
giving
\begin {equation}
  {\cal I}(x^+,\ub) \simeq
  e^{-i\pi/4} \sqrt{\tfrac{\pi}{2}} \, (16\ub\wbig)^{1/4}
  \int_0^\infty \frac{dk}{2\pi} \,
  k^{1/4}
  e^{-{\cal S}(k)} ,
\end {equation}
where
\begin {equation}
  {\cal S}(k) =
  -i \sqrt{16\ub \wbig k}
  +i k x^+
  +(k L)^2
  .
\label {eq:calSzero}
\end {equation}
We now perform steepest descent by finding the zero $k{=}k_\star$
of $\partial{\cal S}/\partial k$.  Formally expanding the solution
in powers of $L$ and keeping only terms through $L^2$, this extremum is at
\begin {equation}
  k_\star \simeq
  \frac{4\ub\wbig}{(x^+)^2}
  \left[
     1 + i \, \frac{16\ub\wbig}{(x^+)^3} \, L^2
  \right] .
\label {eq:kstar}
\end {equation}
The condition for treating $L$ as small in this way is that
the magnitude of the $L^2$ term be small compared to that of
the $L^0$ term, which is the condition
\begin {equation}
  \ub \ll \frac{(x^+)^3}{\wbig L^2}
     =  \frac{(x^+)^2}{\wbig L} \, \frac{x^+}{L}
  .
\label {eq:smallL}
\end {equation}
The value of ${\cal S}$ corresponding to (\ref{eq:kstar}) is
\begin {equation}
  {\cal S}(k_\star)
  \simeq
  -i\,\frac{4\ub\wbig}{x^+}
  + \frac{(4\ub\wbig L)^2}{(x^+)^4} \,,
\end {equation}
giving
\begin {equation}
  e^{-{\cal S}(k_\star)} \simeq
  e^{i/s} \exp\left[ - \frac{(4\ub\wbig L)^2}{(x^+)^4} \right] .
\label {eq:steep}
\end {equation}
The $e^{i/s}$ factor is just the oscillatory factor in the
earlier result (\ref{eq:Iresult}) for $I(s)$.
The size $|{\cal S}(k_\star)|$ of the exponent
will be large, justifying the steepest descent approximation,
when $s$ is small.  By considering the first correction to the
exponent in powers of $L$, we have now found in (\ref{eq:steep})
that there is a decreasing exponential that kicks in and cuts
off the large $\ub$ behavior.  This large-$\ub$ suppression factor
starts to kick in when
\begin {equation}
  \ub \sim \frac{(x^+)^2}{\wbig L} ,
\label {eq:zscale2}
\end {equation}
which is the second scale for $\ub$ previewed in
(\ref{eq:uscale2}).
At this $\ub$, the small-$L$ approximation
(\ref{eq:smallL}) is valid (far away from the source, $x^+ \gg L$), and so
our treatment of $L^2$ effects as a perturbation to
the evaluation of ${\cal S}(k_\star)$ is justified.

The reason we could not just stick with the leading order term
$-i/s$ in the exponent ${\cal S}(k_\star)$ is because it was
pure imaginary: we had to go to first order in $L^2$ to
find the leading contribution to the {\it real part}\/ of the exponent.
In contrast, a leading-order evaluation of the algebraic
prefactor of the exponential is good enough, and so we can
ignore the effects of $L$ on that prefactor.  The upshot is
that (\ref{eq:steep}) tells us to modify our previous analysis
of section \ref{sec:approx3} by simply replacing
\begin {equation}
  I(s) \to
  I(s) \, \exp\left[ - \frac{(4\ub\wbig L)^2}{(x^+)^4} \right] 
  .
\label {eq:Ifix}
\end {equation}
Alternatively, one may obtain the same result at small $s$ by
explicitly finishing the steepest descent analysis
by expanding ${\cal S}(k)$ to quadratic order about ${\cal S}(k^*)$
and doing the Gaussian integral to get the prefactor.

From (\ref{eq:F0approxG}), (\ref{eq:Iresult}), and (\ref{eq:Ifix}),
the final expression for
$\Field$ is
\begin {subequations}
\label{eq:Ffinalboth}
\begin {equation}
  \Field(x,\ub) \simeq
  -i 2\pi^{1/2} \,
  \frac{4\ub\wbig L}{(x^+)^2} \,
  e^{i\wbig x^-} e^{-(x^-)^2/4L^2} \,
  e^{i4\ub\wbig/x^+}
  \exp\left[ - \frac{(4\ub\wbig L)^2}{(x^+)^4} \right] \,
  \theta(x^+)
  .
\label {eq:FfinalG}
\end {equation}
The generalization of the above analysis to generic source envelopes
simply replaces the large-$\ub$ suppression factor above by
the envelope function $\envelope_L^{(2)}(q_+;x^-)$ evaluated
at the saddle point $q_+=-k_\star$:
\begin {equation}
  \Field(x,\ub) \simeq
  -i \,
  \frac{4\ub\wbig}{(x^+)^2} \,
  e^{i\wbig x^-} \,
  e^{i4\ub\wbig/x^+}
  \envelope_L^{(2)}\Bigl( -\frac{4\ub\wbig}{(x^+)^2} ; x^- \Bigr) \,
  \theta(x^+)
  .
\label {eq:Ffinal}
\end {equation}
\end {subequations}


\subsubsection{Final result for current response}

Inserting (\ref{eq:Ffinal}) for $\Field$ into the expression
(\ref{eq:Rapprox2}) for $\jmu$, and changing integration
variable from $\ub$ to $q_+ \equiv - 4\ub\wbig/(x^+)^2$,
gives the final zero-temperature result
\begin {subequations}
\label {eq:FinalzeroAll}
\begin {equation}
  \jmu
  \simeq
  2 \kbig^\mu \, \frac{\Aamp^2}{\gSG^2} \, \theta(x^+)
  \int_{-\infty}^0 dq_+ \>
  |q_+| \, \bigl|\envelope_L^{(2)}(q_+;x^-)\bigr|^2
\label{eq:Finalzero}
\end {equation}
for $|x^+| \gg L$, with $\gSG$ given by (\ref{eq:gSG}).
This result is independent of $x^+$ after the source turns off,
as required by the discussion
of current conservation surrounding (\ref{eq:djzero}) [which we
remind the reader is subject to the resolution requirement
(\ref{eq:resolve2}) for making our approximations].
For the case of the Gaussian envelope (\ref{eq:Genvelope2}),
the result (\ref{eq:Finalzero}) is
\begin {align}
  \jmu
  \simeq
  2\pi \kbig^\mu \, \frac{\Aamp^2}{\gSG^2} \,
  e^{-(x^-)^2/2 L^2} \,
  \theta(x^+) .
\end {align}
\end {subequations}

The total charge density per unit transverse area
at late times can be taken from
(\ref{eq:Finalzero}) as
\begin {align}
   \Chargebar &=
   \int dx_3 \jo \biggl|_{x^+ \gg L}
\nonumber\\
   &\simeq
   2 \wbig \, \frac{\Aamp^2}{\gSG^2}
   \int dx^- \int_{-\infty}^0 dq_+
   |q_+| \, \bigl|\envelope_L^{(2)}(q_+;x^-)\bigr|^2 ,
\end {align}
which may be rewritten as
\begin {equation}
   \Chargebar
   \simeq
   8 \pi \wbig \, \frac{\Aamp^2}{\gSG^2}
   \int \frac{2\,dq_+ \, dq_-}{(2\pi)^2} \>\theta(-q_+) \,
   |q_+| \, \bigl|\tilde\envelope_L^{(2)}(q_+,q_-)\bigr|^2
\label {eq:charge}
\end {equation}
[from where one may now see where the form of the normalizing denominator in
(\ref{eq:formL}) comes from].
In the case of the Gaussian envelope (\ref{eq:Genvelopetilde}),
\begin {equation}
   \Chargebar = 
   (2\pi)^{3/2} \wbig L \, \frac{\Aamp^2}{\gSG^2} \,.
\end {equation}
As a simple check of our calculation, in Appendix \ref{app:charge}
we make a completely independent
computation of the total charge produced by the source
by using Ward identities.
The result is the same as (\ref{eq:charge}).


\subsection {Schr\"odinger interpretation \textit{\`a la} Hatta, Iancu, and
             Mueller}
\label {sec:HIM}

In 4-momentum space, the zero-temperature, linearized
equation of motion $d \, {}^\star\!F=0$ for the
transverse components of the 5-dimensional vector field is
\begin {equation}
   \left( \partial_{\ub}^2 - \frac{q^2}{\ub} \right) A_\perp(q,\ub)
   = 0 .
\end {equation}
The solution, appropriately normalized, is just the transverse
bulk-to-boundary propagator (\ref{eq:Gperpzero}).
If we approximate $q^2 \simeq 4\wbig q_+$ and Fourier transform
$q_+$ to $x^+$, the above equation can be rewritten as
\begin {equation}
   \left( \partial_{\ub}^2 + i\, \frac{4\wbig}{\ub} \partial_+ \right)
   A_\perp(x^+,\ub) \simeq 0 .
\label {eq:calAeom}
\end {equation}
Our solution (\ref{eq:Ffinalboth}) for $\Field(x,\ub)$
(approximately) solves
(\ref{eq:calAeom}).%
\footnote{
   By direct substitution of (\ref{eq:calAxpu}) in (\ref{eq:calAeom}), one may
   double-check that it is an approximate solution everywhere in the region
   $\ub \lesssim (x^+)^2/EL$,
   which is the region where $\Field$ is non-negligible.
}
Focusing on the Gaussian source case for concreteness,
the $x^+$ and $\ub$ dependence of $\Field$ is
\begin {equation}
  \Field \propto
  \frac{\ub\wbig L}{(x^+)^2} \,
  e^{i4\ub\wbig/x^+}
  \exp\left[ - \frac{(4\ub\wbig L)^2}{(x^+)^4} \right]
  .
\label {eq:calAxpu}
\end {equation}

Hatta, Iancu, and Mueller \cite{HIM} noted that the equation
(\ref{eq:calAeom})
(as well as its finite-temperature version in the case of small $u$)
can be recast in the form of a Schr\"odinger-like equation by
changing variables from $\ub$ to $z = 2\sqrt{\ub}$ and
redefining
\begin {equation}
   A_\perp(x^+,\ub) = \sqrt{z} \, \phi(x^+,z) .
\end {equation}
The equation of motion (\ref{eq:calAeom}) then becomes
\begin {equation}
   2i \partial_+ \phi
   = \left( - \frac{\partial_z^2}{2\wbig} + \frac{3}{8\wbig z^2} \right) 
     \phi ,
\end {equation}
which is a Schr\"odinger equation with potential energy $\propto z^{-2}$,
provided $x^+$ is interpreted as time.  The conserved ``probability''
of this Schr\"odinger equation is
\begin {equation}
   \int dz \> |\phi|^2 =
   \tfrac12 \int \frac{d\ub}{\ub} \> |\Field|^2 .
\end {equation}
From our discussion surrounding (\ref{eq:djzero}), 
we see that, for zero temperature,
conservation of probability in this Schr\"odinger problem is
equivalent to  conservation of charge in the original field theory problem.

\begin {figure}
\begin {center}
  \includegraphics[scale=0.8]{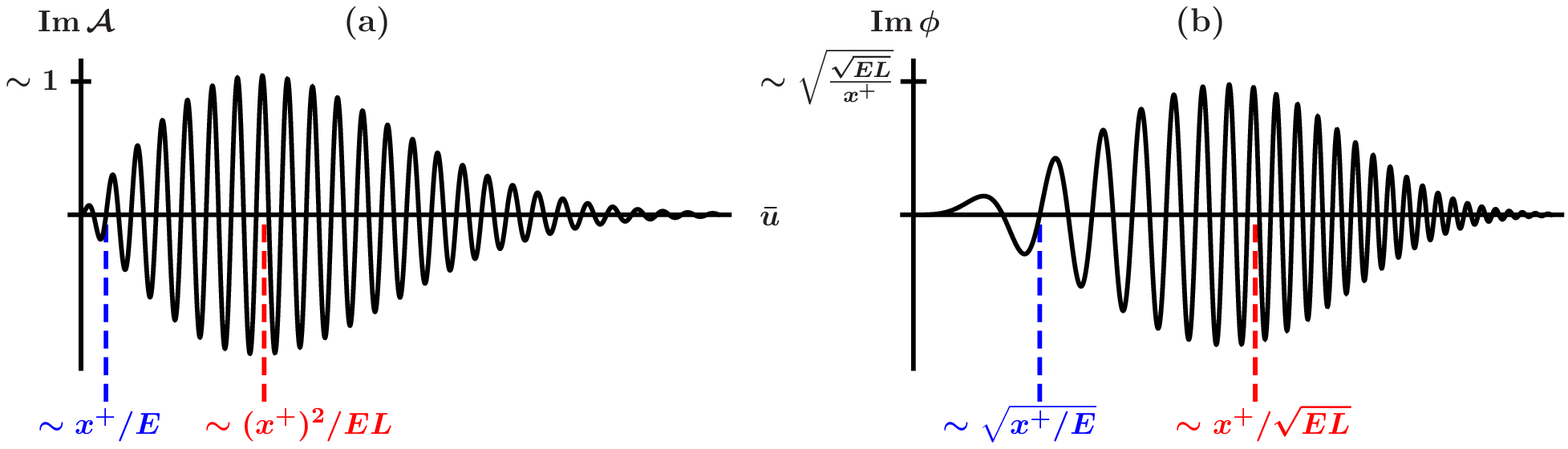}
  \caption{
     \label{fig:calA}
     Qualitative pictures of the real or imaginary parts of (a)
     $\Field$ versus $\ub$ and (b) $\phi = \Field/\sqrt{z}$ versus
     $z = 2\sqrt{\ub}$.
  }
\end {center}
\end {figure}

Qualitative sketches of our result (\ref{eq:calAxpu})
for $\Field$ and $\phi$ are given
in fig.\ \ref{fig:calA}.  Both (i) the
wavelength $z \sim \sqrt{x^+/E}$ of the first oscillation and
(ii) the location $z \sim x^+/\sqrt{EL}$ of the bulk of the probability
grow with time ($x^+$).
The substantive difference with the
study of Hatta et al.\ \cite{HIM} is that they studied
non-localized solutions,%
\footnote{
  There are some other differences.
  The $x^+$ in our discussion plays the role of $2t$ in theirs.
  Similarly, our $A_+$ plays the role of their $A_0$ in what follows.
  They focus more on the longitudinal mode $A_+$ than the
  transverse mode (\ref{eq:calAhatta}).
  The two are qualitatively similar, as can
  be seen from our (\ref{eq:calGzero}), except
  that Hatta et al.\ choose to work with $a \equiv \partial_{\ub} A_+$
  instead of $A_+$ directly.
  They choose boundary conditions so that
  $a \propto (x^+)^{-1} e^{i4\ub\wbig/x^+}$ [see their (3.13)], whereas
  the boundary conditions determined by our type of field theory problem
  would give a different
  solution
  (in the $L{\to}0$ limit)
  to the same second-order equation [their (3.11), dropping the
  $K^2$ term, with $\psi$ and $a$ related by their (2.8)]:
  $a = \partial_{\ub} \Field
  \propto (x^+)^{-2} (1 + i 4\ub\wbig/x^+) e^{i4\ub\wbig/x^+}$
  from our (\ref{eq:calAhatta}).
}
\begin {equation}
  \Field \propto
  \frac{\ub}{(x^+)^2} \,
  e^{i4\ub\wbig/x^+} \,,
\label {eq:calAhatta}
\end {equation}
which do not decay at large $\ub$ and have infinite
normalization $\int dz \> |\phi|^2$.  These solutions
correspond to taking $L\to 0$ in (\ref{eq:calAxpu}) and so,
in our context,
making the failed approximation (\ref{eq:approx3}).
Correspondingly, the only $z$ scale that Hatta et al.\ identify
at zero temperature is the scale
$z \sim \sqrt{x^+/E}$
of (\ref{eq:uscale1}),
associated with the first oscillation, which they call the ``diffusion''
distance in $z$.  However, the bulk of the probability density is
instead characterized by the larger scale
$z \sim x^+/\sqrt{E L}$ of
(\ref{eq:uscale2}),
which grows faster with time.

When we go to finite temperature, we will correspondingly
find two time scales for how long it takes $\Field$
to fall into the event horizon.
The time scale for the first oscillation of $\Field$ to fall into the
event horizon will turn out to be the scale
$(x_3)_{\rm max} \sim \wbig^{1/3}/T^{4/3}$
of (\ref{eq:xmax}),
as found by Hatta et al.
But the time scale for the bulk of the ``probability'' density to
fall into the horizon will turn out to be the shorter
time scale $(x_3)_{\rm dominant} \sim (\wbig L)^{1/4}/T$
of (\ref{eq:xdominant}).

\begin {figure}
\begin {center}
  \includegraphics[scale=0.6]{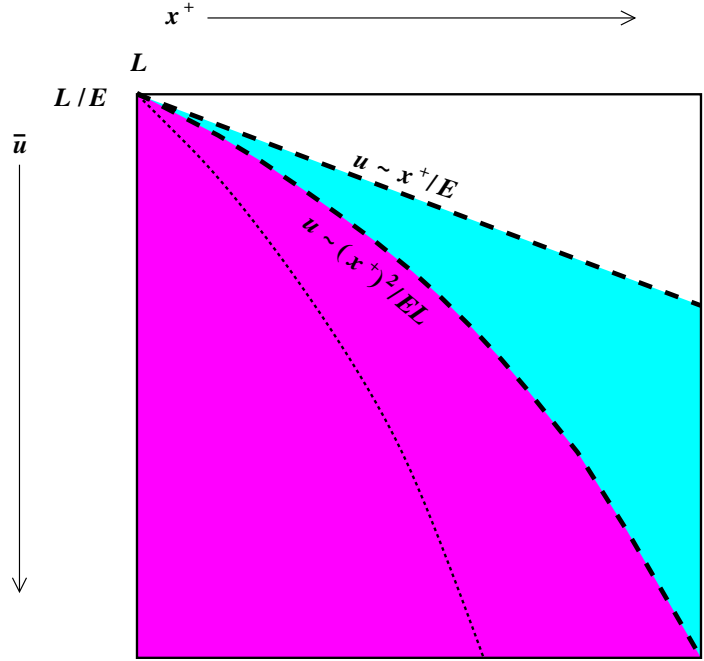}
  \caption{
     \label{fig:SteepDescent0}
     Qualitative picture of the two $\ub$ scales of fig.\ \ref{fig:calA}
     versus $x^+$.  The top of the plot represents $\ub$ very close to
     the boundary $(\ub \sim L/E)$; lower points are deeper in the
     fifth dimension.  The dotted line represents
     $\ub \sim (x^+)^3/\wbig L^2$, above which $L$ can be treated
     as perturbatively small in a steepest descent analysis.
  }
\end {center}
\end {figure}

To facilitate later comparison with the finite temperature case,
we repackage in fig.\ \ref{fig:SteepDescent0}
the qualitative information from fig.\ \ref{fig:calA}.
The $\ub$ scales marked by the dashed lines in fig.\ \ref{fig:calA}
are represented on a plot of $\ub$ versus $x^+$ by the dashed curves in
fig.\ \ref{fig:SteepDescent0}.  The region parametrically
below the top dashed curve
is where $\Field$ has many oscillations and steepest descent is a
useful method of approximation.  The (magenta) region parametrically below
the lower dashed curve is where the amplitude of $\Field$
is exponentially suppressed.

\section{The finite-temperature calculation}
\label {sec:finiteT}

We now apply to the finite-temperature case the methods just
presented for the zero-temperature case.  In this section, we
will use units where $2\pi T = 1$.
Powers of $2\pi T$ may be restored by dimensional
analysis, replacing
\begin {align}
   x^\mu &\to (2\pi T) x^\mu ,
\\
   q^\mu &\to (2\pi T)^{-1} q^\mu ,
\\
   L &\to (2\pi T)L ,
\\
   \wbig &\to (2\pi T)^{-1} \wbig
\end {align}
throughout this section.

The AdS${}_5$-Schwarzschild metric is
\begin {equation}
  ds^2 = \frac{R^2}{4} \left[ \frac{1}{u}(-f \, dt^2 + d\x^2)
         + \frac{1}{u^2 f} \, du^2 \right] ,
\label {eq:metricT}
\end {equation}
and the basic equations (\ref{eq:basic}) for our problem become
\begin {subequations}
\label {eq:basicT}
\begin {multline}
  \jmu
\\
  \simeq
  2 i \Aamp^2 \eta^{\mu\nu}
  \int_{Q_1Q_2}
  G^{\rm R}_{\perp\perp\nu}(Q_1,Q_2;Q) \,
  \tilde\envelope^\econj_L(Q_1-\kbig) \,
  \tilde\envelope^\econj_L(Q_2+\kbig) \,
  e^{-iQ_1\cdot x}
  e^{-iQ_2\cdot x}
  \biggl|_{Q=-Q_1-Q_2} ,
\label {eq:RfT}
\end {multline}
and
\begin {equation}
   G^{\rm R}_{\perp\perp\mu}
   =
   i(Q_2-Q_1)_\rho \,
   \frac{1}{2\gSG^2} \int_0^1 \frac{du}{u} \>
   \finv^{\rho\sigma} \,
     \calGAup_{\perp}(Q_1,u) \,
     \calGAup_{\perp}(Q_2,u) \,
     \calGRup_{\sigma\mu}(Q,u) ,
\label {eq:epsGRT}
\end {equation}
with
\begin {equation}
   \finv^{\rho\sigma} \equiv
   \frac{R^2 g^{\rho\sigma}}{4u} =
   \left(
   \begin{smallmatrix}
      -\tfrac{1}{f}&&&\\
      &1&&\\
      &&1&\\
      &&&1
   \end{smallmatrix}
   \right)^{\rho\sigma} .
\end {equation}
\end {subequations}


\subsection {The crucial approximation}

\subsubsection {The small $Q$ limit}
\label {sec:crucialT}

The first thing we need is the low momentum approximation for
the bulk-to-boundary propagator $\calGRup_{\sigma\mu}(Q,u)$
associated with the measurement point $x$ in (\ref{eq:basicT}).
The issue of resolution scale will be much more straightforward
than in the zero-temperature case: Here we will simply restrict
attention to distance and time scales large compared to $1/T$,
and so small $Q$ will mean that all components of $Q$ are
small compared to $T$.

Because our problem is transverse-translational invariant,
$\Q_\perp = 0$ and we focus on $\omega \equiv Q^0$ and $k \equiv Q^3$.
In Appendix \ref{app:smallQ},
we show that
the small-$Q$ form of the bulk-to-boundary propagator in $A_5{=}0$ gauge is,
to {\it leading}\/ order in the size of $\omega$ and $k^2$, given by%
\footnote{
  We are interested in leading order in $\omega$ and $k^2$, and not
  simply leading order in $\omega$ and $k$, because we are interested
  in studying diffusion, for which $\omega \sim k^2$.
}
\begin {subequations}
\label {eq:calGsmallQ}
\begin {align}
   \calGRup_{0\mu}(\omega,k) \, \eta^{\mu\nu} a_\nu
   &\simeq \phantom{-}\frac{\omega}{i\omega-k^2} (i a_0 + k a_3)
   - \frac{k}{i\omega-k^2} (1-u)^{1-i\omega/2}(k a_0 + \omega a_3) ,
\\
   \calGRup_{3\mu}(\omega,k) \, \eta^{\mu\nu} a_\nu
   &\simeq - \frac{k}{i\omega-k^2} (i a_0 + k a_3)
   + \frac{i}{i\omega-k^2} (1-u)^{-i\omega/2}(k a_0 + \omega a_3) ,
\\
\noalign{\noindent and (irrelevant here except for
comparison)}
   \calGRup_{\perp\mu}(\omega,k) \, \eta^{\mu\nu} a_\nu
   &\simeq
   (1-u)^{-i\omega/2} a_\perp ,
\label {eq:calGperpsmallQ}
\end {align}
\end {subequations}
over the entire range of $u$.
The arbitrary 4-vector $a_\mu$ above
represents the source on the boundary and is introduced here
as a notational device for displaying the
individual components $\calGRup_{\sigma\mu}$ of $\calGRup$.
Note that the above expressions have the property that
\begin {equation}
  \calGRup_{\sigma\mu}(\omega,k) \, \eta^{\mu\nu} Q_\nu
  = Q_\sigma ,
\end {equation}
which is a general property of the bulk-to-boundary propagator
in $A_5{=}0$ gauge, arising from gauge invariance
[see (\ref{eq:calGdotQ}) in appendix \ref{app:conservation}].
The small-$Q$ form (\ref{eq:calGperpsmallQ}) of $\calGRup_\perp$
is irrelevant simply because the transverse-translation invariance
of our source implies that there will be no expectation of
$j^{\perp}$ (which involves transverse derivatives of the fields)
in the response.

Eqs.\ (\ref{eq:calGsmallQ}) are a bit cumbersome to deal with, and
it will greatly help to first make some additional approximations.
We will see later that the terms with factors of $(1-u)^{-i\omega/2}$
will never be important for small $1-u$ in our evaluation of $\jmu$.
Because of this, we can approximate $(1-u)^{-i\omega/2} \simeq 1$
in these terms, to leading order in the size of $\omega$.
Then (\ref{eq:calGsmallQ}) simplifies to
\begin {subequations}
\label {eq:calGsmallQ2}
\begin {align}
   \calGRup_{0\mu}(\omega,k) \, \eta^{\mu\nu} a_\nu
   &\simeq \frac{\omega}{i\omega-k^2} (i a_0 + k a_3)
   - \frac{k}{i\omega-k^2} (1-u)(k a_0 + \omega a_3) ,
\label {eq:calG0smallQ2}
\\
   \calGRup_{3\mu}(\omega,k) \, \eta^{\mu\nu} a_\nu
   &\simeq a_3 .
\end {align}
\end {subequations}
Of the terms remaining, we will later see that the one which
dominates the calculation of the charge deposition
$\dep(x)$ of (\ref{eq:final}) is the fist term in
(\ref{eq:calG0smallQ2}), provided we only wish to resolve
$\dep(x)$ on scales large compared to the source size
$L$ as in (\ref{eq:final}).
For the sake of simplifying the presentation, we will
ignore the other terms for now and replace
(\ref{eq:calGsmallQ}) by
\begin {subequations}
\label {eq:calGsmallQ3}
\begin {align}
   \calGRup_{0\mu}(\omega,k) \, \eta^{\mu\nu} a_\nu
   &\to \frac{\omega}{i\omega-k^2} (i a_0 + k a_3) ,
\label {eq:calG0smallQ3}
\\
   \calGRup_{3\mu}(\omega,k) \, \eta^{\mu\nu} a_\nu
   &\to 0 .
\end {align}
\end {subequations}
Once we see how the calculation works out, we will
return in sec.\ \ref{sec:revisitsmallQ}
to see why the $(1-u)^{-i\omega/2}$ factors in
(\ref{eq:calGsmallQ}) and the
other terms in (\ref{eq:calGsmallQ2}) are unimportant.

Note that the relationship
$\calGRup_{\sigma3} = ik \calGRup_{\sigma 0}$ in the
approximation (\ref{eq:calGsmallQ3}) implies via
(\ref{eq:basicT}) that
\begin {equation}
   \langle j^{i}(x) \rangle
   \simeq
   -\partial_i \langle j^{0}(x) \rangle ,
\end {equation}
which is the standard relationship between current and charge densities
in a diffusive process (in units where the diffusion constant is 1).


\subsubsection {Factorizing the calculation}

When the formulas for $\calGRup_{\sigma\mu}(Q,u)$ taken from
(\ref{eq:calGsmallQ3}) are used in (\ref{eq:basicT}),
the $Q_1$ and $Q_2$ integrals do not factorize like they did
in our zero-temperature calculation.
That's because of the $\omega/(i\omega-k^2)$ factor.
We can get rid of the $i\omega-k^2$ denominator by studying
the charge deposition function
\begin {equation}
   \Chargebar \, \dep(x)
   \equiv
   (\partial_t - \grad^2)
   \jo
\end {equation}
of (\ref{eq:depdef}) instead of directly calculating the
current response $\jmu$.
It will be even more convenient to first calculate
the time integral
\begin {equation}
   \depint(x) \equiv \Chargebar \int_{-\infty}^t dt' \> \dep(t',\x)
\label {eq:depintdef}
\end {equation}
of the charge deposition function, which is related to
the charge response by
\begin {equation}
   (\partial_t - \grad^2)
   \jo
   =
   \partial_t \depint(x) .
\end {equation}
In Fourier space,
\begin {equation}
   \langle j^{0} \rangle
   =
   \frac{i\omega}{i\omega-k^2} \, \depint .
\label {eq:depintj}
\end {equation}
The factor of $i\omega/(i\omega-k^2)$ above
will cancel the similar factor from
(\ref{eq:calGsmallQ3}) so that (\ref{eq:basicT}) leads to
an expression for $\depint$ where the $Q_1$ and $Q_2$ integrations
factorize.  Specifically, combining (\ref{eq:basicT}),
(\ref{eq:calGsmallQ3}), and (\ref{eq:depintj}),
\begin {subequations}
\label {eq:approx1T}
\begin {equation}
  \depint(x)
  \simeq
  \frac{\Aamp^2}{\gSG^2}
  \int_{0}^1 \frac{du}{uf} \,
  \Field(x,u)^* i \tensor\partial_t \Field(x,u) ,
\label {eq:RfF0T}
\end {equation}
where
\begin {equation}
  \Field(x,\ub)
  \equiv
  \int_q 
  \calGRup_{\perp}(q,\ub) \,
  \tilde\envelope_L(q-\kbig) \,
  e^{iq\cdot x} .
\label {eq:F0T}
\end {equation}
\end {subequations}
This is very similar in form to the
zero-temperature expression (\ref{eq:approx1}) for
$\jo$ except for the important distinction
that (\ref{eq:RfF0T}) gives $\depint(x)$
instead of the charge density.

The divergence at the horizon of
the factor $1/f$ in the integrand
of (\ref{eq:RfF0T}) will turn out to be crucial for getting
a physically sensible result for $\depint(x)$,
and we will later see that the terms of
$\calGRup_{\sigma\mu}(Q,u)$ in
(\ref{eq:calGsmallQ2}) that we dropped
in (\ref{eq:calGsmallQ3}) are ignorable because
they do not generate a similar divergent factor
as $u{\to}1$.


\subsection{What will \boldmath$\depint(x)$ look like?}

\begin {figure}
\begin {center}
  \includegraphics[scale=0.4]{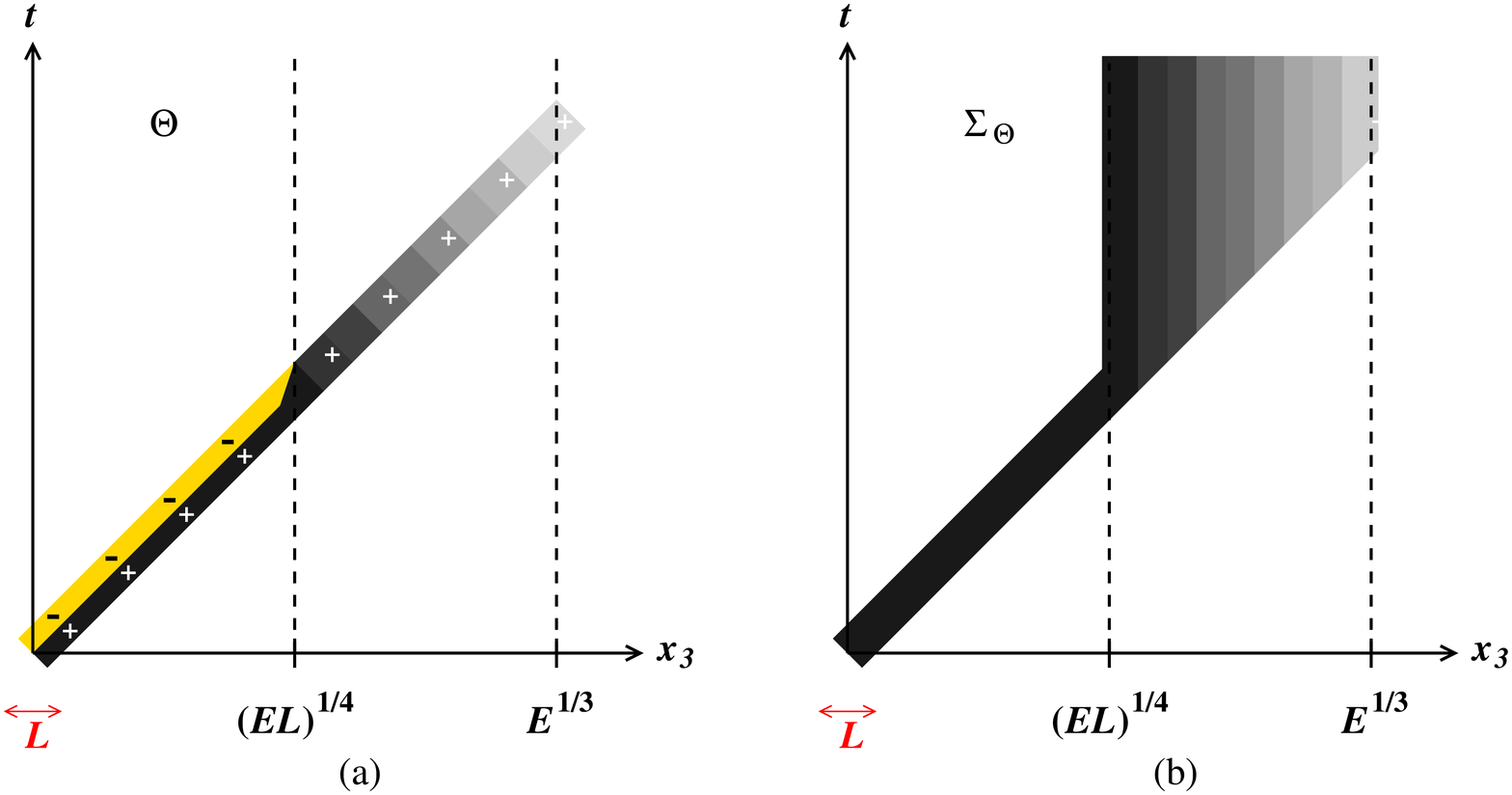}
  \caption{
     \label{fig:deposit2}
     The space-time distribution of
     (a) $\Chargebar\,\dep(x) = \partial_t\depint(x)
          = (\partial_t-\grad^2)j^0(x)$ and
     (b) its time integral $\depint(x)$.
     Picture (a) is like fig.\ \ref{fig:deposit}b but resolved down to
     the distance scale $L$. $\dep$ is negative in the gold region
     (marked $-$) and positive in the other shaded regions (marked $+$).
  }
\end {center}
\end {figure}

Before we discuss the calculation of $\depint(x)$, it will be
helpful to have in advance a qualitative picture of what the result
should look like.
In fig.\ \ref{fig:deposit}b, we gave a pictorial representation of
the final result (\ref{eq:final}) that we will find for
the charge deposition function $\Chargebar\dep(x) = \partial_t \depint(x)$
when resolved on scales large compared to $L$.
The picture is that the excitation initially moves ballistically at the
speed of light, and no charge is deposited until the jet reaches
its stopping distance, which we will find is stretched out between
the scales $(EL)^{1/4}$ and $E^{1/3}$.
But now consider the case where we choose $L \gg T$ (but still small
compared to $E^{1/3}$), and consider what happens if, unlike
fig.\ \ref{fig:deposit}b, we resolve the
charge density and $\Chargebar\dep \equiv (\partial_t - \grad^2) \jo$
down to the scale $L$ itself.
At early times, before the earliest stopping time scale $(EL)^{1/4}$,
the charge density will evolve like the left
half side of Figs.\ \ref{fig:evolution}a and b:
the charge density will be a narrow, positive function of $x^-$
of width $L$, independent of $x^+$, just as in the zero-temperature
result (\ref{eq:FinalzeroAll}).
But then, at these times,
\begin {equation}
   \Chargebar \, \dep(x)
   = (\partial_t - \grad^2) j^0(x^-)
   = -\partial_- (1+ \partial_-) j^0(x^-)
\end {equation}
is the derivative $\partial_-$ of a function that is localized
in $x_-$, and so $\dep(x^-)$ is a localized function whose integral
vanishes.  When resolving down to the scale $L$, the picture
of
fig.\ \ref{fig:deposit}b for $j^0(x)$ therefore becomes
fig.\ \ref{fig:deposit2}a for $\dep(x)$.
The canceling positive and negative edges
at early times blur together and
disappear when we only resolve scales large compared
to $L$.
The time integral of fig.\ \ref{fig:deposit2}a, which
defines $\depint$ as in (\ref{eq:depintdef}), is shown in
fig.\ \ref{fig:deposit2}b.

When we return to resolution scales large compared to $L$, where
the positive and negative regions cancel at early times, then we
approximate the $x^-$ dependence of $\dep(x)$ by $\delta_L(x^-)$
as in our final result
(\ref{eq:final}).  The coefficient of that $\delta$ function
will be the integral of $\dep(x)$ over $x^-$.  Since $\depint$ was
defined as the time integral of $\dep(x)$, this approximation
is then just
\begin {equation}
   \Chargebar \, \dep(x) \simeq \delta_L(x^-) \,
   \depint(t{=}\infty,x_3) .
\label {eq:deprelation}
\end {equation}
Our goal in what follows will be to use (\ref{eq:approx1T}) to compute
$\depint(x)$ at $t=\infty$, corresponding to the very top of
fig.\ \ref{fig:deposit2}b.


\subsection {WKB approximations to $\calGRup_\perp$}

Before we can evaluate the 4-momentum integral (\ref{eq:F0T}) that
gives $\Field(x,u)$, we first need formulas for the
transverse bulk-to-boundary propagator
$\calGRup(q,u)$.
This propagator is the solution to the linearized classical 5-dimensional
equation of motion $\nabla_I F^{I\perp}=0$, which is
\begin {equation}
   \left[
      \partial_u^2
       + \frac{f'}{f} \, \partial_u 
       - \frac{f \q^2 - \omega^2}{u f^2}
   \right] {\cal G}_\perp(\omega,\q,u)
   = 0 .
\end {equation}
It will be useful to rewrite this equation as
\begin {equation}
   \left[
      \partial_u^2
       + \frac{f'}{f} \, \partial_u 
       - \frac{q^2 - u^2 \q^2}{u f^2}
   \right] {\cal G}_\perp(\omega,\q,u)
   = 0 ,
\label {eq:Gperpeom}
\end {equation}
where (as previously)
$q^2 \equiv \eta^{\mu\nu}q_\mu q_\nu = -\omega^2 + \q^2$.
In the high-energy limit,
this classical equation can be solved using methods analogous to
the semi-classical (WKB) approximation in quantum mechanics, as
discussed in Refs.\ \cite{AdSminkowski,AdSWKB} and in particular
for the light-like case of $q^2=0$ by Caron-Huot
et al.\ \cite{dilepton}.%
\footnote{
   Ref.\ \cite{dilepton}
   present their solutions in terms of the
   electric field
   $E_\perp$ instead of
   ${\cal G}_\perp \propto A_\perp$.
   In our transverse-translation invariant problem, the
   relation is $E_\perp = i\omega A_\perp$.
   There is also an additional difference in overall normalization:
   they do not normalize their solutions on the boundary
   like ${\cal G}_\perp$.
}
Here, we will need to examine small
non-zero $q^2$ (with $|q^2| \ll \omega^2 \sim \wbig^2$).
In order to carefully understand the various scales at which
different approximations are valid, we will go through
the WKB approximation from the beginning.

Treat $\omega \sim k$ and substitute
\begin {equation}
   {\cal G}_\perp
   = e^{i (S_{-1} + S_0 + \cdots)}
\end {equation}
in (\ref{eq:Gperpeom}), where the exponent has been expanded formally
in powers of $1/\omega$ for fixed $u$ (with $S_n$ of order $\omega^{-n}$).
This gives
\begin {equation}
   \calGRup_\perp \simeq
    C(q) \, \left[ \frac{u}{u^2\q^2-q^2} \right]^{1/4}
   e^{iS(q,u)} ,
\label {eq:GperpWKB}
\end {equation}
where we now use $S$ as a short-hand notation for $S_{-1}$, given by
\begin {equation}
  S(q,u) =
  \int_0^{u} du' \>
      \frac{ [{u'}^2\q^2 - q^2]^{1/2} }{ {u'}^{1/2} \, f(u') } .
\label {eq:S}
\end {equation}
$C(q)$ is an overall normalization factor not determined by the equation
of motion.
The approximation (\ref{eq:GperpWKB}) is valid when the
exponent $S$ has large magnitude.

We've written the answer in a form that's convenient for
the time-like case $q^2<0$, which will be the most important later
on.  The choice of retarded propagator corresponds to taking the
positive sign on the square root in (\ref{eq:S}) in this case.
For space-like momenta $q^2>0$, the $e^{iS}$ analytically
continues to an exponential
suppression factor $e^{-|S|}$.

Useful approximations to the integral (\ref{eq:S}) for $S$ depend
on whether or not $u$ is small enough that
${u'}^2\q^2$ can be treated as a perturbation
in $[{u'}^2\q^2 - q^2]^{1/2}$.  The scale $u_\star$
that separates different
qualitative behaviors of $S$ is therefore
\begin {equation}
   u_\star \sim \sqrt{\frac{|q^2|}{\q^2}} \sim \sqrt{\frac{|q_+|}{\wbig}} .
\label {eq:ustar}
\end {equation}
We discuss various expansions of (\ref{eq:S}) in Appendix
\ref{app:S}.  The important limits for the present discussion
are that
\begin {equation}
  S =
  2 (-u q^2)^{1/2}
  \left[1 + O(u^2) + O\bigl(\frac{u^2\omega^2}{q^2}\bigr)\right]
\label {eq:Ssmall}
\end {equation}
for use when $u \ll u_\star \ll 1$, and
\begin {equation}
  S = \omega \, \tauo(u)
      + \tfrac43 \num \omega^{-1/2} (-\tfrac14 q^2)^{3/4}
          \left[1 + O\bigl(\frac{q^2}{\omega^2}\bigr)\right]
      + O\bigl(\frac{q^2}{u^{1/2}\omega}\bigr)
\label {eq:Slarge}
\end {equation}
for use in the case $u \gg u_\star$.
The constant $\num$ is defined by (\ref{eq:num}).
$\omega \, \tau(u)$ is defined as the result for $S$
for the case $\omega{=}|\q|$, which gives
\begin {equation}
   \tauo(u) =
   \Tanh^{-1}\sqrt{u} - \Tan^{-1}\sqrt{u}
   =
   \frac12 \ln \left( \frac{1+\sqrt{u}}{1-\sqrt{u}} \right)
   - \Tan^{-1}\sqrt{u}
\end {equation}
with limiting cases
\begin {equation}
   \tauo(u) \simeq
   \begin {cases}
      \tfrac23 \, u^{3/2} , & u \ll 1; \\
      -\frac12 \ln(1-u) ,   & u \to 1.
   \end {cases}
\label {eq:taulims}
\end {equation}

\begin {figure}
\begin {center}
  \includegraphics[scale=0.6]{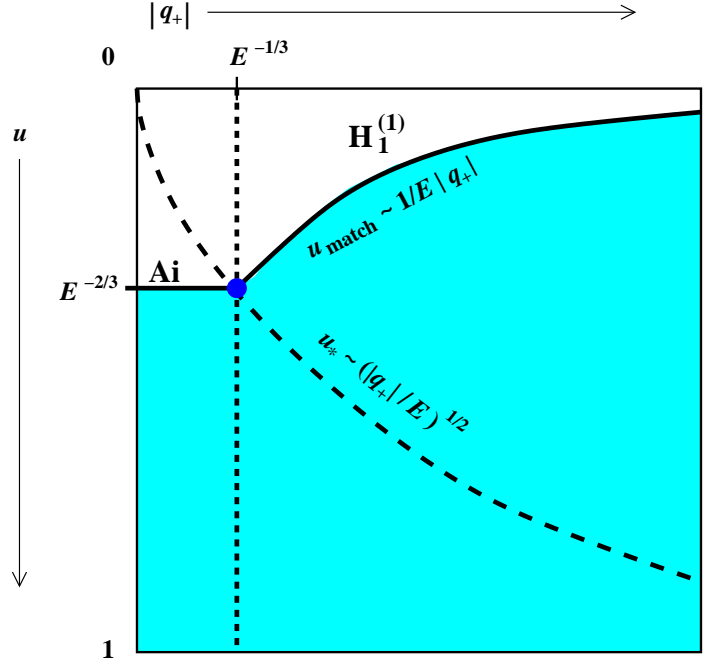}
  \caption{
     \label{fig:WKB}
     The WKB approximation to $\calGRup_\perp(q,u)$ is valid
     in the shaded region parametrically far below the solid curve.
     The solid curve corresponds to the matching scale $\umatch$
     discussed in the text, while the dashed line is the scale
     $u_\star$ of (\ref{eq:ustar}).  $\Ai$ and $H_1^{(1)}$ indicate
     the type of solution valid in the matching region for
     $|q_+| \ll E^{-1/3}$ and $|q_+| \gg E^{-1/3}$ respectively.
  }
\end {center}
\end {figure}

The overall normalization $C(q)$ of (\ref{eq:GperpWKB})
is fixed by the boundary condition ${\cal G}_\perp(q,0) = 1$,
which lies outside of the region of validity for
(\ref{eq:GperpWKB}) since $S(q,0)=0$.
Determining $C(q)$ requires matching
(\ref{eq:GperpWKB}) to a small-$u$ solution of the equation
of motion (\ref{eq:Gperpeom}).  We find $C(q)$ in
two complementary limits $|q_+| \ll \wbig^{-1/3}$ and
$|q_+| \gg \wbig^{-1/3}$,
which will be adequate for deriving our final result
(\ref{eq:final}).
We will discuss the appropriate scale for matching in each
case below, but the result is summarized in fig.\ \ref{fig:WKB}.
The WKB approximation (\ref{eq:GperpWKB}) is a good approximation
in the shaded region, parametrically far below the solid curve,
which we will call $\umatch(q_+)$.


\subsubsection {$\wbig^{-1/3} \ll |q_+| \ll \wbig$}

The case $\wbig^{-1/3} \ll |q_+| \ll \wbig$
(corresponding to $\wbig^{2/3} \ll |q^2| \simeq 4\wbig |q_+| \ll \wbig^2$)
will be
the most important for studying the vast majority of charge
deposition, which (as previewed in the introduction) will turn out
to take place
at distance scales $x^+ \ll \wbig^{1/3}$.

For a given $q_+$, let $u \gg \umatch$ define the region
of $u$ where $|S|$ is large and so the WKB result (\ref{eq:GperpWKB})
is applicable.  In order to determine $C(q)$, we need to solve
the equation of motion in the region $u \sim \umatch$
where the WKB approximation is marginal, which is the region
where $|S| \sim 1$.  If $\umatch \ll u_\star$, then
eq.\ (\ref{eq:Ssmall}) will give
\begin {equation}
   \umatch \sim \frac{1}{|q^2|} \sim \frac{1}{\wbig |q_+|}
   \ll u_\star ,
\end {equation}
which will be consistent with (\ref{eq:ustar}) for $u_\star$ precisely
when $|q_+| \gg \wbig^{-1/3}$.

Because $u \ll u_\star \ll 1$ where we need to do the matching, we can
approximate
\begin {equation}
   q^2 - u^2 \q^2 \simeq q^2
\end {equation}
in the equation of motion (\ref{eq:Gperpeom}).
The complete approximation in this region is
\begin {equation}
   \left[
      \partial_u^2
       - \frac{q^2}{u}
   \right] {\cal G}_\perp(\omega,\q,u)
   \simeq 0 .
\label {eq:Gperpeomsmall}
\end {equation}
The boundary-normalized solution is just the vacuum solution
(\ref{eq:Gperpzero}), which we will write here in the form
\begin {equation}
   \calGRup_\perp \simeq
   i\pi \sqrt{-u q^2} \, H_1^{(1)}(\sqrt{-4u q^2})
   \qquad
   (u \ll u_\star) .
\label {eq:Gperpzero2}
\end {equation}
Using the asymptotic formula for the Hankel function to
match to the WKB formula (\ref{eq:GperpWKB}) in the range
$\umatch \ll u \ll u_\star$ where both are valid
determines
\begin {equation}
   C(q) \simeq e^{-i\pi/4}(-\pi q^2)^{1/2} .
\label {eq:Cq1}
\end {equation}


\subsubsection {$|q_+| \ll \wbig^{-1/3}$}

When $|q_+| \ll \wbig^{-1/3}$, we will see that
$|S|\sim 1$ at $\umatch \gg u_\star$, and so we turn to
eq.\ (\ref{eq:Slarge}) for the WKB exponent $S$.
Note that the second term in (\ref{eq:Slarge}) is of order
$\omega^{-1/2} |q^2|^{3/4} \sim \wbig^{1/4} |q_+|^{3/4} \ll 1$
when $|q_+| \ll \wbig^{-1/3}$, and so it can be ignored,
leaving $S \simeq \omega \, \tauo(u)$.  Then $|S|\sim 1$ at
\begin {equation}
   \umatch \sim \omega^{-2/3} \sim \wbig^{-2/3} \gg u_\star .
\end {equation}
This is consistent with (\ref{eq:ustar}) precisely when
$|q_+| \ll \wbig^{-1/3}$.

Because $\umatch \gg u_\star$ where we need to do the matching,
we can approximate
\begin {equation}
   q^2 - u^2 \q^2 \simeq -u^2 \omega^2
\end {equation}
in the equation of motion (\ref{eq:Gperpeom}).
Since also $\umatch \ll 1$,
the complete approximation in this region is
\begin {equation}
   \left[
      \partial_u^2
       + u\omega^2
   \right] {\cal G}_\perp(\omega,\q,u)
   \simeq 0 .
\label {eq:Gperpeomlarge}
\end {equation}
The retarded, boundary-normalized solution is
\begin {equation}
   \calGRup_\perp \simeq
   \frac{\Ai(e^{-i\pi/3} u \omega^{2/3})}
        {\Ai(0)}
   \qquad
   (u \ll 1) ,
\label {eq:GperpAi}
\end {equation}
where $\Ai$ is the Airy function and
$\Ai(0) = 3^{-2/3}/\Gamma(\frac23)$.

One might worry that the analysis that led to
(\ref{eq:GperpAi}) breaks down at
$u \lesssim u_\star \ll \umatch$.  But in that
$u$ range, $\calGRup$ is very close to its boundary value
$1$ and so the deviation from (\ref{eq:Gperpeomlarge})
there will not affect the approximation (\ref{eq:GperpAi})
at leading order in $u_\star \ll 1$.

Using the asymptotic formula for the Airy function to
match to the WKB formula (\ref{eq:GperpWKB}) in the range
$\umatch \ll u \ll 1$ where both are valid
determines
\begin {equation}
   C(q) \simeq \frac{e^{i\pi/12}\omega^{1/3}}{2\pi^{1/2} \Ai(0)}
\end {equation}
and so
\begin {equation}
   \calGRup_\perp \simeq
   \frac{e^{i\pi/12}}{2\pi^{1/2}\Ai(0)} \,
   u^{-1/4} \omega^{-1/6}
   e^{iS(q,u)}
   \qquad
   (u \gg \umatch).
\label {eq:calGAiryWKB}
\end {equation}
Except for issues of overall normalization convention, this
matching calculation is the same as the $q^2=0$ analysis
of Ref.\ \cite{dilepton}.%
\footnote{
  There is a typographic error in Eq.\ (A6)
  of Ref.\ \cite{dilepton}: The factor in big parenthesis in the
  left-hand equation should be raised to the $2/3$ power.
}


\subsection {Steepest descent analysis of \boldmath$\Field$}
\label{sec:SteepDescent}

\begin {figure}
\begin {center}
  \includegraphics[scale=0.6]{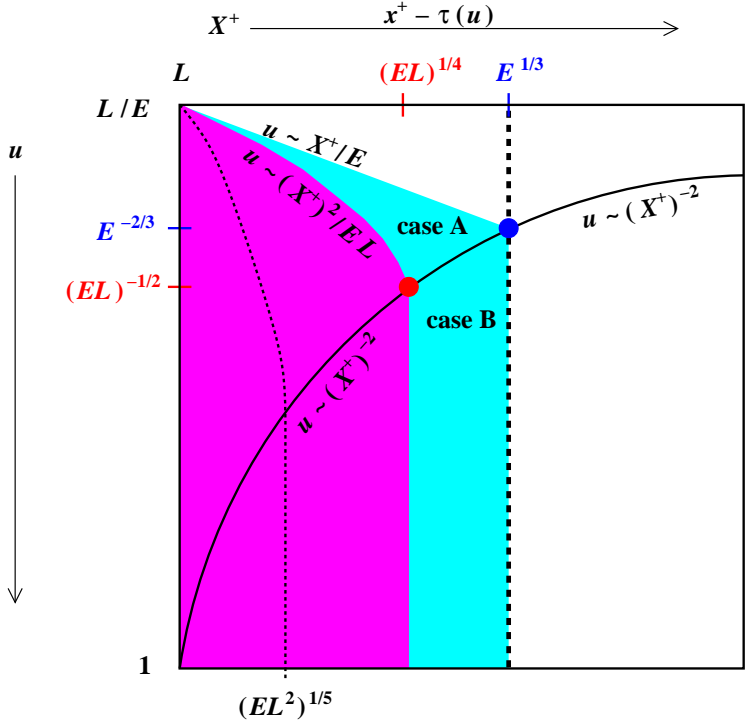}
  \caption{
     \label{fig:SteepDescent}
     Qualitative picture of scales determining the behavior of
     $\Field(x,u)$ at finite temperature.  The horizontal axis
     is $X^+ \equiv x^+ - \tauo(u)$, which is the same as $x^+$
     except at the very bottom ($u{\to}1$) of the figure.  The top curve
     $u \sim x^+/E$ indicates where the first wiggle is in
     $\Field$ as a function of $u$, similar to the zero-temperature
     case of fig.\ \ref{fig:SteepDescent0}.  The steepest descent
     approximation is valid in the shaded region (blue and magenta)
     below this curve ($u \gg x^+/E$ and $X^+ \ll E^{1/3}$).
     In the lower shaded
     (magenta) region below the curve $u \sim (x^+)^2/EL$, the
     field $\Field$ is exponentially suppressed.  The field is also
     exponentially suppressed in the region $X^+ \gg E^{1/3}$ to the
     right of the vertical black dashed line.
  }
\end {center}
\end {figure}

We now turn to using steepest descent methods to evaluate
the integral (\ref{eq:F0T}) that gives $\Field(x,u)$, just as we
did at zero temperature in section \ref{sec:fix}.
As a qualitative
preview of what we will find, fig.\ \ref{fig:SteepDescent}
is the finite-temperature version of fig.\ \ref{fig:SteepDescent0}.
The horizontal axis is
\begin {equation}
   X^+ \equiv x^+ - \tauo(u) .
\end {equation}
Given our interest in distances $x^+ \gg 1$, eq.\ (\ref{eq:taulims})
for $\tauo(u)$ means that the difference between $X^+$ and $x^+$ is
insignificant unless $u$ is {\it extremely}\/ close
to the horizon.  However, we will see that the behavior of
$\Field(x,u)$ as $u{\to}1$ is precisely what we want to get
the large-time limit $\depint(t{=}\infty,x_3)$ that determines
the charge deposition function via (\ref{eq:deprelation}).
We will see below that the $x^-$ dependence of $\Field$ is
localized to $x^- \simeq -\tauo(u)$ and so
\begin {equation}
   X^+ \simeq x^+ + x^- = 2 x_3 .
\label {eq:Xphorizon}
\end {equation}

There will be two different cases we will need to explore,
corresponding to whether the saddle point of the $q_+$ integration
probes the bulk-to-boundary propagator $\calGRup(q,u)$ above
or below the curve $u \sim u_\star$ of
(\ref{eq:ustar}) and fig.\ \ref{fig:WKB}.


\subsubsection {Case A: Just like zero temperature}

At early times, we expect that the physics should be approximately
the same as the zero-temperature case analyzed in section \ref{sec:fix}.
The zero-temperature bulk-to-boundary propagator (\ref{eq:Gperpzero})
corresponds to
the finite-temperature one when (i) $u \ll u_\star$
and (ii) $|q_+| \gg \wbig^{-1/3}$, so that (\ref{eq:Gperpzero2}) applies.
When these two conditions are satisfied,
we may just take over the zero-temperature result
(\ref{eq:Ffinal}) for $\Field$,
\begin {equation}
  \Field(x,u) \simeq
  -i \,
  \frac{4u\wbig}{(x^+)^2} \,
  e^{i\wbig x^-} \,
  e^{i4u\wbig/x^+}
  \envelope_L^{(2)}\Bigl( -\frac{4u\wbig}{(x^+)^2} ; x^- \Bigr) \,
  \theta(x^+)
  .
\label {eq:FfinalA}
\end {equation}

From (\ref{eq:kstar}), the saddle point is at
\begin {equation}
   |q_+| = k_\star \sim \frac{u\wbig}{(x^+)^2} \,.
\label {eq:qstarA}
\end {equation}
Combining this with
$u_\star \sim \sqrt{|q_+|/\wbig}$ from (\ref{eq:ustar}),
the first condition $u \ll u_\star$ is then
\begin {equation}
   u \ll \frac{1}{(x^+)^2} \,,
\label {eq:ustarA}
\end {equation}
which corresponds to the region above the solid curve in
fig.\ \ref{fig:SteepDescent}.  One may ignore the difference
between $X^+=x^+ - \tauo(u)$ and $x^+$ here because of
$\tauo$'s relative insignificance away from the horizon.
Using (\ref{eq:qstarA}),
the second condition $|q_+| \gg \wbig^{-1/3}$ will be satisfied
if
\begin {equation}
   u \gtrsim \frac{1}{E}
   \qquad \mbox{and} \qquad
   x^+ \ll \wbig^{1/3} .
\end {equation}
So the vacuum saddle-point result applies to the shaded region of
fig.\ \ref{fig:SteepDescent} that is above the solid curve
and to the left of the black dashed line.


\subsubsection {Case B: Falling into the black brane}

As we will see,
the shaded region below the solid curve $u \sim (x^+)^{-2}$ in
fig.\ \ref{fig:SteepDescent} will be determined by a saddle
point with (i) $u \gg u_\star$, and (ii) $\wbig^{-1/3} \ll |q_+| \ll \wbig$.
Combining (\ref{eq:GperpWKB}), (\ref{eq:Slarge}), and
(\ref{eq:Cq1})
with the fact that $u^2\q^2-q^2 \simeq u^2\q^2$ when
$u \gg u_\star$, these two conditions give
\begin {subequations}
\label {eq:calGB}
\begin {align}
   \calGRup_\perp &\simeq 
   e^{-i\pi/4} u^{-1/4} \left( - \frac{\pi q^2}{|\q|} \right)^{1/2}
   e^{i\,S(q,u)}
\nonumber\\
   &\simeq
   e^{-i\pi/4} u^{-1/4} (- 4\pi q_+)^{1/2} \,
   e^{i\,S_+(q_+,u)} e^{i q_- \, \tauo(u)} ,
\end {align}
where $S(q,u) \simeq q_- \, \tauo(u) + S_+(q_+,u)$ with
\begin {equation}
   S_+(q,u) \simeq
   - q_+ \, \tauo(u)
      + \tfrac43 \num \wbig^{1/4} (-q_+)^{3/4} .
\label {eq:Splus}
\end {equation}
\end {subequations}
The $e^{i q_- \, \tauo(u)}$ factor is important, and cannot be
ignored, when $u$ is very close to the horizon so that
$\tauo(u)$ is large.  When Fourier transforming from $q_-$
to $x^-$, the effect of this factor will be to shift $x^-$
by $\tauo(u)$.
The high-energy approximation
(\ref{eq:HEapprox}) that we made in the zero-temperature case is
modified to
\begin {equation}
   \calGRup_{\perp}(q_+,q_-,u)
   \simeq
   e^{i q_- \, \tauo(u)} \, \hatcalGRup_{\perp}\bigl(q_+,u)
\label {eq:secondT}
\end {equation}
with
\begin {subequations}
\label {eq:Field99}
\begin {equation}
   \hatcalGRup_\perp(q_+,u) \equiv
   e^{-i\pi/4} u^{-1/4} (- 4\pi q_+)^{1/2} \,
   e^{i\,S_+(q_+,u)} ,
\end {equation}
giving
\begin {equation}
  \Field(x,u)
  \simeq
  e^{i\wbig [x^-+\tauo(u)]}  \,
  \int \frac{dq_+}{2\pi} \,
  \hatcalGRup_{\perp}(q_+,u) \,
  \envelope_L^{(2)}\bigl(q_+;x^-{+}\tauo(u)\bigr) \,
   e^{iq_+x^+} .
\label {eq:F0approxT}
\end {equation}
\end {subequations}
Eq.\ (\ref{eq:F0approxT}) and
the finite size $L$ of the source region in $x^-$ imply that
$\Field$ is localized to
\begin {equation}
   |x^- + \tauo(u)| \lesssim L ,
\label {eq:xmlocalized}
\end {equation}
and so $x^- \simeq -\tauo(u)$ near the horizon.

For a Gaussian source envelope (\ref{eq:Genvelope2}), approximating
the integral (\ref{eq:F0approxT}) by steepest descent requires
extremizing
\begin {equation}
   {\cal S}(q_+,u) = -i S_+(q_+,u) - iq_+ x^+ + (q_+ L)^2 ,
\end {equation}
analogous to the zero-temperature case (\ref{eq:calSzero}).
We will again treat $L$ perturbatively and so find the extremum of
\begin {equation}
   {\cal S}_0 \equiv -i S_+ - iq_+ x^+
   \simeq 
      - i q_+ X^+
      - i \tfrac43 \num \wbig^{1/4} (-q_+)^{3/4} ,
\end {equation}
which is at
\begin {equation}
   q_+^\star \simeq -\frac{\num^4\wbig}{(X^+)^4} \,.
\end {equation}
One may now verify that the location of the saddle point
satisfies the two requirements
$u \gg u_\star$ and $\wbig^{-1/3} \ll |q_+| \ll \wbig$
assumed for the propagator (\ref{eq:calGB})
provided one is in the Case B region of
$u \gg (X^+)^{-2}$ and $1 \ll X^+ \ll \wbig^{1/3}$.
Also note that the requirement $|{\cal S}_0(q_+^\star)| \gg 1$
is $\wbig/(X^+)^3 \gg 1$, which is also satisfied when
$X^+ \ll \wbig^{1/3}$.

Expanding ${\cal S}_0$ to second order in small fluctuations
$q_+{-}q_+^\star$ about the
saddle point, and then doing the Gaussian integral from
(\ref{eq:Field99}) and (\ref{eq:Splus}), yields
\begin {align}
  \Field(x,u)
  &\simeq
  e^{i\wbig [x^-+\tauo(u)]}  \,
  e^{-i\pi/4} u^{-1/4} (- 4\pi q_+^\star)^{1/2} \,
  \envelope_L^{(2)}\bigl(q_+^\star;x^-{+}\tauo(u)\bigr) \,
  \left(
    2\pi \, \frac{\partial^2 S_0}{\partial q_+^2}
  \right)^{-1/2}_{q_+^\star}
  e^{-S_0(q_+^\star,u)}
\nonumber\\
  &\simeq
   -i e^{i\wbig [x^-+\tauo(u)]}
   \, \frac{2^{3/2}\num^4\wbig}{u^{1/4} (X^+)^{9/2}}
   \, \exp\left( i \, \frac{\num^4\wbig}{3 (X^+)^3} \right)
   \, \envelope_L^{(2)}
      \left(-\frac{\num^4\wbig}{(X^+)^4}; x^-{+}\tauo(u)\right)
  .
\label {eq:calAB}
\end {align}
To do a saddle point analysis, one should verify that there is
a choice of integration contour that makes the neighborhood of
the saddle point the dominant contribution to the integral.
Having an explicit contour also helps one sort out exactly which
branch one is on when evaluating the various roots in the
derivation of (\ref{eq:calAB}).  We discuss the choice of
integration contour
in Appendix \ref{app:contour}.


\subsection{Final result for \boldmath$x_3 \ll \wbig^{1/3}$}

We are now ready to assemble our final result (\ref{eq:final}) for
charge deposition up to distances of order $\wbig^{1/3}$.
(The exponential tail at larger distances will be discussed
in section \ref{sec:tail}.)
As in the zero-temperature case, the derivative on $\Field$ in
eq.\ (\ref{eq:RfF0T}) for $\depint(x)$ will be dominated by the
term that hits $e^{i\wbig x^-}$, and so
\begin {equation}
  \depint(x)
  \simeq
  2 \wbig \, \frac{\Aamp^2}{\gSG^2}
  \int_{0}^1 \frac{du}{uf} \,
  |\Field(x,u)|^2 ,
\label {eq:depint2}
\end {equation}
analogous to (\ref{eq:Rapprox2}).
We are interested in this result for $t \to \infty$, which means
arbitrarily large $x^-$ and $x^+$.
Because of the localization (\ref{eq:xmlocalized}) of
$\Field(x,u)$, non-negligible contributions at large $x^-$
can only come from the near-horizon part of the $u$ integration
in (\ref{eq:depint2}), where $\tauo(u)$ is large.
Very large $\tauo(u)$ represents an exponentially-small region of $u$,
and its contribution to the integral would be negligible if not for
the $1/f$ factor in (\ref{eq:depint2}).

For $t \to \infty$ and $u \to 1$, our result (\ref{eq:calAB}) for
$\Field$ becomes
\begin{equation}
   |\Field|^2 \to
   \frac{8\num^8\wbig^2}{(X^+)^9}
   \left| \envelope_L^{(2)}
      \left(-\frac{\num^4\wbig}{(X^+)^4}; x^-{+}\tauo(u)\right)
   \right|^2 .
\label {eq:FieldSqr}
\end {equation}
Using (\ref{eq:taulims}) to rewrite
$du/uf \simeq d\tauo$ in the $u{\to}1$ limit,
and also using (\ref{eq:Xphorizon}) in the same limit,
\begin {equation}
  \depint(t{=}\infty,\x)
  \simeq
  \frac{\Aamp^2}{\gSG^2} \,
  \frac{16 \num^8\wbig^3}{(2x_3)^9}
  \int_{0}^\infty d\tauo \>
   \left| \envelope_L^{(2)}
      \left(-\frac{\num^4\wbig}{(2x_3)^4}; x^-{+}\tauo(u)\right)
   \right|^2 .
\end {equation}
Shifting integration variable from $\tauo$ to $x^-+\tauo$, and
using the fact that $|x^-+\tauo|$ is localized to
$L \ll |x^-|$ in the limit $|x^-|{\to}\infty$ of interest, we can
use the definition (\ref{eq:formL}) and the result
(\ref{eq:charge}) for the total charge $\Chargebar$ per unit
transverse area produced by the source (see also Appendix
\ref{app:charge}) to rewrite the last equation as%
\footnote{
  Given our definitions (\ref{eq:envelope2}) and (\ref{eq:envelope2xm}) of
  $\tilde\envelope_L^{(2)}(q_+,q_-)$ and
  $\envelope_L^{(2)}(q_+;x^-)$, then
  $
     \int dx^- \> |\envelope_L^{(2)}(q_+;x^-)|^2
     = \frac{2}{\pi} \int dq_- \> |\tilde\envelope_L^{(2)}(q_+,q_-)|^2
  $.
}
\begin {equation}
  \depint(t{=}\infty,\x)
  \simeq
  2 \Chargebar \, 
  \frac{(4 \num^4 \wbig L)^2}{(2x_3)^9} \,
  \formL\left( -\frac{c^4 E L}{(2x_3)^4} \right) .
\label{eq:depintfinal}
\end {equation}
Using the relationship (\ref{eq:deprelation}) with the charge
deposition function $\dep(x)$ finally produces our result
(\ref{eq:final}).
As a check of the calculation, one may verify that this result
satisfies
$\int dt\>dx_3\>\dep(x) = \frac12 \int dx^- \> dx^+ \>\dep(x) = 1$,
as it should given the definition
(\ref{eq:depdef}) of $\dep(x)$.


\subsection {The exponential tail of \boldmath$\dep(x)$}
\label {sec:tail}

\subsubsection{Relation to poles of $\calGRup_\perp$}

We will now discuss the exponential fall-off of our result
(\ref{eq:final}) for charge deposition for $x^+ \gg \wbig^{1/3}$.
This requires evaluating $\Field$ near the horizon
for $X^+ \gg \wbig^{1/3}$, corresponding to the bottom of the white
region in fig.\ \ref{fig:SteepDescent}.
Near the horizon, we can always use the WKB formulas for
$\calGRup_\perp$ (see fig.\ \ref{fig:WKB}), and so we can use
the integral expression (\ref{eq:Field99}) for
$\Field$, which we now find convenient to rewrite as
\begin {equation}
  \Field(x,u)
  \simeq
  e^{i\wbig [x^-+\tauo(u)]}  \,
  \int \frac{dq_+}{2\pi} \,
  \hathatcalGRup_{\perp}(q_+,u) \,
  \envelope_L^{(2)}\bigl(q_+;x^-{+}\tauo(u)\bigr) \,
   e^{iq_+X^+}
\label {eq:theintegral}
\end {equation}
where $\hathatcalGRup_\perp$ is defined by
\begin {equation}
   \calGRup_{\perp}(q_+,q_-,u)
   \simeq
   e^{i \omega \, \tauo(u)} \, \hathatcalGRup_{\perp}\bigl(q_+,u) .
\label {eq:hathatdef}
\end {equation}
Attempting to evaluate this
integral for large $X^+$ with saddle point methods will fail.
But we can instead use the fact that the large-argument ($X^+$) behavior of
a Fourier transform is determined by the singularities of that
function in the complex plane ($q_+$).

\begin {figure}
\begin {center}
  \includegraphics[scale=0.3]{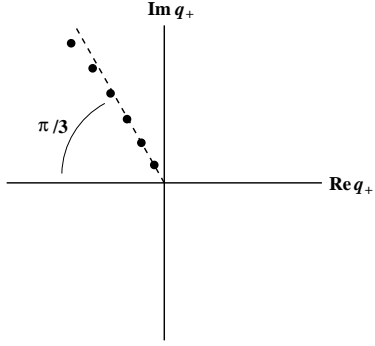}
  \caption{
     \label{fig:poles}
     A qualitative plot of the pole positions of $\calGRup_\perp$ in the
     complex $q_+$ plane.  The dashed line is proportional to
     $e^{i2\pi/3}$.
  }
\end {center}
\end {figure}

In particular, the singularities of the bulk-to-boundary
propagator are poles corresponding to quasi-normal modes
of a vector field in the AdS-Schwarzschild background
\cite{AdSminkowski,Starinetspoles},
shown qualitatively in fig.\ \ref{fig:poles} for $\calGRup_\perp$.
Remember that our convention is that $q_+ \equiv \tfrac12(q^3-q^0)$,
and so the lower-half frequency plane (where the singularities
of a retarded propagator should be) corresponds to the upper-half
$q_+$ plane.

We will review the origin of these poles in a moment, but first let's
examine the consequence.  For $X^+>0$, we can close the
$q_+$ integration contour in (\ref{eq:theintegral}) in the upper-half plane,
giving a sum of residues from the poles $q_+{=}q_+^{(n)}$,
each exponentially suppressed by a factor of $e^{-\Im(q_+^{(n)})X^+}$.
The dominant contribution at large $X^+$ will be from
the pole $q_+ {=} q_+^{(1)}$ closest to the real axis, giving
\begin {equation}
  \Field(x,u)
  \simeq
  i e^{i\wbig [x^-+\tauo(u)]}
  \Res\left[\hathatcalGRup_{\perp}(q_+^{(1)},u)\right]
  \envelope_L^{(2)}\bigl(q_+^{(1)};x^-{+}\tauo(u)\bigr) \,
   e^{-\Im(q_+^{(1)}) X^+} .
\end {equation}
$q_+^{(1)}$ will turn out to be small in the high-energy limit,
so that it may be replaced by zero in the evaluation of the
envelope function.  Then
\begin {equation}
  |\Field|^2
  \simeq
  \Bigl|
    \Res\bigl[\hathatcalGRup_{\perp}(q_+^{(1)},u)\bigr]
    \envelope_L^{(2)}\bigl(0;x^-{+}\tauo(u)\bigr)
  \Bigr|^2
  e^{-2\Im(q_+^{(1)}) X^+} .
\end {equation}
The same steps that we took from (\ref{eq:FieldSqr}) to
(\ref{eq:depintfinal}) then give
\begin {equation}
  \depint(t{=}\infty,\x)
  \simeq
  4
  \Chargebar L^2
  \left|
    \Res \hathatcalGRup_{\perp}(q_+^{(1)},1)
  \right|^2
   e^{-2\Im(q_+^{(1)})\,(2x_3)}
  \formL(0) .
\label {eq:depintdecay1}
\end {equation}

For comparison, note that the zero temperature formula
(\ref{eq:Gperpzero}) for ${\cal G}_\perp$ has a branch point singularity at
$q_+{=}0$.  This is why the zero-temperature result (\ref{eq:Ffinal}) for
$\Field(x,u)$ falls algebraically rather than exponentially at large
$x^+$.%
\footnote{
  On the gauge theory side, the origin of singularities at
  $q^2=0$ is the presence of long-lived massless excitations.
  At finite temperature, however, excitations generally have finite
  life-times due to interactions with the plasma.  Only long-wavelength
  hydrodynamic excitations can have arbitrarily long life times.
  However, these will only couple indirectly to the large-momentum modes
  being considered here.  On the gravity side, the coupling would be
  through loops, which are $1/\Nc^2$ suppressed, as in the
  discussion of long-time hydrodynamic tails of Refs.\
  \cite{HydroTails,HydroTails1}.
}


\subsubsection{Scaling of poles with energy}

Poles in $\calGRup_\perp(q,u)$ occur when $q$ is such that the
normalization condition $\calGRup_\perp(q,0)=1$ at $u{=}0$ causes
$\calGRup(q,u)$ to be infinite for all other $u$.  Turning
this around, poles occur for $q$ where finite solutions
$A_\perp(q,u)$ to the equation
of motion with retarded boundary conditions at the horizon
vanish at the boundary, $A_\perp(q,0)=0$.
For $u \sim 1$, we can use WKB methods to investigate
solutions, but WKB breaks down and requires matching as $u\to 0$.
When will this matching solution for $u \ll 1$ cause
$A_\perp(q,0)=0$?

For $u\ll1$, the equation of motion (\ref{eq:Gperpeom}) becomes
\begin {equation}
   \left[
      \partial_u^2
       - \frac{4\wbig q_+ - u^2 \wbig^2}{u}
   \right] A_\perp
   \simeq 0
\end {equation}
in the high energy limit.  By the WKB analysis, the behavior of the
retarded solution for $u \gg \umatch$ is proportional to
$e^{i\wbig\,\tauo(u)}$, which for $u \ll 1$ is
\begin {equation}
   e^{i \frac23 u^{3/2} \wbig} .
\end {equation}
It will be simpler to analyze the question of when
$\Field(q,0)$ vanishes if we can look at
purely real solutions rather than complex ones.
To this end, change variables to
$U \equiv e^{-i\pi/3} \wbig^{2/3} u$ so that
the asymptotic behavior is
\begin {equation}
   A_\perp \sim e^{- \frac23 U^{3/2}}
\end {equation}
and the equation of motion is
\begin {equation}
   \left[
      - \partial_U^2
      + \left(U -\frac{a}{U}\right)
   \right] A_\perp \simeq 0 ,
\label {eq:poleeq}
\end {equation}
where 
\begin {equation}
   a \equiv 4 \wbig^{1/3} e^{-i2\pi/3} q_+ .
\label {eq:polea}
\end {equation}
This solution will have $A_\perp(q,0)=0$ when the
Schr\"odinger-like equation (\ref{eq:poleeq}) has a zero-energy
bound state solution that vanishes at the origin.
That can happen for real positive $a$, which we label
$a_1,a_2,\cdots$ starting from the smallest value that works.
The corresponding pole
locations are then
\begin {equation}
   q_+^{(n)} \simeq \tfrac14 \wbig^{-1/3} e^{i2\pi/3} a_n .
\end {equation}

Solving (\ref{eq:poleeq}) numerically
to find $a_1$, we obtain $a_1 \simeq 2.141$.%
\footnote{
  We have double checked our analysis by also calculating the full,
  un-approximated bulk-to-boundary propagator numerically and verifying
  that we get the same scaling and pole locations at large $\omega$.
  We did it by brute force, and one could likely find the poles
  more efficiently using the method of Ref.\ \cite{Starinetspoles}.
}
Correspondingly, the exponential decay factor in
(\ref{eq:depintdecay1}) is
\begin {equation}
   e^{-2\Im(q_+^{(1)})\,(2x_3)}
   \equiv e^{-\numpole(2x_3)/\wbig^{1/3}} ,
\end {equation}
where
\begin {equation}
   \numpole = \frac{\sqrt{3}\,a_1}{4} \simeq 0.927 \,.
\end {equation}

Given the analytic structure of fig.\ \ref{fig:poles}, readers may
wonder what has become of the cut associated with the
$\wbig^{1/4} (-q_+)^{3/4}$ term in the WKB formula
exponent (\ref{eq:Splus}).  The formula for any contribution to
a WKB exponent can only be trusted if its magnitude is large
compared to 1, which for $\wbig^{1/4} (-q_+)^{3/4}$ means when
$|q_+| \gg \wbig^{-1/3}$.  But that means that $|q_+|$ is large
compared to the separation between poles in fig.\ \ref{fig:poles},
and so the dense line of poles can approximate a cut.


\subsubsection{The residue}

Unlike the pole position, the residue will depend on $u$,
and we are interested in the value near the horizon.
From (\ref{eq:calGAiryWKB}), we know that in this case there is a
prefactor of order $\wbig^{-1/6}$ when $|q_+| \ll \wbig^{-1/3}$.
Parametrically, the prefactor should be of the same order
when $|q_+| \sim |q_+^{(1)}| \sim \wbig^{-1/3}$.
So the behavior of the propagator near the pole should scale as
\begin {equation}
   |\hathatcalGRup_\perp| \propto
   \left| \frac{\wbig^{-1/6}}{a-a_1} \right| ,
\end {equation}
which we will use (\ref{eq:polea}) to write as
\begin {equation}
   |\hathatcalGRup_\perp| \simeq
   \frac{\numres \wbig^{-1/2}}{|q_+-q_+^{(1)}|}
\label {eq:residue}
\end {equation}
for some constant $c_2$.  By numeric evaluation of the full propagator
(\ref{eq:hathatdef}) for smaller and smaller values of $|q_+-q_+^{(1)}|$
and $u$ closer and closer to 1,
we find
\begin {equation}
   \numres \simeq 3.2 \,.
\end {equation}
Using the residue from (\ref{eq:residue}) in the result
(\ref{eq:depintdecay1}) for $\depint(t{=}\infty,\x)$ then yields
\begin {equation}
  \depint(t{=}\infty,\x)
  \simeq
  4 \Chargebar
   \frac{(\numres L)^2}{E} \,
   \formL(0) \,
   e^{-\numpole(2x_3)/\wbig^{1/3}}
   .
\label {eq:depintdecay2}
\end {equation}
Using (\ref{eq:deprelation}), this gives the $x^+ \gg \wbig^{1/3}$ case of
our final result (\ref{eq:final}).


\subsection {Revisiting the small-\boldmath$Q$ expression for
             \boldmath$\calGRup_{\sigma\mu}(Q,u)$}
\label {sec:revisitsmallQ}

We now return to discuss in hindsight the terms that we
dropped in the small-$Q$ form of $\calGRup_{\sigma\mu}(Q,u)$ when
replacing
\begin {align}
   \calGRup_{0\mu}(\omega,k) \, \eta^{\mu\nu} a_\nu
   &\simeq \phantom{-}\frac{\omega}{i\omega-k^2} (i a_0 + k a_3)
   - \frac{k}{i\omega-k^2} (1-u)^{1-i\omega/2}(k a_0 + \omega a_3) ,
\label {eq:checking0}
\\
   \calGRup_{3\mu}(\omega,k) \, \eta^{\mu\nu} a_\nu
   &\simeq - \frac{k}{i\omega-k^2} (i a_0 + k a_3)
   + \frac{i}{i\omega-k^2} (1-u)^{-i\omega/2}(k a_0 + \omega a_3) ,
\label {eq:checking3}
\end {align}
by
\begin {align}
   \calGRup_{0\mu}(\omega,k) \, \eta^{\mu\nu} a_\nu
   &\to \frac{\omega}{i\omega-k^2} (i a_0 + k a_3) ,
\label {eq:drop0}
\\
   \calGRup_{3\mu}(\omega,k) \, \eta^{\mu\nu} a_\nu
   &\to 0
\label {eq:drop3}
\end {align}
in section \ref{sec:crucialT}.
We have seen in the transition from (\ref{eq:depint2}) to
(\ref{eq:depintfinal}) how the $1/f$ factor was crucial to
get a non-negligible result from near-horizon contributions,
which in turn were crucial to get something for
$\depint(t{=}\infty,\x)$.  That $1/f$ came from a factor
$g^{\rho\mu}$ of the inverse metric in our evaluation
(\ref{eq:basicT}) of the 3-point function.  It is only
present for $g^{00}$, not for $g^{33}$.  As a result,
the $\calGRup_{3\mu}$ of (\ref{eq:checking3}) does not
produce anything significant and can be dropped as in
(\ref{eq:drop3}).
The terms of $\calGRup_{0\mu}$ are multiplied by the
$1/f$ factor in $g^{00}$, but the second term in (\ref{eq:checking0})
contains an explicit factor of $1{-}u$, which cancels the
near-horizon enhancement.  So it too can be dropped, which is
how we arrive at (\ref{eq:drop0}).


\section {Conclusion}
\label {sec:conclusion}

We have shown how to use gauge-gravity duality for retarded
3-point correlators
to solve a well-formulated gauge theory problem for studying the
stopping of high-energy jets in strongly-coupled ${\cal N}{=}4$ super
Yang Mills theory.  Focusing on jets that carry R charge, we found
more than one scale associated with the stopping of that charge, as
described in the introduction.  It would be interesting to check whether
our conclusions depend on the details of what we chose to study.
In future work, one could study different observables, such as energy
rather than R charge, and/or different sources, such as external
gravitational fields rather than external R-charge fields.
It would also be interesting to consider a source that is localized
in the transverse direction and so look at the transverse spreading
of the jet.

Our final result (\ref{eq:final}) for charge deposition is
exponentially suppressed at early times, before the first stopping scale.
In giving (\ref{eq:final}), we ignored details on distance scales
$\lesssim L$.  At early times
there is a dipole contribution $\sim \delta'_L(x^-)$,
as depicted in fig.\ \ref{fig:deposit2}.
A moving, time-dependent dipole source might possibly produce
a response that is not exponentially suppressed.
In our approximations, the early time behavior was
just the vacuum propagation of the
excitation, which does not produce any charge diffusion.
But there may be parametrically small corrections, dropped
in approximations like (\ref{eq:calGsmallQ3}), that might produce
dipole sources that do produce a small amount of diffusion originating
from early times---that is, that produce effects which are suppressed
but not exponentially suppressed.
This is another possibility for further study.


\begin{acknowledgments}

We are indebted to Austen Lamacraft for useful discussions.
This work was supported, in part, by the U.S. Department
of Energy under Grant No.~DE-FG02-97ER41027 and by a
Jeffress research grant, GF12334.

\end{acknowledgments}

\appendix

\section {Current conservation and total charge}
\label{app:conservation}

In this appendix, we will review how the Ward identity implies that
currents are conserved outside of the source region.
(As discussed in the main text, the current anomaly is not
relevant to the particular field theory problem we have set up,
and so we will ignore the anomaly in what follows.)
We will also use the Ward identity to give an independent
calculation of the total charge created by the source.
Finally, we will show in detail how the Ward identity is
respected by some of our main formulas and approximations
in the gravity calculation.  (We found such derivations very useful
in the early stages of research as a debugging tool for our
calculations.)


\subsection{General}

Start from the basic formula (\ref{eq:Rf0}) for the response in
terms of the 3-point function.
To investigate $\partial_\mu \resp$, we need
to know $Q^\mu G^{\rm R}_{\perp\perp\mu}$.
The Ward identity tells us that
\begin {equation}
  \partial_\mu G_{\rm R}^{(abc)\alpha\beta\mu}(x_1,x_2;x)
  =
   f^{abc} \Bigl[
      \delta^{(4)}(x-x_1) \, G_{\rm R}^{\alpha\beta}(x_1-x_2)
      -
      \delta^{(4)}(x-x_2) \, G_{\rm R}^{\beta\alpha}(x_2-x_1)
   \Bigr]
\label {eq:Ward}
\end {equation}
or equivalently [using $G_{\rm R}(-x)=G_{\rm A}(x)$]
\begin {equation}
  i Q_\mu G_{\rm R}^{(abc)\alpha\beta\mu}(Q_1,Q_2;Q)
  = f^{abc} \Bigl[
      G_{\rm A}^{\beta\alpha}(Q_2)
      -
      G_{\rm A}^{\alpha\beta}(Q_1)
  \Bigr] ,
\label {eq:Ward2}
\end {equation}
where both sides are of course multiplied by a momentum-conserving
$\delta^{(4)}(Q_1{+}Q_2{+}Q)$.
The form (\ref{eq:Ward2})
of the Ward identity is familiar except perhaps for the
details of retarded versus advanced prescriptions.  These can
be quickly deduced by starting from the imaginary-time Ward
identity and then analytically continuing in frequency according to
(\ref{eq:G3continue}).%
\footnote{
  Alternatively, the derivation of (\ref{eq:Ward}) directly in
  real time follows by applying $\partial_\mu$ to
  $
  i^2 G_{\rm R}^{(abc)\alpha\beta\mu}(x_1,x_2;x)
  = \theta(t-t_2) \, \theta(t_2-t_1) \,
       \bigl\langle [[j^{c\mu}(x),j^{b\beta}(x_2)],j^{a\alpha}(x_1)]
       \bigr\rangle
  + \theta(t-t_1) \, \theta(t_1-t_2) \,
       \bigl\langle [[j^{c\mu}(x),j^{a\alpha}(x_1)],j^{b\beta}(x_2)]
       \bigr\rangle
  $,
  using $\partial_0\theta(t-t_i)=\delta(t-t_i)$ and
  the operator identity $\partial_\mu j^\mu = 0$ (ignoring
  the anomaly), and using the equal-time current algebra commutation
  relations
  $
   [j^{a0}(t,\x), j^{b\mu}(t,\y)]
   = i f^{abc} \, j^{c\mu}(t,\y) \, \delta^{(3)}(\x-\y)
  $.
}

Because of the $\delta$ functions in (\ref{eq:Ward}),
\begin {equation}
  \partial_\mu \response
  =
  \tfrac12
  \int d^4x_1 \> d^4x_2 \>
  \partial_\mu G_{\rm R}^{(ab3)\alpha\beta\mu}(x_1,x_2;x) \,
  A^a_{\alpha,\rm cl}(x_1) \,
  A^b_{\beta,\rm cl}(x_2)
\end {equation}
will vanish for $x$ outside of the source region.


\subsection {The total charge created}
\label {app:charge}

The total charge created is given by
\begin {equation}
   \chargec = \int d^4x \> \partial_\mu \responsec .
\label {eq:dQstart}
\end {equation}
Using (\ref{eq:Rf0}) and the Ward identity (\ref{eq:Ward2}),
\begin {equation}
  \partial_\mu \responsec
  =
  \tfrac12
  f^{abc}
  \int_{Q_1 Q_2}
  \bigl[ 
     G_{\rm A}^{\beta\alpha}(Q_2) - G_{\rm A}^{\alpha\beta}(Q_1)
  \bigr]
  A^{a\econj}_{\alpha,\rm cl}(Q_1) \, A^{b\econj}_{\beta,\rm cl}(Q_2) \,
  e^{-i Q_1\cdot x} e^{-i Q_2\cdot x} .
\end {equation}
Then
\begin {equation}
  \chargec =
  - \tfrac12 f^{abc} \int_{Q_1} 
  \bigl[ 
     G_{\rm A}^{\alpha\beta}(Q_1) - G_{\rm A}^{\beta\alpha}(-Q_1)
  \bigr]
  A^{a\econj}_{\alpha,\rm cl}(Q_1) \, A^{b}_{\beta,\rm cl}(Q_1) .
\end {equation}
Now specialize to
\begin {equation}
  A^a_{\alpha,\rm cl}(q) \equiv \pol_\alpha \, A^a_{\perp,\rm cl}(q)
\end {equation}
to get
\begin {align}
  \chargec &=
  - \tfrac12 f^{abc}\int_{Q_1} 
  \bigl[ 
     G^{\rm A}_\perp(Q_1) - G^{\rm A}_\perp(-Q_1)
  \bigr]
  A^{a\econj}_{\perp,\rm cl}(Q_1) \, A^b_{\perp,\rm cl}(Q_1)
\nonumber\\
  &=
  - \tfrac{i}{2} f^{abc} \int_{Q_1} 
  \rho^\NR_\perp(Q_1)
  A^{a\econj}_{\perp,\rm cl}(Q_1) \, A^b_{\perp,\rm cl}(Q_1) ,
\label {eq:charge2}
\end {align}
where we have used the relations $G^{\rm A}(-q) = G^{\rm A}(q)^*$
and $\Im G^{\rm A}(q) = \tfrac12 \rho^\NR(q)$.  Here
$\rho^\NR$ is the spectral density with non-relativistic (NR) sign
convention, related to the standard relativistic sign convention
by
\begin {equation}
  \rho^\NR(q) = \sgn(q^0) \, \rho^{\rm rel}(q) .
\end {equation}

To extract $\rho_\perp$, we need the Green function $G_\perp$.
Since the source momenta $Q_1$ are very large (of order $\kbig$ and
so with components $\gg T$), we may use the vacuum result for
the Green function, which is
\begin {equation}
   G_\perp
   = - \frac{1}{\gSG^2} \, \lim_{z\to0} z^{-1} \partial_z{\cal G}_\perp
   = - \frac{1}{2\gSG^2} \, \lim_{\ub\to0} \partial_{\ub}{\cal G}_\perp
\label {eq:Glimz}
\end {equation}
with (\ref{eq:Gperpzero}) for ${\cal G}_\perp$.
This gives
\begin {equation}
   G_\perp(q)
   = - \frac{q^2}{2\gSG^2} \bigl[
        \ln(\ub q^2) + 2\gammaE
     \bigr] .
\end {equation}
The advanced prescription is $q^2 \to q^2 + i\epsilon\sgn(q^0)$,
giving%
\footnote{
  A good check of overall sign is that the spectral density $\rho^\NR$
  should be
  positive for positive frequency.
}
\begin {equation}
   \rho^\NR(q) = 2 \,\Im G^{\rm A}_\perp(q)
   = \frac{\pi (-q^2)}{\gSG^2} \,\theta(-q^2) \sgn(q^0) .
\end {equation}
Using this in (\ref{eq:charge2}),
\begin {equation}
  \chargec =
  \frac{i\pi}{2\gSG^2} \, f^{abc} \int_{Q_1} 
  Q_1^2 \sgn(\omega_1) \, \theta(-Q_1^2) \,
  A^{a\econj}_{\perp,\rm cl}(Q_1) \, A^b_{\perp,\rm cl}(Q_1) .
\end {equation}
Now use the explicit form (\ref{eq:source}) for the source
and $f^{{-}{+}3}=2i$:
\begin {align}
  \charge
  &\simeq
  - \frac{\pi\Aamp^2}{\gSG^2} \int_{Q_1} 
  Q_1^2 \sgn(\omega_1) \, \theta(-Q_1^2) \,
  \left[
    |\tilde\envelope(Q_1-\kbig)|^2 - |\tilde\envelope(Q_1+\kbig)|^2
  \right]
\nonumber\\
  &=
  - \frac{2\pi\Aamp^2}{\gSG^2} \int_{Q_1} 
  Q_1^2 \sgn(\omega_1) \, \theta(-Q_1^2) \,
  |\tilde\envelope(Q_1-\kbig)|^2 .
\end {align}
Since
$\kbig$ is large and $\tilde\envelope(Q_1-\kbig)$ localizes $Q_1$ to be
near $\kbig$, we may set $\sgn(\omega_1) = +1$ and
$Q_1^2 \simeq 4\wbig q_+$, giving
\begin {equation}
  \charge
  \simeq
  \frac{8\pi\wbig\Aamp^2}{\gSG^2} \int_{q}
  \theta(-q_+) \, |q_+| \,
  |\tilde\envelope(q-\kbig)|^2
  \simeq
  \frac{8\pi\wbig\Aamp^2}{\gSG^2} \int_{q}
  \theta(-q_+) \, |q_+| \,
  |\tilde\envelope(q)|^2 .
\end {equation}
Then, for a transverse-translational invariant source,
\begin {equation}
   \chargebar \equiv \frac{\charge}{V_\perp}
   \simeq
   \frac{8\pi\wbig\Aamp^2}{\gSG^2} \int \frac{2 dq_+ \, dq_-}{(2\pi)^2}
   \> \theta(-q_+) \, |q_+|
   \bigl|\tilde\envelope^{(2)}(q_+,q_-)\bigr|^2 ,
\label {eq:chargeA}
\end {equation}
in agreement with (\ref{eq:charge}).


\subsection{Ward identity for (\ref{eq:basic})}

Here we will check that the basic formula (\ref{eq:epsGR})
for the 3-point function satisfies the Ward identity.
To investigate $Q^\mu G^{\rm R}_{\perp\perp\mu}$ we need
$Q^\mu \calGRup_{\sigma\mu}(Q,x^5)$.  The latter is
the response $A_\sigma(Q,x^5)$ to a boundary perturbation
that is $A_\mu(Q,0) = Q_\mu$.  But this boundary perturbation can be
gauged away ($A_I \to A_I - \partial_I \lambda$)
while remaining in $A_5{=}0$ gauge by the $x^5$-independent
transformation $\tilde\lambda(Q,x^5) = i$.  The response to zero boundary
perturbation is $A_I = 0$.  Gauge transforming back gives the
response $A_\mu = Q_\mu$, independent of $x^5$.  Therefore
\begin {equation}
   Q^\mu \calGRup_{\sigma\mu}(Q,x^5) = Q_\sigma
   \qquad \mbox{($A_5{=}0$ gauge)} .
\label {eq:calGdotQ}
\end {equation}
Then, using $Q=-Q_1-Q_2$, (\ref{eq:epsGR}) gives
\begin {equation}
   i Q^\mu G^{\rm R}_{\perp\perp\mu}
   =
   - \frac{1}{\gSG^2 R} \int d(x^5)\> \sqrt{-g} \,
   g^{\perp\perp} \, g^{\rho\sigma} 
     (Q_{1\rho} Q_{1\sigma} - Q_{2\rho} Q_{2\sigma}) \,
     \calGAup_{\perp}(Q_1,x^5) \,
     \calGAup_{\perp}(Q_2,x^5)
     ,
\label {eq:QG1}
\end {equation}
where $g^{\perp\perp} \equiv \pol_I g^{IJ} \pol_J$.

Now consider the transverse equation of motion $\nabla_I F^{I\perp}=0$.
For transverse-translational invariant sources,
we will only need the Green function
for $\Q_{1\perp}{=}\Q_{2\perp}{=}0$.  Using the fact that the
metric coefficients depends only on $u$, we can write the equation
of motion as
\begin {align}
   0
   &= \frac{1}{\sqrt{-g}} \, \partial_I \Bigl(
        \sqrt{-g} \, g^{\perp\perp} g^{IJ} 
        (\partial_J A_\perp - \partial_\perp A_J)
      \Bigr)
\nonumber\\
   &= \frac{1}{\sqrt{-g}} \, \partial_5 \Bigl(
        \sqrt{-g} \, g^{\perp\perp} g^{55} \partial_5 A_\perp
      \Bigr)
      -
      g^{\perp\perp} g^{\mu\nu} q_\mu q_\nu A_\perp
      ,
\end {align}
and so
\begin {equation}
   \sqrt{-g} \, g^{\perp\perp} g^{\mu\nu} q_\mu q_\nu \, {\cal G}_\perp(q,x^5)
   =
   \partial_5 \Bigl(
        \sqrt{-g} \, g^{\perp\perp} g^{55} \partial_5 \,
        {\cal G}_\perp(q,x^5)
      \Bigr)
   .
\end {equation}
Using this in (\ref{eq:QG1}) gives
\begin {align}
   i Q^\mu G^{\rm R}_{\perp\perp\mu}
   &=
   - \frac{1}{\gSG^2 R} \int d(x^5) \Bigl[
      \partial_5 \Bigl(
        \sqrt{-g} \, g^{\perp\perp} g^{55} \partial_5 \,
        \calGAup_\perp(Q_1,x^5)
      \Bigr) \,
      \calGAup_{\perp}(Q_2,x^5)
\nonumber\\
  & \qquad
      -
      \calGAup_\perp(Q_1,x^5) \,
      \partial_5 \Bigl(
        \sqrt{-g} \, g^{\perp\perp} g^{55} \partial_5 \,
        \calGAup_\perp(Q_2,x^5)
      \Bigr)
   \Bigr]
\nonumber\\
  & =
   \frac{1}{\gSG^2 R} \int d(x^5) \>
      \partial_5 \Bigl(
        \sqrt{-g} \, g^{\perp\perp} g^{55}\,
        \calGAup_\perp(Q_1,x^5) \,
        \tensor\partial_5 \,
        \calGAup_{\perp}(Q_2,x^5)
      \Bigr)
\nonumber\\
  & =
   - \frac{1}{\gSG^2 R} \Bigl(
        \sqrt{-g} \, g^{\perp\perp} g^{55}\,
        \calGAup_\perp(Q_1,x^5) \,
        \tensor\partial_5 \,
        \calGAup_{\perp}(Q_2,x^5)
      \Bigr)_{\rm boundary} ,
\end {align}
where the last step implicitly assumes that the integral is sufficiently
convergent that there is no contribution from the horizon
(or $\ub{\to}\infty$ in the zero temperature case).
At the boundary,
\begin {equation}
   {\cal G}(q,x^5) \to 1
\end {equation}
and
\begin {equation}
   \frac{1}{\gSG^2R} \,
   \sqrt{-g} \, g^{\perp\perp} g^{55} \partial_5 {\cal G}(q,x^5)
   \to -G(q) .
\end {equation}
[For instance, in the AdS metric (\ref{eq:metricz}), the last is
the usual expression (\ref{eq:Glimz}).]
So we recover the Ward identity
\begin {equation}
   i Q^\mu G^{\rm R}_{\perp\perp\mu}
   = G^{\rm A}_\perp(Q_2) - G^{\rm A}_\perp(Q_1) .
\end {equation}


\subsection {Current conservation of (\ref{eq:approx1})}

As a final example, consider the zero temperature expression
(\ref{eq:RfF0}) in terms of $\Field(x,\ub)$.
Starting from (\ref{eq:RfF0}),
\begin {equation}
  \partial_\mu \jmu
  \propto
  \int_{0}^\infty \frac{d\ub}{\ub} \, \left[
    (\eta^{\mu\nu} \partial_\mu \partial_\nu \Field)^* \Field
    -
    \Field^* (\eta^{\mu\nu} \partial_\mu \partial_\nu \Field_0)
  \right] .
\end {equation}
The equation of motion $\nabla_I F^{I\perp}=0$ for $\Field$ is
\begin {equation}
  \eta^{\mu\nu} \partial_\mu \partial_\nu \Field
  = - \ub \partial_{\ub}^2 \Field .
\label {eq:F0diff}
\end {equation}
Combining the last two equations,
\begin {align}
  \partial_\mu \jmu
  &\propto
  \int_{0}^\infty d\ub \left[
    -(\partial_{\ub}^2 \Field^* ) \Field
    +
    \Field^* \partial_{\ub}^2 \Field
  \right]
\nonumber\\
  &= \int_{0}^\infty d\ub \>
    \partial_{\ub} \left(
      \Field^* \tensor\partial_{\ub} \Field
    \right)
\nonumber\\
  &= - \left[ \Field^* \tensor\partial_{\ub}\Field \right]_{\ub\to0}
\nonumber\\
  &= \Field(x,0) \, \partial_{\ub} \Field^*(x,0)
     - {\rm h.c.}
\label {eq:F0conserve}
\end {align}
Since $\Field$ is proportional to the source on the boundary, the first
factor $\Field(x,0)$ vanishes outside the source region, verifying that
current conservation holds there.


\section {Evaluation of \boldmath$I(s)$}
\label {app:I}

The first thing to notice is that the definition
\begin {equation}
  I(s) \equiv
  \int \frac{d\kappa}{2\pi} \,
  \calGRup_{\perp}(\kappa) \,
   e^{-\epsilon \kappa^2} e^{i\kappa s}
\label {eq:appIdef}
\end {equation}
of $I(s)$ gives zero for $s < 0$.  The argument is basically to close
the integration contour in the lower half complex $\kappa$ plane,
where $\calGRup_\perp$ has no singularities.
Technically, one has to be a little careful because the convergence factor
$e^{-\epsilon\kappa^2}$ does not converge for
$-3\pi/4 < \arg(\kappa) < -\pi/4$.
This problem can be avoided by
first deforming the integration contour to run, for
example, from $e^{-i7\pi/8} \infty$ to the origin to $e^{-i\pi/8}\infty$.
At that stage, the $e^{i\kappa s}$
factor produces a convergent integrand, and one may drop the
the now superfluous $e^{-\epsilon \kappa^2}$ convergence factor.
Then one can close the integration contour at infinity,

For $s>0$, it's possible to evaluate the integral (\ref{eq:appIdef}) defining
$I(s)$ directly, by
various contour deformation arguments and series expansions of
Bessel functions.  However, there is a simpler way using the
equation of motion
\begin {equation}
   \Bigl(
      \partial_{\ub}^2
       - \frac{q^2}{\ub}
   \Bigr) {\cal G}_\perp
   = 0
\end {equation}
satisfied by ${\cal G}_\perp$ at zero temperature.
In terms of $\kappa \equiv \ub q^2$, this equation is
\begin {equation}
   {\cal G}_\perp =
   \kappa \partial_\kappa^2 {\cal G}_\perp .
\end {equation}
We can use this to rewrite (\ref{eq:appIdef}) as
\begin {equation}
  I(s) =
  \int \frac{d\kappa}{2\pi} \,
  \kappa \bigl[\partial_\kappa^2 \calGRup_{\perp}(\kappa)\bigr]
   e^{-\epsilon \kappa^2} e^{i\kappa s} .
\end {equation}
Integrating by parts twice,
\begin {equation}
  I(s) =
  \int \frac{d\kappa}{2\pi} \,
  \calGRup_{\perp}(\kappa) \,
  (-s^2\kappa +2 i s) e^{-\epsilon \kappa^2} e^{i\kappa s} ,
\end {equation}
which can be rewritten
\begin {equation}
  I(s) = i (s^2\partial_s + 2 s) \,I(s) .
\end {equation}
This differential equation is trivial to solve, giving
\begin {equation}
   I(s) \propto \frac{1}{s^2} \, e^{i/s} .
\end {equation}
All that remains is to fix the overall proportionality constant
by evaluating the original integral for some convenient value of
$s$.  This can be done for small $s$ by the saddle point method
of section \ref{sec:steep0} (taking $L=0$), or it can be done
by evaluating the integral (\ref{eq:appIdef}) for large $s$ by
changing integration variables from $\kappa$ to
$\lambda \equiv \kappa s$ and then expanding the integrand in powers
of $1/s$.  Either way, one obtains the result (\ref{eq:Iresult}).


\section {Small \boldmath$Q$ form of \boldmath$\calGRup$}
\label {app:smallQ}

The low-$Q$ behavior of the vector bulk-to-boundary propagator
has been analyzed previously \cite{ViscReview} but not put into exactly
the form that we need.  What has generally been presented are the
derivatives
$\partial_u \calGRup_{0\mu}$ and $\partial_u \calGRup_{3\mu}$,
whereas in this paper we want $\calGRup_{0\mu}$ and
$\calGRup_{3\mu}$ directly, in $A_5{=}0$ gauge.
It is easy to integrate, however,
and determine the constants of integration.
From Ref.\ \cite{ViscReview},%
\footnote{
  See specifically eqs.\ 73--78 of Ref.\ \cite{ViscReview}.
}
keeping only the leading-order terms and
for the sake of notational brevity writing $A_\sigma$ for our
$\calGRup_{\sigma\mu}(\omega,k) \, \eta^{\mu\nu} a_\nu$,
\begin {equation}
   \partial_u A_0
   \simeq
   \frac{k}{i\omega-k^2} (1-u)^{-i\omega/2}(k a_0 + \omega a_3) ,
\end {equation}
and
\begin {align}
   \partial_u A_3
   =
   - \frac{\omega}{kf} \, \partial_u A_0
   &\simeq
   -\frac{\omega}{i\omega-k^2} \frac{(1-u)^{-i\omega/2}}{1-u^2}
       (k a_0 + \omega a_3) .
\nonumber\\
   &=
   -\frac{\omega}{2(i\omega-k^2)}
   \left[
       (1-u)^{-1-i\omega/2}
       + \frac{(1-u)^{-i\omega/2}}{1+u}
   \right]
   (k a_0 + \omega a_3) .
\label {eq:smallQa}
\end {align}
Integration gives
\begin {equation}
   A_0
   \simeq
   C_0(\omega,k) -
   \frac{k}{i\omega-k^2} (1-u)^{1-i\omega/2}(k a_0 + \omega a_3) ,
\end {equation}
and
\begin {equation}
   A_3
   \simeq
   C_3(\omega,k) +
   \frac{i}{i\omega-k^2} (1-u)^{-i\omega/2}
       (k a_0 + \omega a_3) ,
\end {equation}
where the integral of the $(1-u)^{-i\omega/2}/(1+u)$ term from
(\ref{eq:smallQa}) has been dropped because that term's
integral is sub-leading in
powers of $\omega$ and $k^2$.
The integration constants $C_0$ and $C_3$ are
constrained by the fact that $A_0$ and $A_3$ must satisfy the
$A_0$ equation of motion, which is
\begin {equation}
   \partial_u^2 A_0 - \frac{k}{uf} (k A_0 + \omega A_3) = 0 ,
\end {equation}
and so $k C_0 + \omega C_3 = 0$.  They are also constrained
by the boundary normalization that $A_\mu \to a_\mu$.
These constraints give (\ref{eq:calGperpsmallQ}).


\section {The WKB exponent \boldmath$S$ for  $\calGRup_\perp$}
\label {app:S}

Separately expand in powers of $u'$ the
factors of $f^{-1}$ and $[{u'}^2\q^2 - q^2]^{1/2}$ in the
second form of the integrand in
(\ref{eq:S}).  Integrating term by term then yields
\begin {subequations}
\label {eq:SsmallAll}
\begin {equation}
  S = 2 u^{1/2} (-q^2)^{1/2} \,
      F_1\Bigl(
        \tfrac14;-\tfrac12,1;\tfrac54;
        \frac{u^2\q^2}{q^2}, u^2
      \Bigr) ,
\end {equation}
where
\begin {equation}
   F_1(\alpha,\beta,\beta';\gamma;x,y) =
   \sum_{m=0}^\infty \sum_{n=0}^\infty
   \frac{(\alpha)_{m+n}(\beta)_m(\beta')_n}{(\gamma)_{m+n}m!n!} x^m y^n
\label {eq:AppellF1}
\end {equation}
\end {subequations}
is the Appell hypergeometric function of two
variables.
The first term in this expansion gives (\ref{eq:Ssmall}).

Rewriting the $m$ sum in (\ref{eq:AppellF1}) as a hypergeometric
function $F\equiv{}_2F_1$ gives the expansion
\begin {equation}
   S =
   \tfrac12 (-u q^2)^{1/2} \sum_{n=0}^\infty
     \frac{
       \hyperF\bigl(-\tfrac12,n+\tfrac14;n+\tfrac54;\frac{u^2\q^2}{q^2}\bigr)
     }{n+\tfrac14} \, u^{2n} .
\label {eq:Sexpansion1}
\end {equation}
The standard hypergeometric transformation
\begin {equation}
   \hyperF(\alpha,\beta;\gamma;z) =
   \frac{\Gamma(\gamma)\,\Gamma(\beta{-}\alpha)}
        {\Gamma(\beta)\,\Gamma(\gamma{-}\alpha)} \,
   (-z)^{-\alpha} \,
   \hyperF(\alpha,\alpha{+}1{-}\gamma;\alpha{+}1{-}\beta; \frac{1}{z} )
   + (\alpha{\leftrightarrow}\beta)
\end {equation}
gives
\begin {multline}
   \hyperF\bigl(-\tfrac12,n+\tfrac14;n+\tfrac54;\frac{u^2\q^2}{q^2}\bigr)
\\
   =
   \frac{n+\tfrac14}{n+\frac34} \,
   \hyperF\bigl(-\tfrac12,-n-\tfrac34;-n+\tfrac14;\frac{q^2}{u^2\q^2}\bigr)
     \left( \frac{u^2\q^2}{-q^2} \right)^{1/2}
   - \frac{\Gamma(n+\tfrac54)\Gamma(-n-\tfrac34)}{2\pi^{1/2}}
     \left( \frac{-q^2}{u^2\q^2} \right)^{n+\tfrac14} ,
\end {multline}
with which we can rewrite (\ref{eq:Sexpansion1}) as
\begin {multline}
   S =
   \tfrac12 u^{3/2} |\q| \sum_{n=0}^\infty
     \frac{
      \hyperF\bigl(-\tfrac12,-n-\tfrac34;-n+\tfrac14;\frac{q^2}{u^2\q^2}\bigr)
     }{n+\tfrac34} \, u^{2n}
\\
   - \frac{|\q|}{4\pi^{1/2}} 
       \left( \frac{-q^2}{\q^2} \right)^{\tfrac34}
       \sum_{n=0}^\infty
       \Gamma(n+\tfrac14)\Gamma(-n-\tfrac34)
       \left( \frac{-q^2}{\q^2} \right)^n .
\end {multline}
Expand the hypergeometric function as
\begin {multline}
   \tfrac12 u^{3/2} |\q|
   \sum_{n=0}^\infty
     \frac{
      \hyperF\bigl(-\tfrac12,-n-\tfrac34;-n+\tfrac14;\frac{q^2}{u^2\q^2}\bigr)
     }{n+\tfrac34} \, u^{2n}
\\
   =
   \tfrac12 u^{3/2} |\q|
   \sum_{m=0}^\infty
   \sum_{n=0}^\infty
   \frac{(-\frac12)_m}{m!(n-m+\frac34)}
   \left(\frac{q^2}{u^2\q^2}\right)^m u^{2n}
   .
\end {multline}
Then rewrite the $n \ge m$ part of the $n$
sum as a sum over $r \equiv n-m$, and use
\begin {equation}
   |\q|
   \sum_{m=0}^\infty
   \frac{(-\frac12)_m}{m!}
   \left(\frac{q^2}{\q^2}\right)^m
   =
   \sqrt{\q^2 - q^2}
   = \omega
\end {equation}
and
\begin {equation}
   \tfrac12 u^{3/2}
   \sum_{r=0}^\infty
   \frac{u^{2r}}{r+\frac34}
   =
   \int_0^{u} du' \>
      \frac{{u'}^{1/2}}{1-{u'}^2}
   = \tauo(u)
\end {equation}
to obtain
\begin {align}
  S =
  \omega \, \tauo(u)
  & + \tfrac43 \, \num \,
      |\q| \left( \frac{-q^2}{4\q^2} \right)^{3/4}
      \hyperF\bigl( \tfrac14, 1; \tfrac74; \frac{q^2}{\q^2} \bigr)
\nonumber\\
  & + \tfrac12 u^{3/2} |\q|
      \sum_{m=1}^\infty
      \frac{(-\frac12)_m}{m!}
      \left(
        \sum_{n=0}^{m-1}
        \frac{u^{2n}}{n-m+\frac34}
      \right)
      \left(\frac{q^2}{u^2\q^2}\right)^m
      .
\label {eq:SlargeAll}
\end {align}
The first terms of this expansion give (\ref{eq:Slarge}).

We have presented the expansion (\ref{eq:SlargeAll}), useful when
$u \gg u_\star$, as a series of tricks starting from the
complementary expansion (\ref{eq:SsmallAll}) useful when
$u \ll u_\star \ll 1$.  It is also possible to derive
(\ref{eq:SlargeAll}) directly from the integral (\ref{eq:S})
by appropriate expansions for $u \gg u_\star$.  We will not
reproduce here the full derivation of (\ref{eq:SlargeAll}) from this
starting point, but the origin of the first terms (\ref{eq:Slarge})
is easy to explain.  Rewrite (\ref{eq:S}) as
\begin {equation}
  S =
  \int_0^{\mu} du' \>
      \frac{ [{u'}^2\q^2 - q^2]^{1/2} }{ {u'}^{1/2} \, f(u') }
  + \int_\mu^{u} du' \>
      \frac{ [{u'}^2\q^2 - q^2]^{1/2} }{ {u'}^{1/2} \, f(u') } ,
\end {equation}
where $\mu$ is an arbitrary scale with
$u_\star \ll \mu \ll u$.  Then approximate as
\begin {align}
  S &\simeq
  \int_0^{\mu} du' \>
      \frac{ [{u'}^2\q^2 - q^2]^{1/2} }{ {u'}^{1/2} }
  + \int_\mu^{u} du' \>
      \frac{ [{u'}^2\q^2]^{1/2} }{ {u'}^{1/2} \, f(u') }
\nonumber\\
  &=
  S_{\rm nonanalytic}
  + \int_0^{u} du' \>
      \frac{ [{u'}^2\q^2]^{1/2} }{ {u'}^{1/2} \, f(u') }
\nonumber\\
  &=
  S_{\rm nonanalytic}
  + |\q| \, \tauo(u) ,
\label {eq:Sint1}
\end {align}
where
\begin {equation}
  S_{\rm nonanalytic} =
  \int_0^{\mu} du' \>
      \frac{ [{u'}^2\q^2 - q^2]^{1/2} - [{u'}^2\q^2]^{1/2}}{ {u'}^{1/2} }
\end {equation}
will not be analytic in $q^2$ (because a naive expansion of the
integrand in $q^2$ leads to integrals with $u{\to}0$ divergences).
A simple way to evaluate $S_{\rm nonanalytic}$ is
to evaluate its derivative with respect to $q^2$ and then
integrate back:
\begin {align}
  \frac{\partial S_{\rm nonanalytic}}{\partial(q^2)} &=
  -\tfrac12 \int_0^{\mu} du' \>
      \frac{ [{u'}^2\q^2 - q^2]^{-1/2}}{ {u'}^{1/2} }
\nonumber\\
  &\simeq
  -\tfrac12 \int_0^{\infty} du' \>
      \frac{ [{u'}^2\q^2 - q^2]^{-1/2}}{ {u'}^{1/2} }
\nonumber\\
  &=
  - 2^{-3/2} \num |\q|^{-1/2} (-q^2)^{-1/4} ,
\end {align}
and so
\begin {equation}
  S_{\rm nonanalytic} \simeq
  \tfrac43 \num |\q|^{-1/2} (-\tfrac14 q^2)^{3/4} .
\label {eq:Snonanal}
\end {equation}
Approximating $|\q| \simeq \omega$ in (\ref{eq:Sint1}) and
(\ref{eq:Snonanal}) then gives the leading terms shown in
(\ref{eq:Slarge}).


\section {Saddle point integration contours}
\label {app:contour}

\begin {figure}
\begin {center}
  \includegraphics[scale=0.4]{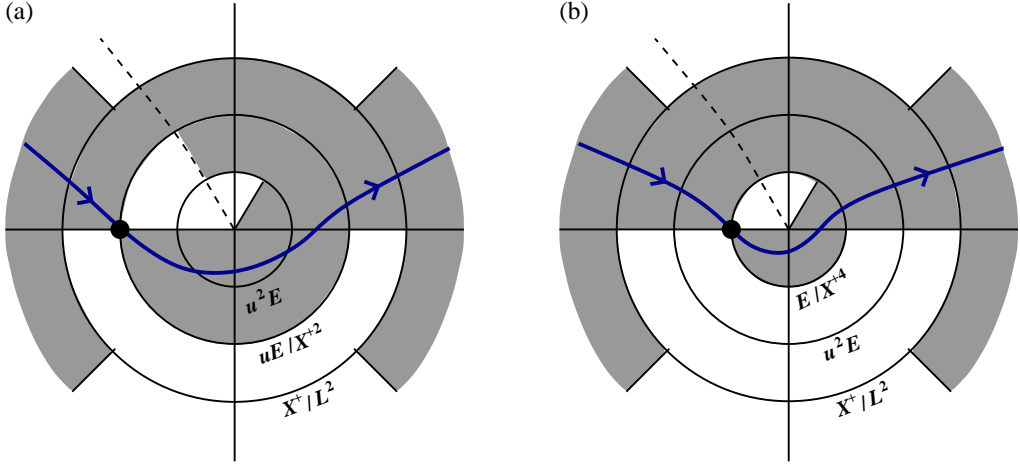}
  \caption{
     \label{fig:contour}
     Integration contours in the $q_+$ complex plane
     for saddle point approximations in (a) Case A
     and (b) Case B.  The location of the saddle point is
     marked by the large dot.
  }
\end {center}
\end {figure}

When making a saddle point approximation, one should choose an
integration contour such that the contributions to the integral
are negligible everywhere except in the neighborhood of the
saddle point.
Fig.\ \ref{fig:contour} shows examples of integration contours
that do the job for Case A and Case B analyzed in section
\ref{sec:SteepDescent}.  (We consider here a Gaussian envelope function
for the sake of concreteness.)  The dashed line depicts the line of poles
discussed in section \ref{sec:tail}, which also serve as the
location for the cut of the $(-u \wbig q_+)^{1/2}$ and
$\wbig^{1/4} (-q_+)^{3/4}$ terms in the WKB exponents
(\ref{eq:Ssmall}) and (\ref{eq:Slarge}).  The integrand
of the $q_+$ integral in (\ref{eq:F0approxT}) will be
exponentially suppressed compared to the saddle point
(shown by the large dot) in the interior of the shaded region.
The various circles indicate
different scales for $|q_+|$, as labeled.

The reasons for the suppression are
different in different regions.  We will go through Case B as an
example.
We get exponential suppression when
\begin {equation}
   {\cal S}(q_+,u) \simeq
      - i \tfrac43 \num \wbig^{1/4} (-q_+)^{3/4}
      - iq_+ X^+ + (q_+ L)^2
\label {eq:ScontourB}
\end {equation}
has a positive real part $\gg 1$.
In the interior of the smallest circle
($|q_+|\ll |q_+^\star| \sim \wbig/(X^+)^4$)
in fig.\ \ref{fig:contour}b, the
$-i \wbig^{1/4} (-q_+)^{3/4}$ term
dominates, and the shaded part shows where its real part is positive.
In the next annulus ($\wbig/(X^+)^4 \ll |q_+| \ll u^2 \wbig$),
the $-iq_+ X^+$ term dominates, and the shaded part shows when
its real part is positive.  In the next annulus out
($u^2 \wbig \ll |q_+| \ll X^+/L^2$), $|q_+|$ is large enough
that the expansion (\ref{eq:ScontourB}) is no longer appropriate,
and we should switch from (\ref{eq:Slarge}) to (\ref{eq:Ssmall}),
giving
\begin {equation}
   {\cal S}(q_+,u) \simeq
      - i 4( - u \wbig q_+)^{1/2}
      - iq_+ X^+ + (q_+ L)^2
\end {equation}
as in (\ref{eq:calSzero}).  But the $-iq_+ X^+$ term still dominates.
Finally, beyond the outermost circle
($|q_+| \gg X^+/L^2$), the $(q_+ L)^2$ term dominates.

\end {document}